\documentclass[a4paper,UKenglish,authorcolumns,cleveref, autoref, thm-restate]{lipics-v2021}\nolinenumbers

\usepackage{booktabs}   \usepackage{subcaption} \usepackage{bussproofs}
\usepackage[cal=boondoxo]{mathalfa}
\DeclareMathAlphabet{\mathpzc}{OT1}{pzc}{m}{it}
\usepackage{amsmath}
\usepackage{color}
\usepackage{xspace}

\newcommand{\ignore}[1]{{}}

\newcommand{\syntaxDef}[3]{\rulebox{\syntaxKeyword$#1\mathrel{::=}{#2}$ \ifthenelse{\equal{#3}{}}{}{[#3]}}}

\makeatletter
\newcommand{\shorteq}{\settowidth{\@tempdima}{-}\resizebox{\@tempdima}{\height}{=}}
\makeatother

  \newcommand{\dom}[1]{\ensuremath{\mathtt{dom}(#1)}}

 \usepackage{bm}

\usepackage{xcolor}
\usepackage{colortbl}

\usepackage[utf8]{inputenc}

\usepackage{amsmath}
\usepackage{stmaryrd}
\usepackage{hyperref}

\usepackage{subcaption}
\usepackage{wrapfig}

\usepackage{xspace}
\usepackage{float}
\usepackage{balance}

\usepackage{textcomp} 

\usepackage{changepage}
\usepackage{array,etoolbox}

\preto\tabular{\setcounter{magicrownumbers}{0}}
\newcounter{magicrownumbers}

\usepackage{cleveref} 

\usepackage{wrapfig}
\usepackage{caption}

\usepackage{longtable}
\usepackage{tabu}

\usepackage{wasysym}
\usepackage{mathpartir}

\usepackage{mathtools} \usepackage{xfrac}

\usepackage[qm,braket]{qcircuit}

\newcommand{\qass}[2]{{#1}\;{\leftarrow}\;{#2}}
\newcommand{\qfor}[4]{\texttt{{for} }{#1}\texttt{ \(\in\) }[{#2},{#3})\texttt{ {\&\&} }{#4}}
\newcommand{\qif}[2]{\texttt{{if}\,(}{#1}\texttt{)\,\{}{#2}\texttt{\}}}
\newcommand{\sifq}[2]{\texttt{if}~\cn{(}{#1}\cn{)}~{#2}}
\newcommand{\qbool}[4]{\cn{(}#1\,\cn{#2}\,#3\cn{)\,@\,}#4}
\newcommand{\qboola}[3]{#1\,\cn{#2}\,#3}
\newcommand{\mmod}{\cn{\%}}
\newcommand{\Hassert}[1]{\{\ #1\ \}}
\colorlet{kwd}{black!80!green}
\definecolor{spec1}{RGB}{78, 131, 162}
\definecolor{spec0}{RGB}{66, 102, 136}
\definecolor{lespec}{RGB}{30, 80, 180}
\colorlet{spec}{lespec}
\colorlet{auto}{lespec!35!lightgray}
\colorlet{stack}{magenta}

\newcommand{\qafny}{\rulelab{Qafny}\xspace}
\newcommand{\sep}{\rulelab{Sep}\xspace}

\newcommand{\sourcelang}{\ensuremath{\mathcal{O}\textsc{qimp}}\xspace}
\newcommand{\oqasm}{\textsc{oqasm}\xspace}

\newcommand{\sqir}{SQIR\xspace}
\newcommand{\coqq}{{CoqQ}\xspace}
\newcommand{\qwire}{{Qwire}\xspace}
\newcommand{\qbricks}{{QBricks}\xspace}

\newcommand{\myparagraph}[1]{\noindent\paragraph{\textbf{#1}}}
\newcommand{\qket}[1]{\ket{\Phi(#1)}}
\usepackage{tikz}

\newcommand{\cmode}{\texttt{C}}
\newcommand{\mmode}{\texttt{M}}
\newcommand{\qmodename}{\texttt{Q}}
\newcommand{\qmode}[1]{\texttt{Q}~#1}
\newcommand{\topt}[1]{#1\;\texttt{opt}}
\newcommand{\slen}[1]{|#1|}

\newcommand{\CC}{\ensuremath{\mathbb{C}}\xspace}

\usepackage{pgf}

\usepackage{tikz} \usetikzlibrary{arrows,shapes.misc,shapes.geometric, shapes.arrows,shapes.callouts,
  shapes.gates.logic.US,
  chains,matrix,positioning,scopes,decorations.pathmorphing,decorations.text,
  decorations.pathreplacing, shadows,automata,
  fit, calc, arrows.meta
}

\tikzset{ machine/.style={
rectangle,
minimum width=25mm,
    minimum height=18mm,
    text width=24mm,
align=center,
very thick,
    draw=black,
color=black,
    fill=white,
}
}

\newcommand\wideparen[1]{\tikz[baseline=(wideArcAnchor.base)]{
    \node[inner sep=0] (wideArcAnchor) {$#1$}; 
    \coordinate (wideArcAnchorA) at ($0.9*(wideArcAnchor.north west) + 0.1*(wideArcAnchor.north east)+(0.0em,0.75ex)$);
    \coordinate (wideArcAnchorB) at ($0.1*(wideArcAnchor.north west) + 0.9*(wideArcAnchor.north east)+(0.0em,0.75ex)$);
\draw[line width=0.1ex,line cap=round] 
        ($(wideArcAnchor.north west)+(0.0em,0.1ex)$) 
            .. controls (wideArcAnchorA) and (wideArcAnchorB) ..
        ($(wideArcAnchor.north east)+(0.0em,0.1ex)$)        
    ;
}}
\newcommand{\cmsg}[1]{\wideparen{#1}}

\DeclarePairedDelimiter\abs{\lvert}{\rvert}
\DeclarePairedDelimiter\norm{\lVert}{\rVert}

\makeatletter
\let\oldabs\abs
\def\abs{\@ifstar{\oldabs}{\oldabs*}}
\let\oldnorm\norm
\def\norm{\@ifstar{\oldnorm}{\oldnorm*}}
\makeatother

\makeatletter
\DeclareRobustCommand{\vardivision}{\mathbin{\mathpalette\@vardivision\relax}}
\newcommand{\@vardivision}[2]{\reflectbox{$\m@th\smallsetminus$}}
\makeatother

\DeclareFontFamily{U} {MnSymbolC}{}
\DeclareFontShape{U}{MnSymbolC}{m}{n}{
  <-6> MnSymbolC5
  <6-7> MnSymbolC6
  <7-8> MnSymbolC7
  <8-9> MnSymbolC8
  <9-10> MnSymbolC9
  <10-12> MnSymbolC10
  <12-> MnSymbolC12}{}
\DeclareFontShape{U}{MnSymbolC}{b}{n}{
  <-6> MnSymbolC-Bold5
  <6-7> MnSymbolC-Bold6
  <7-8> MnSymbolC-Bold7
  <8-9> MnSymbolC-Bold8
  <9-10> MnSymbolC-Bold9
  <10-12> MnSymbolC-Bold10
  <12-> MnSymbolC-Bold12}{}

\DeclareSymbolFont{MnSyC} {U} {MnSymbolC}{m}{n}

\DeclareMathSymbol{\sqcupplus}{\mathbin}{MnSyC}{70}

\newcommand{\ie}{\emph{i.e.,}\xspace}

\usepackage{booktabs}

\usepackage{alltt}
\usepackage{listings,lstcoq}
\definecolor{ltblue}{rgb}{0,0.4,0.4}
\definecolor{dkblue}{rgb}{0,0.1,0.6}
\definecolor{dkgreen}{rgb}{0,0.35,0}
\definecolor{dkviolet}{rgb}{0.3,0,0.5}
\definecolor{dkred}{rgb}{0.5,0,0}
\usepackage[export]{adjustbox}

\newcommand{\code}[1]{{\small\texttt{#1}}}

\newcommand{\rulelab}[1]{{\small \textsc{#1}}}

\newcommand{\sexp}[3]{\texttt{let}~{#1}\,\texttt{=}\,{#2}~\texttt{in}~{#3}}

\newcommand{\sskip}{\texttt{\{\}}}

\newcommand{\sifb}[3]{\texttt{if}~{(#1)}~{#2}~\texttt{else}~{#3}}
\newcommand{\sqwhile}[5]{\texttt{for}~{#1 \in [#2,#3)~\texttt{\&\&}~#4}~#5}

\newcommand{\tnor}[1]{\texttt{Nor}~{#1}}
\newcommand{\tnort}{\texttt{Nor}}
\newcommand{\thad}[1]{\texttt{Had}~{#1}}
\newcommand{\thadt}{\texttt{Had}}
\newcommand{\tch}[1]{\texttt{EN}~{#1}}
\newcommand{\tcht}{\texttt{EN}}
\newcommand{\shad}[3]{\frac{1}{\sqrt{#1}}\Motimes_{j=0}^{#2}{(\ket{0}+#3\ket{1})}}
\newcommand{\shadi}[3]{\frac{1}{\sqrt{#1}}\Motimes_{i=0}^{#2}{(\ket{0}+#3\ket{1})}}
\newcommand{\ssum}[3]{\Msum_{#1}^{#2}{#3}}
\newcommand{\sch}[3]{\Msum_{j=0}^{#1}{#2 {#3}}}
\newcommand{\scha}[4]{\Msum_{j=0}^{#1}{#2\ket{#3}#4}}
\newcommand{\schai}[4]{\Msum_{i=0}^{#1}{#2\ket{#3}#4}}

\newcommand{\paren}[1]{\left(#1\right)}
\newcommand{\typing}[5]{{#1},{#2}\vdash_{#3} {#4} \triangleright{#5}}
\newcommand{\smch}[3]{#1\Msum_{k=0}^{#2}{\ket{#3}}}
\newcommand{\smich}[4]{#1\Msum_{k=0}^{#2}{#3\ket{#4}}}
\newcommand{\schk}[3]{\Msum_{k=0}^{#1}{#2\ket{#3}}}
\newcommand{\schii}[3]{\Msum_{i=0}^{#1}{#2\ket{#3}}}
\newcommand{\schd}[4]{\Msum_{#1=0}^{#2}{#3\ket{#4}}}
\newcommand{\schia}[3]{\Msum_{i=0}^{#1}{#2#3}}
\newcommand{\schi}[3]{\Msum^{#1}{#2\ket{#3}}}

\newcommand{\sord}[1]{\texttt{ord}({#1})}
\newcommand{\srnd}[1]{\texttt{rnd}({#1})}
\DeclareMathOperator*{\Motimes}{\text{\raisebox{0.25ex}{\scalebox{0.8}{$\bigotimes$}}}}
\DeclareMathOperator*{\Msum}{\text{\raisebox{0.25ex}{\scalebox{0.8}{$\sum$}}}}
\newcommand{\frz}{\mathpzc{F}}
\newcommand{\mea}{\mathpzc{M}}
\newcommand{\ufrz}{\mathpzc{U}}

\newcommand{\fivepule}[6]{#1;#2\vdash_{#3}\big{\{}#4\big{\}}\, #5 \,\big{\{}#6\big{\}}}

\newcommand{\ssassign}[3]{{#1} \xleftarrow{#2} {#3}}

\newcommand{\sdis}{\texttt{dis}}
\newcommand{\sreduce}[2]{\texttt{reduce}(#1,#2)}

\newcommand{\smea}[1]{\texttt{measure}\cn{(}#1\cn{)}}

\newcommand{\snext}[1]{#1\splus 1}
\newcommand{\sminus}{\texttt{-}}
\newcommand{\splus}{\texttt{+}}
\newcommand{\slt}{\texttt{<}}
\newcommand{\srange}[2]{[#1,#2)}

\newcommand{\sseq}[2]{{#1}\,\texttt{;}\,{#2}}

\newcommand{\inst}[3][ ]{\texttt{#2}^{#1}~{#3}}
\newcommand{\insttwo}[4][ ]{\texttt{#2}^{#1}~{#3}~{#4}}

\newcommand{\iskip}[1]{\inst{ID}{#1}}

\newcommand{\inot}[1]{\inst{X}{#1}}
\newcommand{\ictrl}[2]{\insttwo{CU}{#1}{#2}}
\newcommand{\irz}[3][ ]{\insttwo[#1]{RZ}{#2}{#3}}
\newcommand{\isr}[3][ ]{\insttwo[#1]{SR}{#2}{#3}}

\newcommand{\ilshift}[1]{\inst{Lshift}{#1}}
\newcommand{\irshift}[1]{\inst{Rshift}{#1}}
\newcommand{\irev}[1]{\inst{Rev}{#1}}
\newcommand{\iqft}[3][ ]{\insttwo[#1]{QFT}{#2}{#3}}

\newcommand{\tphi}[1]{\texttt{Phi}~{#1}}

\newcommand{\ihadh}{\texttt{H}}

\newcommand{\hsp}[1]{\mathcal{#1}}
\newcommand{\iseq}[2]{{#1}\,\texttt{;}\,{#2}}

\newcommand{\itext}[1]{\texttt{#1}}

\newcommand{\instr}{\iota}

\newcommand{\app}[3]{#2\texttt{[}{#3}\mapsto{#1}\texttt{]}}
\newcommand{\xsem}{\texttt{xg}}

\newcommand{\qsem}{\texttt{qt}}
\newcommand{\psem}{\texttt{pm}}
\newcommand{\rsem}{\texttt{rz}}
\newcommand{\rrsem}{\texttt{rrz}}

\newcommand{\csem}{\texttt{cu}}

\newcommand{\cn}[1]{\texttt{#1}}
\newcommand{\Omegasz}{\Sigma}
\newcommand{\Omegaty}{\Omega}

\newcommand{\denote}[1]{\llbracket #1 \rrbracket\xspace}
\newcommand{\dabs}[1]{|\!| #1 |\!|}
\newcommand{\tos}[1]{(\!| #1 |\!)}
\newcommand{\tob}[1]{[\!\!( #1 )\!\!]}
\newcommand{\qfun}[2]{#1\langle #2 \rangle}
\newcommand{\tov}[1]{\{\hspace{-0.2em}| #1 |\hspace{-0.2em}\}}

\usepackage{newunicodechar}
\let\Alpha=A
\let\Beta=B
\let\Epsilon=E
\let\Zeta=Z
\let\Eta=H
\let\Iota=I
\let\Kappa=K
\let\Mu=M
\let\Nu=N
\let\Omicron=O
\let\omicron=o
\let\Rho=P
\let\Tau=T
\let\Chi=X

\newunicodechar{Α}{\ensuremath{\Alpha}}
\newunicodechar{α}{\ensuremath{\alpha}}
\newunicodechar{Β}{\ensuremath{\Beta}}
\newunicodechar{β}{\ensuremath{\beta}}
\newunicodechar{Γ}{\ensuremath{\Gamma}}
\newunicodechar{γ}{\ensuremath{\gamma}}
\newunicodechar{Δ}{\ensuremath{\Delta}}
\newunicodechar{δ}{\ensuremath{\delta}}
\newunicodechar{Ε}{\ensuremath{\Epsilon}}
\newunicodechar{ε}{\ensuremath{\epsilon}}
\newunicodechar{ϵ}{\ensuremath{\varepsilon}}
\newunicodechar{Ζ}{\ensuremath{\Zeta}}
\newunicodechar{ζ}{\ensuremath{\zeta}}
\newunicodechar{Η}{\ensuremath{\Eta}}
\newunicodechar{η}{\ensuremath{\eta}}
\newunicodechar{Θ}{\ensuremath{\Theta}}
\newunicodechar{θ}{\ensuremath{\theta}}
\newunicodechar{ϑ}{\ensuremath{\vartheta}}
\newunicodechar{Ι}{\ensuremath{\Iota}}
\newunicodechar{ι}{\ensuremath{\iota}}
\newunicodechar{Κ}{\ensuremath{\Kappa}}
\newunicodechar{κ}{\ensuremath{\kappa}}
\newunicodechar{Λ}{\ensuremath{\Lambda}}
\newunicodechar{λ}{\ensuremath{\lambda}}
\newunicodechar{Μ}{\ensuremath{\Mu}}
\newunicodechar{μ}{\ensuremath{\mu}}
\newunicodechar{Ν}{\ensuremath{\Nu}}
\newunicodechar{ν}{\ensuremath{\nu}}
\newunicodechar{Ξ}{\ensuremath{\Xi}}
\newunicodechar{ξ}{\ensuremath{\xi}}
\newunicodechar{Ο}{\ensuremath{\Omicron}}
\newunicodechar{ο}{\ensuremath{\omicron}}
\newunicodechar{Π}{\ensuremath{\Pi}}
\newunicodechar{π}{\ensuremath{\pi}}
\newunicodechar{ϖ}{\ensuremath{\varpi}}
\newunicodechar{Ρ}{\ensuremath{\Rho}}
\newunicodechar{ρ}{\ensuremath{\rho}}
\newunicodechar{ϱ}{\ensuremath{\varrho}}
\newunicodechar{Σ}{\ensuremath{\Sigma}}
\newunicodechar{σ}{\ensuremath{\sigma}}
\newunicodechar{ς}{\ensuremath{\varsigma}}
\newunicodechar{Τ}{\ensuremath{\Tau}}
\newunicodechar{τ}{\ensuremath{\tau}}
\newunicodechar{Υ}{\ensuremath{\Upsilon}}
\newunicodechar{υ}{\ensuremath{\upsilon}}
\newunicodechar{Φ}{\ensuremath{\Phi}}
\newunicodechar{φ}{\ensuremath{\phi}}
\newunicodechar{ϕ}{\ensuremath{\varphi}}
\newunicodechar{Χ}{\ensuremath{\Chi}}
\newunicodechar{χ}{\ensuremath{\chi}}
\newunicodechar{Ψ}{\ensuremath{\Psi}}
\newunicodechar{ψ}{\ensuremath{\psi}}
\newunicodechar{Ω}{\ensuremath{\Omega}}
\newunicodechar{ω}{\ensuremath{\omega}}

\newunicodechar{ℕ}{\ensuremath{\mathbb{N}}}
\newunicodechar{∅}{\ensuremath{\emptyset}}

\newunicodechar{∙}{\ensuremath{\bullet}}
\newunicodechar{≈}{\ensuremath{\approx}}
\newunicodechar{≅}{\ensuremath{\cong}}
\newunicodechar{≡}{\ensuremath{\equiv}}
\newunicodechar{≤}{\ensuremath{\le}}
\newunicodechar{≥}{\ensuremath{\ge}}
\newunicodechar{≠}{\ensuremath{\neq}}
\newunicodechar{∀}{\ensuremath{\forall}}
\newunicodechar{∃}{\ensuremath{\exists}}
\newunicodechar{±}{\ensuremath{\pm}}
\newunicodechar{∓}{\ensuremath{\pm}}
\newunicodechar{·}{\ensuremath{\cdot}}
\newunicodechar{⋯}{\ensuremath{\cdots}}
\newunicodechar{…}{\ensuremath{\ldots}}
\newunicodechar{∷}{~\mathrel{:\!\!\!:}~}
\newunicodechar{×}{\ensuremath{\times}}
\newunicodechar{∞}{\ensuremath{\infty}}
\newunicodechar{→}{\ensuremath{\to}}
\newunicodechar{←}{\ensuremath{\leftarrow}}
\newunicodechar{⇒}{\ensuremath{\Rightarrow}}
\newunicodechar{↦}{\ensuremath{\mapsto}}
\newunicodechar{↝}{\ensuremath{\leadsto}}
\newunicodechar{∨}{\ensuremath{\vee}}
\newunicodechar{∧}{\ensuremath{\wedge}}
\newunicodechar{⊢}{\ensuremath{\vdash}}
\newunicodechar{⊣}{\ensuremath{\dashv}}
\newunicodechar{∣}{\ensuremath{\mid}}
\newunicodechar{∈}{\ensuremath{\in}}
\newunicodechar{⊆}{\ensuremath{\subseteq}}
\newunicodechar{⊂}{\ensuremath{\subset}}
\newunicodechar{∪}{\ensuremath{\cup}}
\newunicodechar{⋓}{\ensuremath{\Cup}}
\newunicodechar{∉}{\ensuremath{\not\in}}
\newunicodechar{√}{\ensuremath{\sqrt}}

\newunicodechar{⊸}{\ensuremath{\multimap}}
\newunicodechar{⊗}{\ensuremath{\otimes}}
\newunicodechar{⨂}{\ensuremath{\bigotimes}}
\newunicodechar{⊕}{\ensuremath{\oplus}}
\newunicodechar{〈}{\ensuremath{\langle}}
\newunicodechar{⟨}{\ensuremath{\langle}}
\newunicodechar{⟩}{\ensuremath{\rangle}}
\newunicodechar{〉}{\ensuremath{\rangle}}
\newunicodechar{¡}{\ensuremath{\upsidedownbang}}
\newunicodechar{∘}{\ensuremath{\circ}}
\newunicodechar{†}{\ensuremath{\dagger}}
\newunicodechar{⊤}{\ensuremath{\top}}
\newunicodechar{⊥}{\ensuremath{\bot}}

\newunicodechar{〚}{\ensuremath{\llbracket}}
\newunicodechar{〛}{\ensuremath{\rrbracket}}

\usepackage{etoolbox} \newtoggle{comments}
\toggletrue{comments}
\usepackage[normalem]{ulem}

\iftoggle{comments}{
  \newcommand{\fixme}[1]{\textbf{\textcolor{red}{[ Fixme: #1]}}}
  \newcommand{\todo}[1]{\textbf{\textcolor{green}{[ TODO: #1 ]}}}
  \newcommand{\mwh}[1]{\textbf{\textcolor{red}{[ Mike: #1 ]}}}
  
  \newcommand{\khh}[1]{\textbf{\textcolor{orange}{[ Kesha: #1 ]}}}
  \newcommand{\shh}[1]{\textbf{\textcolor{purple}{[ Shih-Han: #1 ]}}}
  \newcommand{\liyi}[1]{\textbf{\textcolor{blue}{[ Liyi: #1 ]}}}
  \newcommand{\oth}[2]{\textbf{\textcolor{red}{[ #1: #2 ]}}}
  \newcommand{\xwu}[1]{\textbf{\textcolor{purple}{[ Xiaodi: #1 ]}}}
  
  \newcommand{\ynote}[1]{\textbf{\textcolor{magenta}{[ Yi: #1 ]}}}

  \colorlet{MZ}{violet!80!pink}

  \newcommand{\mzr}[1]{{\color{MZ}{#1}}}
  \newcommand{\was}[1]{}
  
\NewCommandCopy{\Creff}{\Cref}
  \renewcommand{\Cref}[1]{\mbox{\Creff{#1}}}

  \usepackage[inline]{enumitem}
  \colorlet{LC}{cyan!31!teal}

  \usepackage[inline]{enumitem}
}{
  \newcommand{\fixme}[1]{}
  \newcommand{\todo}[1]{}
  \newcommand{\rnr}[1]{}
  \newcommand{\mwh}[1]{}  
  \newcommand{\khh}[1]{}
  \newcommand{\liyi}[1]{}
  \newcommand{\shh}[1]{}
  \newcommand{\xwu}[1]{}
  \newcommand{\oth}[2]{}
  \newcommand{\mzr}[1]{}

  \newcommand{\ynote}[1]{}
}

\newtoggle{submission}
\toggletrue{submission}

\iftoggle{submission}{
  
}{
  
}

 \usepackage{pifont}\usepackage{multicol,tabularx,capt-of}
\usepackage{multirow}
\newcommand{\cmark}{\text{\ding{51}}}
\newcommand{\xmark}{\text{\ding{55}}}

\title{Qafny: A Quantum-Program Verifier {\large (Extended Version)}}

\author{Liyi Li}{Iowa State University}{liyili2@iastate.edu}{https://orcid.org/0000-0001-8184-0244}{}

\author{Mingwei Zhu}{University of Maryland}{mzhu1@umd.edu}{}{}

\author{Rance Cleaveland}{University of Maryland}{rance@cs.umd.edu}{}{}

\author{Alexander Nicolellis}{Iowa State University}{akn5@iastate.edu}{}{}

\author{Yi Lee}{University of Maryland}{ylee1228@umd.edu}{}{}

\author{Le Chang}{University of Maryland}{lchang21@umd.edu}{}{}

\author{Xiaodi Wu}{University of Maryland}{xwu@cs.umd.edu}{https://orcid.org/0000-0001-8877-9802}{}

\authorrunning{L. Li, M. Zhu, R. Cleaveland, A. Nicolellis, Y. Lee, L. Chang \& X. Wu}

\Copyright{Li et al.} 



\relatedversion{} 

\acknowledgements{We thank Finn Voichick for his helpful comments and contributions during the work's development. This paper is dedicated to the memory of our dear co-author Rance Cleaveland.}\EventEditors{}
\EventNoEds{2}
\EventLongTitle{ECOOP 2024}
\EventShortTitle{ECOOP 2024}
\EventAcronym{ECOOP}
\EventYear{2024}
\EventDate{September 16 - 20, 2024}
\EventLocation{Vienna, Austria}
\EventLogo{}
\SeriesVolume{}
\ArticleNo{}
\begin{document}

\maketitle
\begin{abstract}
 Because of the probabilistic/nondeterministic behavior of quantum programs, it is highly advisable to verify them formally to ensure that they correctly implement their specifications.  Formal verification, however, also traditionally requires significant effort. To address this challenge, we present \qafny, an automated proof system based on the program verifier Dafny and designed for verifying quantum programs. At its core, \qafny uses a type-guided quantum proof system that translates quantum operations to classical array operations modeled within a classical separation logic framework. We prove the soundness and completeness of our proof system and implement a prototype compiler that transforms \qafny programs and specifications into Dafny for automated verification purposes. We then illustrate the utility of \qafny's automated capabilities in efficiently verifying important quantum algorithms, including quantum-walk algorithms, Grover’s algorithm, and Shor’s algorithm. 

 \end{abstract}

\section{Introduction}
\label{sec:intro}

Quantum computers can be used to program substantially faster algorithms compared to those written for classical computers.
For example, Shor's algorithm \cite{shors} can factor a number in polynomial time, which is not known to be polynomial-time-computable in the classical setting.
Developing more and more comprehensive quantum programs and algorithms is essential for the continued practical development of quantum computing \cite{gill2024quantumcomputingvisionchallenges,MattSwayne}.
Unfortunately, because quantum systems are inherently probabilistic and must obey quantum physics laws, traditional validation techniques based on run-time testing are virtually impossible to develop for large quantum algorithms. This leaves \emph{formal methods} as a viable alternative for program checking, and yet these typically require a great effort; for example, four experienced researchers needed two years to formally verify Shor's algorithm \cite{shorsprove}.
To alleviate the effort required for formal verification, many frameworks have been proposed to verify quantum algorithms \cite{qhoreusage,qhoare,qbricks,qsepa,qseplocal,VOQC}
using interactive theorem provers, such as Isabelle, Coq, and Why3, by building quantum semantic interpretations and libraries.
Some attempts towards proof automation have been made by creating new proof systems for quantum data structures such as Hilbert spaces; however,
building and verifying quantum algorithms in these frameworks are still time-consuming and require great human effort.
Meanwhile, automated verification is an active research field in classical computation with many proposed frameworks
\cite{HoareLogic,separationlogic,nat-proof-fun,nat-proof,nat-proof-frame,10.1145/3453483.3454087,arxiv.1609.00919,martioliet00rewriting,rosu-stefanescu-2011-nier-icse,rosu-stefanescu-ciobaca-moore-2013-lics,10.1007/978-3-642-17511-4_20,10.1007/978-3-642-03359-9_2}
showing strong results in reducing programmer effort when verifying classical programs. 
None of the existing quantum verification frameworks utilize these classical verification infrastructures, however.

\begin{figure}[t]
\begin{minipage}[b]{0.57\textwidth}
  \includegraphics[width=0.95\textwidth]{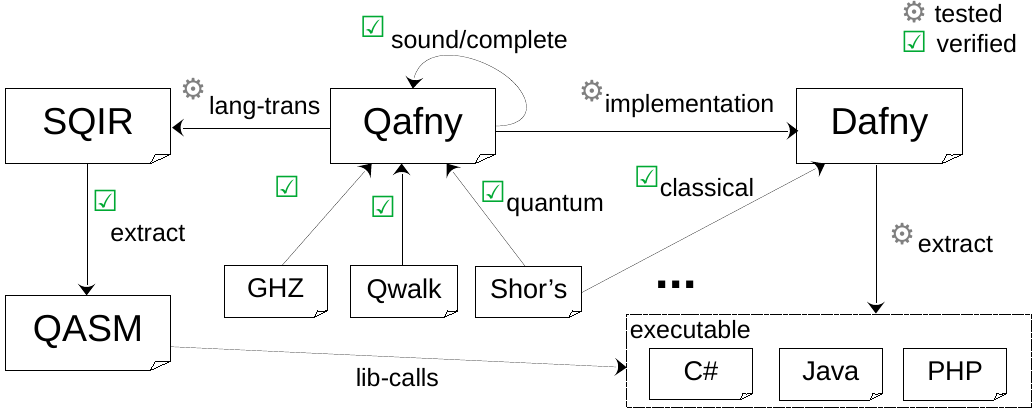}
\vspace*{-0.5em}
  \caption{Dafny Development Stages/Key Aspects}
\label{fig:arch}
\end{minipage}
\hfill
\begin{minipage}[b]{0.4\textwidth}
{\;\scriptsize
  \Qcircuit @C=0.5em @R=0.5em {
 \lstick{\ket{0}}  & \qw    & \gate{H} & \qw & \multigate{3}{{ f(\ket{j})=(-i)^{j}\ket{j}}}   & \qw &\qw & \qw \\
   &  \vdots &          &     &                                          &    \rstick{{ {\Msum_{j=0}^{2^n\sminus 1}(-i)^{j}\ket{j}}}} & &\\
   & \vdots  &          &     &                                          &     &       &  \\
  \lstick{\ket{0}} &  \qw   & \gate{H} & \qw &  \ghost{f(\ket{\kappa})=(-i)^x\ket{\kappa}}         &\qw  &\qw    & \qw
    }
}
\vspace*{-0.5em}
\caption{State Preparation Circuit}
\label{fig:intros-example}
\end{minipage}
\vspace*{-1.4em}
\end{figure}

We present \qafny, a framework that enables programmers to develop and verify \emph{quantum programs} based on quantum program semantics and classical automated verification infrastructure.
It has several elements (\Cref{fig:arch}). The core is a strongly typed, flow-sensitive imperative quantum language \qafny, admitting a classical separation-logic-style proof system, in which users specify quantum programs and input the properties to be verified as pre- and post-conditions and loop invariants, such as GHZ, Quantum Walk, and Shor's algorithm.
\qafny programs and specifications are verified via translation to a classical Hoare/separation logic framework implemented in the Dafny program verifier~\cite{dafnyref}.
\qafny programs may also be compiled into quantum circuits and run on a quantum computer via the \qafny to \sqir and \sqir to OpenQASM 2.0 \cite{Cross2017} compilers (\Cref{sec:vqir-compilation}).
Quantum programs can be components of \emph{hybrid classical-quantum} (HCQ) programs, so the compiled \qafny code can also be a library function called by an HCQ program defined. For example, one can extract the compiled Dafny program to a programming language, such as C\#, PHP, and Java, and utilize quantum programs compiled from \qafny to OpenQASM \cite{10.1007/978-3-642-17511-4_20}.

A key component of the design of \qafny is to encode quantum states as array-like structures and quantum operations as aggregate array operations on the states.
In \Cref{fig:intros-example}, a quantum state in superposition \(\psi =\sum_{j=0}^{2^n\sminus 1} {1 \ket{j}}\) is prepared by applying a Hadamard gate to each qubit.
\qafny treats $\psi$ as an array containing $2^n$ elements, one for each indexed basis element in $\psi$. Each element is a pair of a complex and a natural number (computational basis, essentially a bitstring). For example, Bell pair $\frac{1}{\sqrt{2}}\ket{00} + \frac{1}{\sqrt{2}}\ket{11}$
can be thought of as a two-element array, with two pairs: $(\frac{1}{\sqrt{2}},\ket{00})$ and $(\frac{1}{\sqrt{2}},\ket{11})$, where the first one is a complex number and the second one is a bitstring that can be represented as a natural number.
Applying a quantum oracle ($f(\ket{j})=(-i)^{j}\ket{j}$) on \(\psi\), which evolves each indexed element \(1 \ket{j}\) to \((-i)^{j}\ket{j}\), is similar to an array map function that applies \(f'(\alpha_j, j) = ({(-1)}^j \alpha_j, j)\) to each element $j$ in the \(2^n\)-array.
The design and analysis of many quantum algorithms leverage the representation of different groups of qubits in terms of classical arrays \cite{grover1997,mike-and-ike,Shor94}.
Besides the opportunities for automated reasoning provided by representing quantum states as arrays, \qafny also uses language abstractions such as quantum conditionals and loops,
which generalize quantum controlled gates to enable local reasoning in the presence of quantum entanglement.
In the prior works above, the usual approach in reasoning about quantum controlled gates such as \coqe{CNOT} and controlled-\(U\) gates is to transform these operations into a monolithic representation, such as a unitary matrix, not scaling up well for automated verification, because the relations among different entries, representing inductive relation among program constructs, are largely omitted.
In \qafny, reasoning about comprehensive constructs, such as controlled gates, amount to building a structural inductive relation among different parts, such as a quantum conditional and its subparts, through deliberately designed proof rules in \Cref{sec:guidedtypeexp,sec:semantics-proof}; such design permits automated local reasoning.

These designs pose several challenges: 1) quantum operations can be performed on any qubit positions, e.g., $f$ above could apply on arbitrary bits in every bitstring $j$;
2) performing automated local reasoning requires one to know which states and qubits can be excluded locally, but qubits can form entangled groups that are typically viewed as not separable;
and 3) the \qafny proof system should obey quantum physical laws, such as no-cloning and no quantum observer effects.
To address these problems, we first introduce different types of quantum-state representations and special data structures (\emph{loci} in \Cref{sec:overview}) to partition qubits into disjoint entanglement groups for local reasoning.
We then combine a flow-sensitive type system (\Cref{sec:type-checking}) with our proof system, capable of 1) statically identifying the quantum state types and tracking entanglement group transformations and 2) performing type-guided quantum-state rewrites in the assertions and automated local operation reasoning on canonicalized quantum states, without violating quantum laws.

The paper's contributions are listed as follows.

\vspace*{-0.5em}
\begin{itemize}
\item We present the \qafny language, including a big-step semantics and flow-sensitive type system, which provides a simple way of enforcing quantum program properties, such as no-cloning and no observer breakdown. We also prove type soundness for \qafny in Coq.

\item The \qafny type-guided proof system permits classical-array-operation views of quantum operations and captures the inductive behaviors of quantum conditionals and loops. Soundness and relative completeness are also proved in Coq. To the best of our knowledge, \qafny provides the first proof-rule definitions for quantum conditionals and for-loops.

\item  We exhibit a prototype \qafny to Dafny compiler as evidence of connecting quantum-program verification to classical Hoare/separation logic. We verify a number of quantum algorithms (\Cref{fig:circ-evaluation}) with a high degree of automation. \Cref{sec:implementation,sec:related} compares proof automation in \qafny with other frameworks.

\item We faithfully implement several algorithms, such as GHZ, Shor's, and quantum walk, as case studies in tje paper to demonstrate how \qafny can help programmers to efficiently verify quantum-algorithm implementations. The program operations of these examples are a high-level abstraction of the algorithms' quantum circuit-based description, while the \qafny program state specifications are directly based on the algorithms' state representations based on Dirac notations.

\end{itemize}
\vspace*{-1em}

 \vspace*{-0.5em}
\section{Background}
\label{sec:background}

Here, we provide background information on quantum computing. 

\noindent\textbf{\textit{Quantum Value States}.} A quantum value
state\footnote{Most literature mentioned Quantum value states as \emph{quantum states}. Here, we refer to them as quantum value states or quantum values to avoid confusion between program and quantum states.  } consists of one or more quantum bits (\emph{qubits}), which can be expressed as a two-dimensional vector $\begin{psmallmatrix} \alpha \\ \beta \end{psmallmatrix}$ where the \emph{amplitudes} $\alpha$ and $\beta$ are complex numbers and $|\alpha|^2 + |\beta|^2 = 1$.
We frequently write the qubit vector as $\alpha\ket{0} + \beta\ket{1}$ (the Dirac notation \cite{Dirac1939ANN}), where $\ket{0} = \begin{psmallmatrix} 1 \\ 0 \end{psmallmatrix}$ and $\ket{1} = \begin{psmallmatrix} 0 \\ 1 \end{psmallmatrix}$ are \emph{computational basis-kets}. When both $\alpha$ and $\beta$ are non-zero, we can think of the qubit being ``both 0 and 1 at once,'' a.k.a. in a \emph{superposition} \cite{mike-and-ike}, e.g., $\frac{1}{\sqrt{2}}(\ket{0} + \ket{1})$ represents a superposition of $\ket{0}$ and $\ket{1}$.
Larger quantum values can be formed by composing smaller ones with the \emph{tensor product} ($\otimes$) from linear algebra, e.g., the two-qubit value $\ket{0} \otimes \ket{1}$ (also written as $\ket{01}$) corresponds to vector $[~0~1~0~0~]^T$.
However, many multi-qubit values cannot be \emph{separated} and expressed as the tensor product of smaller values; such inseparable value states are called \emph{entangled}, e.g.\/ $\frac{1}{\sqrt{2}}(\ket{00} + \ket{11})$, known as a \emph{Bell pair}, which can be rewritten to $\Msum_{d=0}^{1}{\frac{1}{\sqrt{2}}}{\ket{dd}}$, where $dd$ is a bit string consisting of two bits, each of which must be the same value (i.e., $d=0$ or $d=1$). Each term $\frac{1}{\sqrt{2}}\ket{dd}$ is named a \emph{basis-ket} \cite{mike-and-ike}, consisting an amplitude ($\frac{1}{\sqrt{2}}$) and a basis vector $\ket{dd}$.

\noindent\textbf{\textit{Quantum Computation and Measurement.}} 
Computation on a quantum value consists of a series of \emph{quantum operations}, each acting on a subset of qubits in the quantum value. In the standard form, quantum computations are expressed as \emph{circuits}, as in \Cref{fig:background-circuit-examplea}, which depicts a circuit that prepares the Greenberger-Horne-Zeilinger (GHZ) state \cite{Greenberger1989} --- an $n$-qubit entangled value of the form: $\ket{\text{GHZ}^n} = \frac{1}{\sqrt{2}}(\ket{0}^{\otimes n}+\ket{1}^{\otimes n})$, where $\ket{d}^{\otimes n}=\Motimes_{d=0}^{n\sminus 1}\ket{d}$.
In these circuits, each horizontal wire represents a qubit, and boxes on these wires indicate quantum operations, or \emph{gates}. The circuit in \Cref{fig:background-circuit-examplea} uses $n$ qubits and applies $n$ gates: a \emph{Hadamard} (\coqe{H}) gate and $n-1$ \emph{controlled-not} (\coqe{CNOT}) gates. Applying a gate to a quantum value \emph{evolves} it.
Its traditional semantics is expressed by multiplying the value's vector form by the gate's corresponding matrix representation: $n$-qubit gates are $2^n$-by-$2^n$ matrices.
Except for measurement gates, a gate's matrix must be \emph{unitary} and thus preserve appropriate invariants of quantum values' amplitudes. 
A \emph{measurement} operation extracts classical information from a quantum value. It collapses the value to a basis state with a probability related to the value's amplitudes (\emph{measurement probability}), e.g., measuring $\frac{1}{\sqrt{2}}(\ket{0} + \ket{1})$ collapses the value to $\ket{0}$ with probability $\frac{1}{2}$,  and likewise for $\ket{1}$, returning classical value $0$ or $1$, respectively. 
A more general form of quantum measurement is \emph{partial measurement}, which measures a subset of qubits in a qubit array;
such operations often have simultaneity effects due to entanglement, \ie{} in a Bell pair $\frac{1}{\sqrt{2}}(\ket{00} + \ket{11})$, measuring one qubit guarantees the same outcome for the other --- if the first bit is measured as $0$, the second bit is too.

\noindent\textbf{\textit{Quantum Conditionals.}}
Controlled quantum gates, such as controlled-not gates (\coqe{CNOT}), can be thought of as quantum versions of classical operations, where we view a quantum value as an array of basis-kets and apply an array map operation of the classical operation to every basis-ket.
This is evident when using Dirac notation. 
For example, in preparing a two-qubit GHZ state (\Cref{fig:background-circuit-examplea}, also a Bell pair) for qubit array $x$, the \coqe{H} gate evolves the value to
$\frac{1}{\sqrt{2}}\ket{00}+\frac{1}{\sqrt{2}}\ket{10}$ (same as $\frac{1}{\sqrt{2}}(\ket{0} + \ket{1})\otimes \ket{0}$).
The quantum conditional maps the classical conditional $\qif{x[0]}{\qass{x[1]}{x[1] \,\splus\, 1}}$ onto the two basis-kets, where the operation $x[1]\, \splus\, 1$ acts as a modulo 2 addition to flip $x[1]$'s bit.
Here, we do not flip the $x[1]$ position in the first basis-ket ($\frac{1}{\sqrt{2}}\ket{00}$) due to $x[0]=0$, and we flip $x[1]$ in the second basis-ket because of $x[0]=1$.
Such behaviors can be generalized to other controlled gates, such as the controlled-$U$ gate appearing in Shor's algorithm (\Cref{fig:shorqafny}), where $U$ refers to a modulo-multiplication operation.
The controlled nodes (Boolean guards) in these quantum conditionals can also be generalized to other types of Boolean expressions, e.g., it can be a quantum inequality ($\qbool{\kappa}{<}{n}{x[i]}$) that compares every basis vector of qubit array $\kappa$'s value state with the number $n$ and stores the result in qubit $x[i]$, and the controlled node queries $x[i]$ to determine if the conditional body is executed, more in \Cref{sec:qafny,sec:quantumwalk}.

\noindent\textbf{\textit{Quantum Oracles.}} Quantum algorithms manipulate input information encoded in ``oracles,'' which are callable black-box circuits. 
Quantum oracles are usually quantum-reversible implementations of classical operations, especially arithmetic operations. Their behavior is defined in terms of transitions between single basis-kets.
We can infer its global state behavior based on the single basis-ket behavior through the quantum summation formula below. This resembles an array map operation in \Cref{fig:intros-example}.
\oqasm \cite{oracleoopsla} is a language that permits the definitions of quantum oracles with efficient verification and testing facilities using the summation formula.

\vspace*{-1.4em}
{\footnotesize
  \begin{mathpar}
 \inferrule[]{ \forall j. \ket{x_j} \longrightarrow f(\ket{x_j})  }{ \Sigma_j \alpha_j \ket{x_j} \longrightarrow \Sigma_j \alpha_j f(\ket{x_j})}
    \end{mathpar}
}
\vspace*{-1em}

\noindent\textbf{\textit{No Cloning and Observer Effect.}} 
The \emph{no-cloning theorem} suggests no general way of copying a quantum value. In quantum circuits, this is related to ensuring the reversible property of unitary gate applications.
For example, the Boolean guard and body of a quantum conditional cannot refer to the same qubits, e.g., $\qif{x[1]}{\qass{x[1]}{x[1]\, \splus\, 1}}$ violates the property as $x[1]$ is mentioned in the guard and body.
The quantum \emph{observer effect} refers to leaking information from a quantum value state. If a quantum conditional body contains a measurement or classical variable updates, the quantum system breaks down due to the observer effect. 
\qafny{} enforces no cloning and no observer breakdown through the syntax and flow-sensitive type system.

 \vspace*{-0.5em}
\section{Qafny Design Principles: Locus, Type, and State}
\label{sec:overview}

\begin{figure}[t]
  \centering
  \captionsetup[subfigure]{justification=centering}
  {
    \begin{tabular}{l}
      \begin{tabular}{c c}
        \hspace*{-1em}
        \begin{tabular}{l}
          \begin{minipage}[t]{.38\textwidth}
            \subcaption{GHZ Circuit}
{\qquad
              \small
              \Qcircuit @C=0.5em @R=0.5em {
                \lstick{\ket{0}} & \gate{H} & \ctrl{1} & \qw &\qw & & \dots & \\
                \lstick{\ket{0}} & \qw & \targ & \ctrl{1} & \qw & &  \dots &  \\
                \lstick{\ket{0}} & \qw & \qw   & \targ & \qw & &  \dots &  \\
                & \vdots &   &  &  & & & \\
                & \vdots &  & \dots & & & \ctrl{1} & \qw  \\
                \lstick{\ket{0}} & \qw & \qw & \qw &\qw &\qw & \targ & \qw
              }
            }

            \label{fig:background-circuit-examplea}
          \end{minipage}
          \\
          \begin{minipage}[t]{.38\textwidth}
            \subcaption{GHZ For-loop Analogy}
            {
              \small
              \includegraphics[width=1\textwidth]{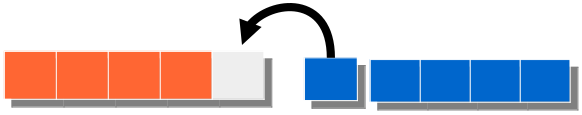}
            }
            \label{fig:background-circuit-analogy}
          \end{minipage}
        \end{tabular}
        &
          \begin{tabular}{l}
            \begin{minipage}[t]{.6\textwidth}
\subcaption{\qafny GHZ Program and Proof}
                         \vspace*{-1em}
              {\small
                \[\hspace*{-2.5em}
                  \begin{array}{r l}
                    \textcolor{blue}{1}
                    &
                      \textcolor{spec}{
                      \Hassert{x\srange{0}{n}\mapsto \ket{\overline{0}}}
                      }
                    \\[0.3em]
                    \textcolor{blue}{2}
                    &
                      \qass{x[0]\;}{\;\cn{H}};
                    \\[0.2em]
                    \textcolor{blue}{3}
                    &
                      \textcolor{auto}{
                      \Hassert{x\srange{0}{1}\mapsto \frac{1}{\sqrt{2}}(\ket{0}+\ket{1}) * x\srange{1}{n}\mapsto \ket{\overline{0}}}
                      }
                    \\[0.3em]

                    \textcolor{blue}{4}
                    &
                      \qfor{j}{1}{n}{x[j\,\sminus \,1]}
\\[0.3em]

                    \textcolor{blue}{5}
                    &
                      \quad
                      \textcolor{spec}{
                      \Hassert{x\srange{0}{j}\mapsto \Msum_{d=0}^{1}{\frac{1}{\sqrt{2}}}{\ket{\overline{d}}} * x\srange{j}{n}
                                      \mapsto \ket{\overline{0}} * j \leq n}
                      }
                    \\[0.3em]

                    \textcolor{blue}{6}
                    &
                      \quad
\qass{x[j]\;}{\;x[j] \cn{ + 1}};
                    \\[0.3em]

                    \textcolor{blue}{7}
                    &
                      \textcolor{auto}{
                      \Hassert{x\srange{0}{n}\mapsto \Msum_{d=0}^{1}{\frac{1}{\sqrt{2}}}
                                {\ket{\overline{d}}} * x\srange{n}{n}\mapsto \ket{\overline{0}}}
                      }
                    \\[0.3em]

                    \textcolor{blue}{8}
                    &
                      \textcolor{spec}{
                      \Hassert{x\srange{0}{n}\mapsto \Msum_{d=0}^{1}{\frac{1}{\sqrt{2}}}{\ket{\overline{d}}}}
                      }
                  \end{array}
                \]
              }
              \label{fig:background-circuit-example-proof}
            \end{minipage}
          \end{tabular}
      \end{tabular}
    \end{tabular}
  }
  \vspace*{-1.2em}
  \caption{GHZ Examples. $x\srange{t_1}{t_2}\mapsto \ket{\overline{0}}$ means $\overline{0}$ is a bitstring of length $t_2\sminus t_1$.
    $x\srange{t_1}{t_2}\mapsto \Msum_{d=0}^{1}{}{\ket{\overline{d}}}$ means $\overline{d}$ is a bitstring of length $t_2\sminus t_1$ and $d\in[0,1]$ is a bit. $*$ is the separation conjunction. In (c), black parts are \qafny programs, while blue and gray parts are \qafny state predicates.}
  \label{fig:background-circuit-example}
 \vspace*{-1.5em}
\end{figure}

Here, we show the \qafny fundamental design principles for quantum program verification.
We use the GHZ example in \Cref{fig:background-circuit-examplea} to highlight these principles, with a proof outline in 
\Cref{fig:background-circuit-example-proof}; $x$ is initialized to an $n$-qubit $\tnort$ typed value $\ket{\overline{0}}$ ($n$ number of $\ket{0}$ qubits).
After preparing a superposition ($\qass{x[0]}{\cn{H}}$) for a single qubit $x[0]$ in line 2,
we execute a \textit{quantum for-loop} that entangles each pair of adjacent qubits in $x$ to prepare the GHZ state.
We can unroll each iteration of the loop as a
quantum conditional ($\qif{x[j\sminus{1}]}{\qass{x[j]}{x[j]\, \splus\, 1}}$).
When verifying the program in \qafny, it is only needed to provide the program and specifications in
\textcolor{spec}{blue}, with the \textcolor{auto}{grayed out} parts automatically inferred.
We show critical features in our type-guided proof system, making the above verification largely automatic.

\vspace*{-0.5em}
\subsection{{Loci, Types, and States}}\label{sec:states}
\vspace*{-0.2em}

\Cref{fig:qafny-state} shows \qafny \emph{loci}, types, and states, which are used for tracking possibly entangled qubits.
In GHZ (\Cref{fig:background-circuit-example-proof}), each loop step in lines 4-6 entangles the qubit $x[j]$ with \(x\srange{0}{j}\), i.e., the entangled qubit group is expanded from \(x\srange{0}{j}\) to \(x\srange{0}{j\,\splus\, 1}\).
However, the entanglement here is implicit: the program syntax does not directly tell if \(x[j]\) is entangled with \(x\srange{0}{j}\) but relies on an analysis to resolve the entanglement scopes, which is captured by introducing
\begin{enumerate*}[label=\arabic*)]
    \item \emph{loci} (\(\kappa\)) to group possibly entangled qubits,
    \item standard \emph{kind environments} (\(\Omega\)) to record variable \emph{kinds} (explained below), and
    \item \emph{locus type environments} (\(\sigma\)) to keep track of both loci and
their quantum state types.
\end{enumerate*}

\(\qafny\) variables may represent one of three \emph{kinds}
\footnote{ $\cmode$ and
  $\mmode$ kinds are also used as context modes in type checking. See
  \Cref{fig:exp-sessiontype}.
} 
of values (\Cref{fig:qafny-state}). $\cmode$ and $\mmode$ kinds are scalars; the former is an integer\footnote{ Any classical values are permitted in our implementation.
For simplicity, we only consider integers here.
}, and the latter is a measurement outcome $(r,n)$ where $r$ is the probability of outcome $n$.
$\qmode{m}$ kind variables represent a physical $m$-length qubit array conceptually living in a heap.
For simplicity, we assume no aliasing in variable names, no overlapping between qubit arrays referred to by any two different variables, and scalar and qubit array variables are always distinct.
Quantum values are categorized into three different types: \(\tnort\),
\(\thadt\) and \(\tcht\).
A \emph{normal} value (\(\tnort\)) is an array (tensor product) of single-qubit values \(\ket{0}\) or \(\ket{1}\).
Sometimes, a (\(\tnort\))-typed value is associated with an amplitude $z$, representing an intermediate partial program state; an example is in \Cref{sec:controled-ghz}.
A \emph{Hadamard} (\(\thadt\)) typed value represents a collection of qubits in superposition but not entangled,
i.e., an $n$-qubit array $\frac{1}{\sqrt{2}}(\ket{0} + \alpha(r_0)\ket{1})\otimes ... \otimes \frac{1}{\sqrt{2}}(\ket{0} + \alpha(r_{n-1})\ket{1})$, can be encoded as $\shad{2^n}{n\sminus 1}{\alpha(r_j)}$, with $\alpha(r_j)=e^{2\pi i r_j}$ ($r_j \in \mathbb{R}$) being the \emph{local phase},
a special amplitude whose norm is $1$, i.e., $\slen{\alpha(r_j)}=1$. 
The most general form of $n$-qubit values is the \emph{entanglement} (\(\tcht\)) typed value,
consisting of a linear combination (represented as an array) of basis-kets, as $\sch{m}{z_j}{\beta_j\eta_j}$, where $m$ is the number of elements in the array.
In \qafny, we \textit{extend} traditional basis-ket structures in the Dirac notation to be the above form, so each basis-ket of the above value contains not only an amplitude $z_j$ and a basis $\beta_j$ but also a frozen basis stack $\eta_j$, storing bases not directly involved in the current computation.
Here, $\beta_j$ can always be represented as a single $\ket{c_j}$ by the equation in \Cref{fig:qafny-state}.
Every $\beta_j$ in the array has the same cardinality, e.g., if $\slen{c_0}=n$ ($\beta_0=\ket{c_0}$), then $\slen{c_i}=n$ ($\beta_j=\ket{c_j}$) for all $j$.

\begin{figure}[t]
  \caption{\footnotesize \qafny element syntax. Each range $x\srange{n}{m}$ in a locus represents the number range $[n,m)$ in physical qubit array $x$. Loci are finite lists, while type environments and states are finite sets. The operations after "concatenated op" are concatenations for loci, type environments, and quantum states.}
\vspace*{-1.4em}
{
\footnotesize
\[\hspace*{-1em}
\begin{array}{l}
\textcolor{blue}{\text{Basic Terms:}}\\
\begin{array}{l@{\;\;\;}l@{\;\;}c@{\;\;}l @{\qquad} l@{\;\;\;}l@{\;\;}c@{\;\;}l @{\qquad} l@{\;\;\;}l@{\;\;}c@{\;\;}l @{\qquad} l@{\;\;\;}l@{\;\;}c@{\;\;}l}
\text{Nat. Num} & m, n & \in & \mathbb{N}
&
      \text{Real} & r & \in & \mathbb{R}
&
      \text{Amplitude} & z & \in & \mathbb{C}
&
\text{Phase} & \alpha(r) & ::= & e^{2\pi i r}

\\
      \text{Variable} & x,y &&

 & \text{Bit} & d & ::= & 0\mid 1       
&
      \text{Bitstring} & c & \in & d^{+}
&
      \text{Basis Vector}& \beta & ::= & (\ket{c})^*
\end{array}
\\[0.7em]
\textcolor{blue}{\text{Modes, Kinds, Types, and Classical/Quantum Values:}}\\
\begin{array}{llcl@{\quad}c@{\quad}l@{\quad}c@{\quad}l} 
      \text{Mode} & g & ::= & \cmode  & \mid & \mmode\\
      \text{Classical Scalar Value} & v & ::= & n & \mid & (r,n)\\
      \text{Kind} & g_k & ::= & g & \mid & \qmode{n} \\
      \text{Frozen Basis Stack} & \gamma & ::= & \tos{\beta} \\
      \text{Full Basis Vector} & \eta & ::= & \beta \gamma \\
      \text{Basic Ket} & w & ::= & z \eta\\
      \text{Quantum Type} & \tau & ::= & \tnort &\mid& \thadt  &\mid& \tcht \\
      \text{Quantum Value (Forms)} & q & ::= & w & \mid& \shad{2^n}{n\sminus 1}{\alpha(r_j)} &\mid& \ssum{j=0}{m}{w_j}
    \end{array}
\\[-0.1em]
\textcolor{blue}{\text{Quantum Loci, Environment, and States}}\\
\begin{array}{llclcl} 
      \text{Qubit Array Range} & s & ::= & x\srange{n}{m} \\
      \text{Locus} & \kappa & ::= & \overline{s}  & \text{concatenated op} & \sqcupplus \\
      \text{Kind Environment} & \Omega & ::= & x \to g_k \\
      \text{Type Environment} & \sigma & ::= & \overline{\kappa : \tau } & \text{concatenated op} & \uplus \\
      \text{Quantum State (Heap)} & \varphi & ::= & \overline{ \kappa : q } & \text{concatenated op} & \uplus \\
    \end{array}
\\[0.7em]
\textcolor{blue}{\text{Syntax Abbreviations and Basis/Locus Equations}}\\
\setlength\arraycolsep{4pt}
  \begin{array}{r @{~} c @{~} l  r@{~}c @{~} l  r @{~}c @{~} l  r @ {~} c @{~} l }
   \Msum_{j=0}^{0}w_j &\simeq& w_0
&
   \Msum_{j=0}^{m}w_j &\simeq& \Msum_{j}w_j
&
  1 \gamma &\simeq& \gamma
&
z\beta\tos{\emptyset} &\simeq& z\beta
 \quad \qquad z\beta\tos{\beta'} \simeq  z\beta\textcolor{stack}{\cmsg{\beta'}}
\\[0.5em]
\ket{c_1}\ket{c_2} &\equiv& \ket{c_1 c_2}
&
x[n,n) & \equiv & \emptyset
&
\emptyset \sqcupplus \kappa & \equiv & \kappa
&
x[n,m)\sqcupplus\kappa & \equiv & x[n,j)\sqcupplus x[j,m) \sqcupplus\kappa
\;\;
\texttt{where}\;\;n \le j \le m

    \end{array}

\end{array}
  \]
}
  \label{fig:qafny-state}
 \vspace*{-2.2em}
\end{figure}

A \qafny quantum state ($\varphi$), representing a quantum heap, maps \emph{loci} to \emph{quantum values}.
Loci in a heap $\varphi$ \emph{partition} it into regions that contain possibly entangled qubits,
with the guarantee that cross-locus qubits are not entangled.
Each locus is a list of \emph{disjoint ranges} ($s$), each represented by \(x\srange{n}{m}\)---an in-place array slice selected
from \(n\) to \(m\) (exclusive) in a physical qubit array \(x\) (always being $\qmodename$ kind).
Ranges in a locus are pairwise disjoint, written as \(s_1 \sqcupplus s_2\).
For conciseness, we abbreviate a singleton range $x\srange{j}{j\,\splus\, 1}$ as $x[j]$.
At the type level, we maintain a locus type environment ($\sigma$) mapping loci to quantum types: any quantum state $\varphi$ always has an entry in $\sigma$, guaranteeing that $\dom{\varphi}=\dom{\sigma}$, i.e., loci mentioned in $\varphi$ and $\sigma$ are the same.
Locus type environments are stateful, i.e., a statement that starts with the environment \(\sigma\)
could end up with a different one \(\sigma'\) because a locus could be modified
during the execution.
In addition to the locus type environment, we keep a kind environment between variables and their kinds. 
However, it is scoped and immutable as the kind of any scoped variable does not change.

On the bottom of \Cref{fig:qafny-state}, we show abbreviations ($\simeq$) rules for presentation purposes; $A\simeq B$ means we write $B$ to mean $A$. 
The left-most rule shows that $\tnort$ typed value is a singleton $\tcht$ typed array; see the type-guided state rewrites in \Cref{sec:monitortypes1}.
We can also omit the $\tos{}$ in a basis-ket presentation and \textcolor{stack}{color} the basis stack with a hat sign $\cmsg{-}$, e.g., ${\frac{1}{\sqrt{2}}}\ket{0} \textcolor{stack}{\cmsg{\ket{1}}}$ means ${\frac{1}{\sqrt{2}}}\ket{0} \tos{\ket{1}}$; additionally, ${\frac{1}{\sqrt{2}}}\ket{0}\ket{\overline{1}}$ means ${\frac{1}{\sqrt{2}}}\ket{0}\ket{\overline{1}}\tos{\emptyset}$.
Below the $\simeq$ rules in the figure, we present structural equations ($\equiv$) among bases and loci:
1) the locus concatenation $\sqcupplus$ holds identity and associativity equational properties; 2) a range ($x\srange{n}{n}$) containing $0$ qubit is empty; and 3) it is free to split a range ($x[n,m)$) into two  ($x[n,j)$ and $x[j,m)$), preserving the disjointness of $\sqcupplus$. 

\begin{figure}
  \vspace*{-0.8em}
{\hspace*{-0.5em}
{\scriptsize
  \begin{mathpar}
\hspace*{-0.5em}
    \mprset{flushleft}
    \inferrule[]{
      \inferrule[]{
      \inferrule[]{ 
        \inferrule[]{ \inferrule[]{\,\,}
          {\quad \fivepule{\Omega}{\{\kappa_2:\tcht\}}{\mmode}
          {\textcolor{spec}{
             \kappa_2 \mapsto {\frac{1}{\sqrt{2}}}\ket{0}\ket{\overline{1}}}
            \textcolor{stack}{
              \cmsg{\ket{1}}
            }
          }{ e }{
            \textcolor{spec}{
              \kappa_2\mapsto {\frac{1}{\sqrt{2}}}\ket{1}\ket{\overline{1}}}
            \textcolor{stack}{
              \cmsg{\ket{1}}
            }
          }}\;\;\rulelab{P-Oracle} }
       { \fivepule{\Omega}{\{\kappa_1:\tcht\}}{\mmode}
          {\textcolor{spec}{
              \kappa_1\mapsto {\frac{1}{\sqrt{2}}}\ket{\overline{1}}\ket{0}}
            \textcolor{stack}{
              \cmsg{\ket{1}}
            }
          }{ e }{
            \textcolor{spec}{
              \kappa_1\mapsto {\frac{1}{\sqrt{2}}}\ket{\overline{1}}\ket{1}}
            \textcolor{stack}{
              \cmsg{\ket{1}}
            }
          }
        } \;\; \rulelab{EQ}}
        { \fivepule{\Omega}{\{\kappa_1:\tcht\}}{\mmode}
          {\textcolor{spec}{
              \frz(x[j\sminus{1}],\kappa_1)\mapsto
              \schd{d}{1}{\frac{1}{\sqrt{2}}}{\overline{d}}\ket{0}
            }
          }{ e }{
            \textcolor{spec}{
              \kappa_1\mapsto {\frac{1}{\sqrt{2}}}\ket{\overline{1}}\ket{1}}
            \textcolor{stack}{
              \cmsg{\ket{1}}
            }
          }
        }\;\; \rulelab{M-\(\frz\)} }
      {
        \fivepule{\Omega}{\{\kappa:\tcht\}}{\cmode}
        {\textcolor{spec}{
            \kappa\mapsto \schd{d}{1}{\frac{1}{\sqrt{2}}}{\overline{d}}\ket{0}
          }
        }{ \sifq{x[j\sminus{1}]}{e} }{
          \textcolor{spec}{
            U ({\neg{x[j\sminus{1}]}})\mapsto \schd{d}{1}{\frac{1}{\sqrt{2}}}{\overline{d}}\ket{0}
            *
            U ({x[j\sminus{1}]})\mapsto {\frac{1}{\sqrt{2}}}\ket{\overline{1}}\ket{1}
          }
          \textcolor{stack}{
            \cmsg{\ket{1}}
          }
        }} \;\rulelab{P-If} } {
      \fivepule{\Omega}{\sigma}{\cmode}
      {\textcolor{spec}{
          x\srange{0}{j}\mapsto \schd{d}{1}{\frac{1}{\sqrt{2}}}{\overline{d}} * x\srange{j}{n}\mapsto \ket{\overline{0}}
        }
      }{ \sifq{x[j\sminus{1}]}{e} }{
        \textcolor{spec}{
          x\srange{0}{j\splus 1}\mapsto \schd{d}{1}{\frac{1}{\sqrt{2}}}{\overline{d}} * x\srange{j\splus 1}{n}\mapsto \ket{\overline{0}}
        } } }
  \end{mathpar}
  \vspace*{-1.8em}
  {
    \begin{center}
      $\begin{array}{l@{~}c@{~}ll@{~}c@{~}ll@{~}c@{~}l}
         \kappa&=&x[j\sminus 1] \sqcupplus \kappa_1
      &
        \kappa_1 &=& x\srange{0}{j\sminus 1}\sqcupplus x[j]
       &
          \kappa_2 &=& x[j]\sqcupplus x[0,j\sminus 1)
         \\
         e&=&\qass{x[j]\;}{\;x[j]\,\splus\, 1};
         &
         \sigma & = &\{x\srange{0}{j}: \tcht\,,\, x\srange{j}{n}: \tnort\}
         &
         U(b) &=& \ufrz(b\,,\,x[j\sminus 1]\,,\,\kappa)
       \end{array}$
     \end{center}
   }
 }
 }
\vspace*{-1.5em}
  \caption{Detailed automated proof for a loop-step in GHZ. Constructed from the bottom up.}
  \label{fig:ghz-proof}
 \vspace*{-1.5em}
\end{figure}

\vspace*{-0.7em}
\subsection{Simultaneity for Tracking Qubit Positions and Entanglement Scopes}\label{sec:monitortypes}
\vspace*{-0.4em}

\qafny uses locus transformations, captured by the type inference on program operations, to track entanglement scopes.
\Cref{fig:ghz-proof} describes the automated proof steps for verifying a loop-step in \Cref{fig:background-circuit-example-proof}.
The bottom pre-condition contains the quantum values for the two disjoint loci $x\srange{0}{j}$ and $x\srange{j}{n}$.
The quantum conditional's Boolean guard and body are applied to the qubits $x[j\sminus 1]$ and $x[j]$, respectively, appearing in the above two loci. The application entangles $x[j]$ with the locus $x\srange{0}{j}$
and transforms the locus to be $x\srange{0}{j\splus 1}$, by appending $x[j]$ to the end of $x\srange{0}{j}$.
The append, as the first (bottom) proof step in \Cref{fig:ghz-proof}, happens \emph{automatically} through rewrites guided by the locus transformations in the type
environment $\sigma$ associated with each proof triple.
After the rewrites, we preserve the property of no entangled cross-locus qubits.

In the above example, the static rewrites of a locus in a type environment simultaneously gear and change the rewrites of the locus value form in the associated state.
We call this manipulation mechanism \emph{simultaneity}.
As shown in \Cref{sec:intro}, quantum operations can apply on arbitrary qubit positions, which might seriously harm proof automation,
based on previous experiments \cite{VOQC,qbricks} (\Cref{sec:implementation}), even if they tried hard for automation tactics.
It is necessary to statically track qubit positions to permit the canonicalization of quantum state rewrites, allowing a uniform way of defining proof rules for operations.

To permit automated proof inference, we design the uniformity in \qafny proof rules to require that the locus fragments for qubits that an operation is directly applied always be prefixed.
For example in \Cref{fig:ghz-proof} \rulelab{P-If}, instead of having locus $x\srange{0}{j\splus 1}$,
we rewrite it further to $\kappa$ ($x[j\sminus 1] \sqcupplus x\srange{0}{j\sminus 1}\sqcupplus x[j]$), so the qubit $x[j\sminus 1]$ that the Boolean guard is applied to appears at the start position.
These rewrites simultaneously and appropriately rearrange the quantum value associated with the loci.
In a \qafny quantum state, qubits in a locus are arranged as a list of indices pointing to qubit positions.
The locus indices point to particular qubits in a $\tnort$ and $\thadt$ typed value since they essentially represent an array of qubits.
An $\tcht$ typed value consists of a list of basis-kets; the locus indices refer to the corresponding bases appearing in each basis-ket.

\vspace*{-0.6em}
{\footnotesize
\tikzstyle{every picture}+=[remember picture]
\tikzstyle{na} = [shape=rectangle,inner sep=0pt,text depth=0pt]
\begin{center}
$\arraycolsep=-1pt
\begin{array}{l}
\begin{array}{l}
$\tikz\node[na](worda){$\textcolor{spec}{{x[j\sminus 1]}}$};$\textcolor{spec}{\sqcupplus}$ 
\tikz\node[na](wordc){$\textcolor{spec}{{x\srange{0}{j\sminus 1}}}$}; $\textcolor{spec}{\sqcupplus}$
 \tikz\node[na](word1){$\textcolor{spec}{{x[j]}}$};
$\textcolor{spec}{\!\mapsto\! \Msum_{d=0}^{1}{\frac{1}{\sqrt{2}}}}$
\tikz\node[na](wordb){$\textcolor{spec}{\ket{d}}\!$}; 
\tikz\node[na](wordd){$ \textcolor{spec}{\ket{\overline{d}}}\!$};
 \tikz\node[na](word2){$\textcolor{spec}{\ket{0}}$};$
\end{array}
$
\begin{tikzpicture}[overlay]
  \draw[->,red,thick](worda) -- ++(up:0.40) -- ++(east:1.0) -| (wordb);
\end{tikzpicture}

\begin{tikzpicture}[overlay]
  \draw[->,red,thick](wordc) -- ++(down:0.60) -- ++(east:1.0) -| (wordd);
\end{tikzpicture}

\begin{tikzpicture}[overlay]
  \draw[->,red,thick](word1) -- ++(down:0.40) -- ++(east:1.0) -| (word2);
\end{tikzpicture}
$
\begin{array}{l}
\textcolor{spec}{{x\srange{0}{j\sminus 1}\sqcupplus}} $\tikz\node[na](word3){$\textcolor{spec}{{x[j]}}$};$\textcolor{spec}{\mapsto\! {\frac{1}{\sqrt{2}}}\ket{\overline{1}}}$\tikz\node[na](word4){$\textcolor{spec}{\ket{0}}$};$ \textcolor{stack}{\cmsg{\ket{1}}}
\end{array}
$
\begin{tikzpicture}[overlay]
  \draw[->,red,thick](word3) -- ++(down:0.5) -- ++(east:1.0) -| (word4);
\end{tikzpicture}
$
\;\;
\begin{array}{l}
\tikzstyle{every picture}+=[remember picture]
\tikzstyle{na} = [shape=rectangle,inner sep=0pt,text depth=0pt]
 $\tikz\node[na](word5){$\textcolor{spec}{{x[j]}}$};$\textcolor{spec}{\sqcupplus {x\srange{0}{j\sminus 1}}}\textcolor{spec}{\mapsto\! \frac{1}{\sqrt{2}}}$\tikz\node[na](word6){$\textcolor{spec}{\ket{0}}$};$\textcolor{spec}{\ket{\overline{1}}}\textcolor{stack}{\cmsg{\ket{1}}}
\end{array}
$
\begin{tikzpicture}[overlay]
  \draw[->,red,thick](word5) -- ++(down:0.5) -- ++(east:1.0) -| (word6);
\end{tikzpicture}
$
\end{array}
$
\end{center}
}
\vspace*{0.3em}

In the three example states above from \Cref{fig:ghz-proof},
the first maps the locus $\kappa$ ($x[j\sminus 1] \sqcupplus x\srange{0}{j\sminus 1}\sqcupplus x[j]$) to the pre-state in the bottom of line \rulelab{P-If}, where $\ket{\overline{d}}$ is expanded to $\ket{d}\ket{\overline{d}}$. 
In each basis-ket ($d$ is $0$ or $1$), the first qubit $x[j\sminus 1]$ of the locus $\kappa$ corresponds to the first basis bit, while the last qubit $x[j]$ corresponds to $\ket{0}$, the last basis bit. 
Applying an operation on $x[j]$ performs the application on the last basis bit $\ket{0}$ for every basis-ket.
We can also refer a consecutive fragment of a locus to its basis bits, e.g., range $x\srange{0}{j\sminus 1}$ refers to $\ket{\overline{d}}$, the middle portion of each basis-ket, provided that they have the same cardinality.
In this paper, we call the corresponding basis bits of qubits or locus fragments for a value (or a basis-ket set) as the \textit{qubit's/locus's position bases} of the value (or the basis-ket set).
A locus's position bases are linked and moved according to the rewrites of the locus, e.g., the middle and right examples above represent the rewrites from locus $\kappa_1 = {x\srange{0}{j\sminus 1}\sqcupplus x[j]}$ to $\kappa_2 = {x[j]\sqcupplus x[0,j\sminus 1)}$ in the pre-states (bottom to upper) of \Cref{fig:ghz-proof} line \rulelab{EQ}.

The rewrite moves $x[j]$ in $\kappa_1$ to the front in $\kappa_2$; correspondingly, $x[j]$'s position basis ($\ket{0}$) is also moved to the front.
These examples show the functionality of \emph{frozen basis stacks}.
The two basis-kets' frozen basis stacks both contain a basis $\textcolor{stack}{\cmsg{\ket{1}}}$, which are not referenced by any part of a locus and therefore unreachable qubits.
As shown in \Cref{sec:intro}, we want local reasoning and preserving quantum theorems, i.e., a quantum state for a program piece does not mention qubits that are not reachable in the piece, e.g., accessing $x[j\sminus 1]$ above inside the conditional body means a violation of no-cloning.
However, quantum states can be entangled, so unreachable qubits cannot be separated from the states.
Instead, we hide the unreachable qubits, such as $x[j\sminus 1]$, in the frozen stack and retrieve it after jumping out of the conditional body.
A comprehensive example is given in \Cref{sec:quantumwalk}.
We show how to unfreeze the frozen bases and explain the motivation for having frozen bases shortly below. 

\vspace*{-0.7em}
\subsection{Rewrites based on Locus Type and State Equivalence Relations}\label{sec:monitortypes1}
\vspace*{-0.4em}

The \qafny type system maintains simultaneity through the type-guided state rewrites, formalized as equivalence relations (\Cref{sec:type-checking}).  Other than the locus qubit position permutation introduced above, the types associated with loci in the environment also play an essential role in the rewrites.
In \qafny, a locus represents a possibly entangled qubit group. From the study of many quantum algorithms \cite{qft-adder,ChildsNAND,mike-and-ike,shors,1366221,Rigolin_2005,10.5555/1070432.1070591,ccx-adder}, we found that the establishment of an entanglement group can be viewed as a loop structure of incrementally adding a qubit to the group at a time, representing the entanglement's scope expansion;
as the analogy in \Cref{fig:background-circuit-analogy},  qubits in the blue part are added to the orange part one by one.
This behavior is similar to splits and joins of array elements if we view quantum states as arrays.
However, joining and splitting two $\tcht$-typed values are hard problems \footnote{The former is a Cartesian product; the latter is $\ge$ $\mathit{NP}$-hard, both unsuitable for automated verification. }.
Another critical observation in studying many quantum algorithms is that the entanglement group establishment usually involves splitting a qubit in a $\tnort$/$\thadt$ typed value and joining it to an existing $\tcht$ typed entanglement group.
We manage these join and split patterns type-guided equations in \qafny, suitable for automated verification.
The GHZ example above (\Cref{fig:background-circuit-example-proof} line 5) is an example of $\tnort$ and an $\tcht$ type state split and join, where in each loop-step in \Cref{fig:background-circuit-example-proof}, a $\tnort$-typed qubit ${x[j]}$ is split from locus ${x[j,n)}$ and moved to the end of the $\tcht$-typed locus ${x[0,j)}$. Details are in \Cref{sec:type-checking}.

\vspace*{-0.8em}
\subsection{The Qafny Proof System Glance Via Quantum Conditional Proofs}\label{sec:guidedtypeexp}
\vspace*{-0.2em}

We integrate our type system with the \qafny proof system, where
\qafny's type-guided proof triple
(\(\fivepule{\Omega}{\sigma}{g}{P}{e}{Q}\)) states that from a pre-condition $P$, executing $e$ results in a post-condition $Q$, provided that $P$ and $Q$ are resp. well-formed w.r.t $\sigma$ and $\sigma'$,  where $\typing{\Omega}{\sigma}{g}{e}{\sigma'}$ is a valid typing judgment (explained in \Cref{sec:type-checking}).
\ignore{
The well-formedness of predicates \(P\) and \(Q\) mandates that each locus
mentioned in them must be in \(\dom{\sigma}\) and \(\dom{\sigma'}\), respectively. 
On the bottom step of \Cref{fig:ghz-proof}, other than the locus rewrites from $x\srange{0}{j}$ to $\kappa$ described above,
a classical separation-logic-styled frame rule (\rulelab{P-Frame} in \Cref{sec:semantics-proof}) 
is applied to ignore the locus $x\srange{j\splus{1}}{n}$ that does not affect the rest of the proof, and the type environment is transformed accordingly.
In line \rulelab{P-Oracle} (\Cref{fig:ghz-proof}), 
we apply an oracle operation ($\qass{x[j]\;}{}{\;x[j]\,\splus\, 1}$),
analogized to an array map function described in \Cref{fig:intros-example},
i.e., for every basis-ket of locus $\kappa_2$, we flip the bit of $x[j]$'s position basis.
}

A key design principle for proof automation rules is compositional and rule generalization, i.e.,
automated proof steps should be compositional, where each proof step is localized regarding a localized state,
and the generalization means that automation should not depend on the specific local states.
The issue with quantum proof rule designs is entanglement,
i.e., a program execution on a local state might have global effects, which force the proof automation system to perform case analyses on the local states to resolve the global effects.
For example, in verifying the conditional $\sifq{x[j\sminus{1}]}{e}$ in the bottom of \Cref{fig:ghz-proof},
$e$ can be applied to an entangled qubit state outside the visibility of qubits mentioned in $e$.
Since $e$ can be arbitrarily complicated, the prior work \cite{qhoare} handles the verification of the quantum conditional by expanding it as a whole matrix applied to a whole quantum state and performing case analyses.
For proof automation, we need to design a uniform procedure, expressed as proof rules, to derive the verification; such a task is handled by predicate transformers and frozen stacks built on our locus structures.

An example is given at the line \rulelab{P-If} in \Cref{fig:ghz-proof}, we utilize two locus predicate transformers $\frz$ and $\ufrz$ to transform the pre- and post-conditions so that they focus on the loci and basis-kets relevant to the current computation.
In verifying the quantum conditional ($\qif{x[j\sminus{1}]}{\qass{x[j]}{}{x[j]\splus 1}}$), we first apply $\frz$ to transform the pre-condition.
For the value $\textcolor{spec}{\frac{1}{\sqrt{2}}\ket{\overline{0}}\ket{0}+\frac{1}{\sqrt{2}}\ket{\overline{1}}\ket{0}}$, we filter out the basis-ket $\textcolor{spec}{\frac{1}{\sqrt{2}}\ket{\overline{0}}\ket{0}}$, as the Boolean guard ($x[j\sminus{1}]$) is not satisfiable for $x[j\sminus 1]$'s position basis ($\textcolor{spec}{\ket{0}}$) of the basis-ket.
For the remaining basis-ket $\textcolor{spec}{\frac{1}{\sqrt{2}}\ket{\overline{1}}\ket{0}}$,
we freeze $x[j\sminus 1]$'s position basis ($\ket{1}$),
by pushing $\ket{1}$ to the frozen stack as an unreachable position, highlighted by \(\textcolor{stack}{\cmsg{\ket{1}}}\), since it represents the qubit appearing in the Boolean guard that should not join any computation in $e$ ($\qass{x[j]}{}{x[j]\splus 1}$).
A frozen stack represents the link between the local state and its entangled global state.
Each basis-ket in a superposition state can be associated with a single frozen stack,
and we utilize a predicate transformer to manipulate all these frozen stacks in a state, recording the side-effects of the entangled global state caused by local state changes.

Notice that the locus is transformed from $\kappa$ to $\kappa_1$ by removing $x[j\sminus 1]$ to preserve simultaneity.
The post-condition $\textcolor{spec}{\kappa_1\mapsto {\frac{1}{\sqrt{2}}}\ket{\overline{1}}\ket{1}}\textcolor{stack}{\cmsg{\ket{1}}}$ contains
only the computation result of the basis-ket $\textcolor{spec}{\frac{1}{\sqrt{2}}\ket{\overline{1}}\ket{0}}$,
and we want the final post-state to contain all other missing pieces, which is the task of the two $\ufrz$ transformers.
$\ufrz(b,x[j\sminus 1],\kappa)$ points to the basis-ket satisfying the Boolean guard ($\textcolor{spec}{\frac{1}{\sqrt{2}}\ket{\overline{1}}\ket{0}}$),
from the above result post-condition.
The transformer transforms $x[j\sminus 1]$'s position basis, currently in the stack, back to its normal position.
$\ufrz(\neg b,x[j\sminus 1],\kappa)$ represents the basis-ket not satisfying the guard ($\textcolor{spec}{\frac{1}{\sqrt{2}}\ket{\overline{0}}\ket{0}}$), where we retrieve it from the pre-condition through the transformer.
Finally, the two transformers transform and assemble the two states into one as the post-condition at the bottom  \Cref{fig:ghz-proof}.
\Cref{sec:semantics-proof} contains more details.

\ignore{
The post-state above the line \rulelab{P-If} 
($\textcolor{spec}{\kappa_1\mapsto {\frac{1}{\sqrt{2}}}\ket{\overline{1}}\ket{1}}\textcolor{stack}{\ket{1}}$)
is the result of executing $s$. We then replace locus $\kappa_1$ with $\ufrz(b,x[j\sminus 1],\kappa)$ ($b=x[j\sminus 1]$),
which contain all basis-kets that $x[j\sminus 1]$'s position basis satisfying $b$ ($x[j\sminus 1]=\ket{1}$).
For the state in the pre-condition ($\textcolor{spec}{\kappa\mapsto \schii{2}{\frac{1}{\sqrt{2}}}{\overline{d}}\ket{0}}$),
we replace $\kappa$ with $\ufrz(\neg b)$ representing the basis-kets whose $x[j\sminus 1]$'s position basis unsatisfying $b$.
The two $\ufrz$ transformers are merged to the post-state for locus $x\srange{0}{j\splus 1}$ at the bottom,
where we assemble the basis-kets unsatisfying $b$ with the basis-kets in the $\ufrz(b,x[j\sminus 1],\kappa)$ portion by unfreezing the basis in the stack \(\textcolor{stack}{\ket{1}}\) to its ordinary location. More details are in \Cref{sec:semantics-proof}.
}

 \vspace*{-0.5em}
\section{Qafny Formalism}
\label{sec:qafny}
\vspace*{-0.5em}

\begin{figure}[t]
  {\footnotesize
    \(
    {\qquad
    1 < a < N \qquad
      E(t) =
      x\srange{t}{n} \mapsto \shadi{2^{n \,\sminus\,  t}}{n \sminus t\sminus 1}{}\;*
      x\srange{0}{t}\sqcupplus y\srange{0}{n} \mapsto \schai{2^t\sminus 1}{\frac{1}{\sqrt{2^t}}}{i}{\ket{a^{i}\;\mmod\;N}}
    }
    \)
\[\hspace*{-0.5em}
      \begin{array}{r l @{\qquad} l}
        \textcolor{blue}{1}
        & 
      \textcolor{spec}{
       \{x\srange{0}{n} \mapsto \shad{2^n}{n\sminus 1}{} * y\srange{0}{n} \mapsto \ket{\overline{0}}\ket{1}\}
          }
&
          \textcolor{spec}{
          \Hassert{x\srange{0}{n}: \thadt\,,\, y\srange{0}{n}: \tnort}
          }
        \\[0.2em]
        \textcolor{blue}{2}
        & 
          \textcolor{auto}{\Hassert{E(0)}}
        &
          \textcolor{auto}{
          \Hassert{x\srange{0}{n}: \thadt\,,\, x\srange{0}{0}\sqcupplus y\srange{0}{n}: \tcht}
          }
        \\[0.2em]
        \textcolor{blue}{3}
        &\qfor{j}{0}{n}{x[j]}
        \\[0.2em]
        \textcolor{blue}{4}
        &
          \quad\textcolor{spec}{\Hassert{E(j)}}
        &
          \textcolor{spec}{\Hassert{x[j,n): \thadt\,,\, x\srange{0}{j}\sqcupplus y\srange{0}{n}: \tcht}}
        \\[0.2em]
        \textcolor{blue}{5}
        &
          \quad\qass{y\srange{0}{n}\;}{\;a^{2^j} \cdot y\srange{0}{n}\;\mmod\;N};
        \\[0.2em]
        \textcolor{blue}{6}
        &
          \textcolor{auto}{
          \Hassert{E(n)}
          }
        &
          \textcolor{auto}{
          \Hassert{x\srange{0}{0}: \thadt\,,\, x\srange{0}{n}\sqcupplus y\srange{0}{n}: \tcht}
          }
        \\[0.2em]
        \textcolor{blue}{7}
        &
          \textcolor{auto}{
\Hassert{x\srange{0}{n}\sqcupplus y\srange{0}{n} \mapsto \schai{2^{n}\sminus 1}{\frac{1}{\sqrt{2^{n}}}}{i}{\ket{a^{i}\;\mmod\;N}}}
          }
        &
          \textcolor{auto}{
          \Hassert{x\srange{0}{n}\sqcupplus y\srange{0}{n}: \tcht}
          }
        \\[0.2em]
        \textcolor{blue}{8}
        & \sexp{{u}}{\smea{{y}}}{...}
        \\[0.2em]
        \textcolor{blue}{9}
        &
          \textcolor{spec}{
          \left\{
          {\begin{array}{l}
             x\srange{0}{n} \mapsto \smch{\frac{1}{\sqrt{r}}}{r\sminus 1}{t\,\splus\,k p} 
             *
             p = \texttt{ord}(a,N)
             \\
             *\;
             u=(\frac{p}{2^n},a^{t}\;\mmod\;N)
             *
             r=\texttt{rnd}(\frac{2^n}{p})
           \end{array}}
          \right\}
          }
        &
          \textcolor{spec}{
          \Hassert{x\srange{0}{n}: \tcht}
          }
      \end{array}
    \]
  }
  \vspace*{-1em}
  \caption{                     Snippets from quantum order finding in Shor's algorithm; full proof in \Cref{appx:shors}.
$\sord{a,N}$ gets the order of $a$ and $N$. $\srnd{r}$ rounds $r$ to the nearest integer. 
We interpret integers as bitstrings in $\ket{i}$ and $\ket{a^{i}\;\mmod\;N}$.
The right column presents the types of all loci involved.
}
  \label{fig:shorqafny}
\vspace*{-1em}
{
  \small
  \setlength\arraycolsep{3.5pt}
  \[\begin{array}{l@{\quad}l@{\;\;}c@{\;\;}llllll} 
      \text{\oqasm Expr} & \mu\\
      \text{Arith Expr} & a & ::= & x & \mid x\srange{i}{j} &\mid v & \mid a_1 \;\splus\; a_2 & \mid a_1 \,\cdot\,a_2 & \mid ... \\
      \text{Bool Expr} & b & ::= & x[a] &\multicolumn{2}{l}{\mid \qbool{a_1}{=}{a_2}{x[a]}} & \multicolumn{2}{l}{\mid \qbool{a_1}{<}{a_2}{x[a]}} & \mid ... \\
      \text{Predicate Locus} & \Kappa & ::= & \kappa & \mid \mea(x,n,\kappa) & \mid \frz(b,\kappa,\kappa) & \mid \ufrz(b,\kappa,\kappa)\\
      \text{Predicate} & P,Q,R & ::= & a_1 = a_2 & \mid a_1 < a_2 & \mid \Kappa \mapsto q  &\mid P \wedge P & \mid P * P & \mid ... \\
      \text{Gate Expr} & op & ::= & \texttt{H} & \mid \iqft[\lbrack -1 \rbrack]{}{} \\
      \text{C/M Kind Expr}& am & ::= & a & \mid \smea{y} \\
      \text{Statement} & e & ::= & \sskip & \mid  \ssassign{\kappa}{}{op} & \mid \ssassign{\kappa}{}{\mu} & \multicolumn{2}{l}{\mid \sexp{x}{am}{e}}
                                 \\ & & \mid & \sseq{e_1}{e_2} &  \mid \sifq{b}{e} && 
                                     \multicolumn{2}{l}{\mid \sqwhile{j}{a_1}{a_2}{b}{\{e\}}}
    \end{array}
  \]
}
\vspace*{-1.4em}
  \caption{Core \qafny syntax. Element syntax is in \Cref{fig:qafny-state} and \oqasm is in \Cref{sec:vqir}. $\texttt{QFT}^{\lbrack -1 \rbrack}$ refers to the \texttt{QFT} and reversed \texttt{QFT}. An arithmetic expression $x$ is a $\cmode$/$\mmode$ kind variable, $x\srange{i}{j}$ is a quantum array range, and $v$ is a $\cmode$/$\mmode$ kind value. $x[a]$ is the $a$-th element of qubit array $x$. }
  \label{fig:vqimp}
 \vspace*{-1em}
\end{figure}

This section formalizes \qafny's syntax,  semantics, type system, proof system, and the corresponding soundness and completeness theorems. Running example in \Cref{fig:shorqafny} describes quantum order finding, the core component of Shor's algorithm (complete one in \Cref{appx:shors}).
The program assumes that an $n$-qubit $\texttt{H}$ gate and an addition ($y[0,n)\splus 1$) respectively applied to ranges $x\srange{0}{n}$ and $y\srange{0}{n}$ before line 3.
The for-loop entangles range $y\srange{0}{n}$ with every qubit in $x\srange{0}{n}$, one per loop step, and applies a modulo multiplication in each step. 
\(\smea{y}\) (partial measurement) in line 8 non-deterministically outputs a classical value $a^{t}\;\mmod\;N$ for $y$, and interconnectively rearranges $x\srange{0}{n}$'s quantum state, with all basis kets' bases related to a period value $p$.
We unveil the details along with the section.
\ignore{
Besides the minimal yet practical proof illustrated in the GHZ example, the
quantum conditional structure cooperates well with complicated oracles, such as
the modulo multiplication that's crucial in implementing quantum order
finding.
Applying similar proof techniques to the program in \Cref{fig:shorqafny},
we can assert that the loop statement on lines 3-5 entangles \(x\srange{0}{n}\) with
\(y\srange{0}{n}\) having an $\tcht$-typed value $\schai{2^{n}}{\frac{1}{\sqrt{2^{n}}}}{i}{\ket{a^{i}\;\mmod\;N}}$.
In the QNP framework, this can be treated as a \(2^n\)-array of pairs where the
\(i\)-th pair represents \(\ket{i}\ket{a^i\, \mmod\, N}\).
Such a representation is helpful to define quantum conditionals, measurements, and other quantum operations in \Cref{fig:vqimp}.
The \(\smea{y}\) expression on line 8 is interpreted as a non-deterministic map
filter function that first samples a pair \(\ket{i}\ket{u}\), with $u=a^i\, \mmod\, N$, from the above array
 and then removes all pairs whose second element disagrees with $u$
and finally keeps only the first element in every
remaining pair.
Because the modulo multiplication function is periodic, there exists a
smallest natural number \(t\) such that $a^t\,\mmod\,N=u$ and 
the filtered array can be further reordered and rewritten
as $\smch{\frac{1}{\sqrt{s}}}{s}{t\,\splus\,k p}$, where \(p\) is the
order.
}

\vspace*{-0.7em}
\subsection {Qafny Syntax}\label{sec:syntax-types}
\vspace*{-0.2em}

\qafny is a C-like flow-sensitive language equipped with quantum heap mutations, quantum conditionals, and for-loops.
We intend to provide users with a high-level view of quantum operations, e.g., viewing \texttt{H} and $\texttt{QFT}^{\lbrack -1 \rbrack}$ gates as state preparation, quantum oracles ($\mu$ in \Cref{sec:exp-mu}) as quantum arithmetic operations, and controlled gates as quantum conditionals and loops.
As in \Cref{fig:vqimp}, aside from standard forms such as sequence \((\sseq{e}{e})\) and \texttt{SKIP} (\(\sskip\)), statements \(e\) also include let binding ($\sexp{x}{am}{e}$), quantum heap mutations ($\ssassign{{\_}}{}{{\_}}$), quantum/classical conditionals ($\sifq{b}{e}$), and loops ($\sqwhile{j}{a_1}{a_2}{b}{\{e\}}$).
The let statement binds either the result of an arithmetic expression (\(a\)) or a \emph{computational basis measurement} operator (\(\smea{y}\)) to an immutable $\cmode$/$\mmode$ kind variable $x$ in the body \(e\).
This design ensures all classical variables are immutable and lexically scoped to avoid quantum observer breakdown due to mutating a classical variable inside a quantum conditional body.

\vspace*{-1em}
{\footnotesize
\[  
\begin{array}{c}
\texttt{(1)}
\;\;
\texttt{int}\;u\;\texttt{=}\;0;
\;
\sifq{x[0]}{u\;\texttt{=}\;1};
\;
\textcolor{red}{\xmark}
\quad
\texttt{(2)}\;\;
\sifq{x[0]}{\sexp{u}{1}{\sskip}};\;
\textcolor{green}{\cmark}
\quad
\texttt{(3)}\;\;
\sexp{u}{1}{\sifq{x[0]}{\sskip}};\;
\textcolor{green}{\cmark}
\end{array}
\]
}
\vspace*{-1em}
 
Here, case \texttt{(1)} declares $u$ as $0$ and changes its value to $1$ inside the quantum conditional, which creates an observer effect because $u$'s value depends on qubit $x[0]$.  
Cases \texttt{(2)} and \texttt{(3)} show that our immutable \texttt{let} binding can avoid the issue because the binding in \texttt{(2)} can be compiled to \texttt{(3)} due to the immutability; thus, $u$'s value does not depend on the qubit.

A quantum heap mutation operation mutates qubit array data by applying to a locus \(\kappa\) either a unitary state preparation operation (${op}$) (one of Hadamard $\texttt{H}$, quantum Fourier transformation $\texttt{QFT}$, and its inverse $\texttt{QFT}^{-1}$) or a unitary oracle operation (${\mu}$). \footnote{$\mu$ can define all quantum arithmetics, e.g., $x[j]\splus 1$ (Fig. \ref{fig:background-circuit-example-proof}) $\&$ $a^{2^j}y\srange{0}{n}\;\mmod\; N$ (Fig. \ref{fig:shorqafny}). See \Cref{sec:exp-mu}.
}  Other unitary operations, including quantum diffusion and amplification operations, are in \Cref{appx:syntax-types}.

Quantum reversible Boolean guards $b$ are implemented as \oqasm oracle operations, expressed by one of $\qbool{a_1}{=}{a_2}{x[a]}$, $\qbool{a_1}{<}{a_2}{x[a]}$, and $x[a]$, which intuitively amounts to computing $a_1 = a_2$, $a_1 < a_2$ and $\texttt{false}$ respectively as \(b_0\) and storing the result of $b_0 \oplus x[a]$ as a binary in qubit $x[a]$.\footnote{$a_1$ and $a_2$ can possibly apply to a range, like $y\srange{0}{n}$, in an entangled locus.
}
In both conditionals and loops, guards \(b\) are used to represent the qubits that are controlling.
In addition to the \texttt{let} bindings, the quantum for-loop also introduces and enumerates $\cmode$-kind value \(j\) over interval $[a_1,a_2)$ with \(j\) visible to both the guard \(b\) and the loop body \(e\).
As a result of immutability, loops in \qafny are guaranteed to terminate.
If all variables in the guard \(b\) are classical, the conditional or loop becomes a standard classical one, which is differentiated and definable by our type system, described in \Cref{sec:qafny-app}. Obviously, users can always view a \qafny program as a quantum sub-component in a Dafny program, which provides better library support for classical conditionals.
Predicates and predicate loci in \Cref{fig:vqimp} describe quantum state properties in the \qafny proof system, explained in \Cref{sec:semantics-proof}.

\vspace*{-0.5em}
\subsection{Qafny Semantics}\label{sec:semantics}

\begin{figure*}[t]
\vspace*{-0.5em}
{\scriptsize
  \begin{mathpar}
      \inferrule[S-ExpC]{(\varphi,e[n/x]) \Downarrow \varphi'  }{ (\varphi,\sexp{x}{n}{e}) \Downarrow \varphi'}

      \inferrule[S-ExpM]{ (\varphi,e[(r,n)/x]) \Downarrow \varphi' }{ (\varphi,\sexp{x}{(r,n)}{e}) \Downarrow \varphi'}

      \inferrule[S-OP]{\circ=op\vee\circ=\mu}{ 
(\varphi\uplus \{\kappa \sqcupplus\kappa':  q\},\ssassign{\kappa}{}{\circ}) \Downarrow \varphi\uplus\{\kappa\sqcupplus\kappa': \denote{\circ}^{\slen{\kappa}}q\} }

      \inferrule[S-If]{  FV(\Omega,b)=\kappa \\\denote{b}^{\slen{\kappa}}q=\qfun{q}{\kappa,b}+\qfun{q}{\kappa,\neg b}
        \\(\varphi\uplus \{\kappa' : S^{\slen{\kappa}}(\qfun{q}{\kappa,b})\},e) 
        \Downarrow \varphi\uplus \{\kappa': q'\} } 
      {(\varphi\uplus \{\kappa\sqcupplus\kappa' : q\},\sifq{b}{e}) \Downarrow 
        \varphi\uplus \{\kappa\sqcupplus\kappa' : P(q')+\qfun{q}{\kappa,\neg b}\} }

    \inferrule[S-Loop]{ n < n' \quad (\varphi,\sifq{b[n_1/j]}{e[n_1/j]}) \Downarrow \varphi'
               \quad (\varphi',\sqwhile{j}{n\splus 1}{n'}{b}{\{e\}}) \Downarrow \varphi''}
                  {(\varphi,\sqwhile{j}{n}{n'}{b}{\{e\}}) \Downarrow \varphi''}
\quad
    \inferrule[S-Loop1]{ n \ge n' }
                  {(\varphi,\sqwhile{j}{n}{n'}{b}{\{e\}}) \Downarrow \varphi}

        \inferrule[S-Seq]{(\varphi,e_1) \Downarrow \varphi' \\(\varphi',e_2) \Downarrow \varphi''}
        {(\varphi,\sseq{e_1}{e_2}) \Downarrow \varphi'' }

    \inferrule[S-Mea]{\kappa=y\srange{0}{n} \\ r= \Msum_j \slen{z_j}^2
   \\ (\varphi\uplus \{\kappa' :  \Msum_j {\frac{z_j}{\sqrt{r}}}{\eta_j}\},e[(r,\tov{c})/x]) \Downarrow \varphi'}
  {(\varphi\uplus \{\kappa\sqcupplus\kappa': \Msum_j z_j\ket{c}{\eta_j}+\qfun{q}{\kappa,c \neq \kappa}\},\sexp{x}{\smea{y}}{e}) \Downarrow \varphi' }
  \end{mathpar}
}
\vspace*{-1.6em}
{\scriptsize
\begin{center}
$
\begin{array}{lclcl}
\denote{\circ}^{n}(\sum_{j}{z_j\ket{c_j}\eta_j})
&\triangleq&
 \Msum_{j}{z_j(\denote{\circ}\ket{c_j})\eta_j} & \texttt{where}& (\circ = \mu\vee \circ = op \vee \circ = b) \wedge \forall j\,\slen{c_j}=n
\\[0.2em]
\qfun{(\Msum_{i}{z_i}{\ket{c_{i}}}{\eta_i}+q)}{\kappa,b} &\triangleq& \Msum_{i}{z_i}{\ket{c_{i}}}{\eta_i}
      &\texttt{where}&\forall i.\,\slen{c_{i}}=\slen{\kappa}\wedge \denote{b[c_{i}/\kappa]}=\texttt{true}
\\[0.2em]
S^n(\Msum_{j}{z_j\ket{c_j}\beta_j\tos{\beta'_j}}) &\triangleq& \Msum_{j}{z_j\beta_j\tos{\ket{c_j}\beta'_j}} &\texttt{where}&\forall j\,\slen{c_j}=n
\\[0.2em]
P(\Msum_{j}{z_j\beta_j\tos{\ket{c_j}\beta'_j}}) &\triangleq& \Msum_{j}{z_j\ket{c_j}\beta_j\tos{\beta'_j}} 
\end{array}
$
\end{center}
}
\vspace*{-1em}
\caption{Selected semantic rules. $\tov{c}$ turns basis $c$ to an integer.}
\label{fig:exp-semantics-1}
\vspace*{-1.5em}
\end{figure*}

The \qafny semantics is formalized as a big-step transition relation $(\varphi,e) \Downarrow \varphi'$, with $\varphi$ / $\varphi'$ being quantum states as described in \Cref{fig:qafny-state}.
The judgment relation states that a program $e$ with the pre-state $\varphi$ transitions to a post-state $\varphi'$.
A selection of the rules defining $\Downarrow$ may be found \Cref{fig:exp-semantics-1}, and the additional rules are in \Cref{sec:add-sem-rules}.
$FV(\Omega, -)$ produces a locus by unioning all qubits in {$-$} with the quantum variable kind information in $\Omega$; its definition is given in \Cref{sec:session-gen}.

\ignore{
\begin{figure}[t]
  {
\begin{minipage}[t]{0.4\textwidth}
  \includegraphics[width=1\textwidth]{../res/conditional}
  \caption{Conditional Analogy}
\label{fig:qafny-con-analog}
\end{minipage}
\qquad
\begin{minipage}[t]{0.35\textwidth}
  \includegraphics[width=1\textwidth]{../res/measure}
  \caption{Measurement Analogy}
\label{fig:qafny-mea-analog}
\end{minipage}
  }
\end{figure}
}

\noindent\textbf{\textit{Assignment and Mutation Operations.}}
Rules \rulelab{S-ExpC} and \rulelab{S-ExpM} define the behaviors for $\cmode$ and $\mmode$ kind classical variable assignments, which perform variable-value substitutions.
Rule \rulelab{S-OP} defines a quantum heap mutation applying a state preparation operation ($op$) or an oracle expression ($\mu$) to a locus $\kappa$ for a \(\tcht\)-typed state.
Here, the locus fragment $\kappa$ to which the operation is applied must be the very first one in the locus \(\kappa \sqcupplus \kappa'\) that refers to the entire quantum state \(q\).
If not, we will first apply equivalence rewrites to be explained in \Cref{sec:type-checking} to move $\kappa$ to the front.
With \(\kappa\) preceding the rest fragment \(\kappa'\), the operation's semantic function $\denote{\circ}^{n}$ ($\circ$ being $\ihadh$ or $\mu$) is then applied to $\kappa$'s position bases in the quantum value \(q\).
More specifically, the function is only applied to the first $n$ (equal to $\slen{\kappa}$) basis bits of each basis-ket in the value while leaving the rest unchanged.
The semantic interpretations of the $op$ and $\mu$ operations are essentially the quantum gate semantics given in Li \textsf{et al.} \cite{oracleoopsla} and \Cref{sec:exp-mu}. 
For example, in \Cref{fig:ghz-proof} line \rulelab{P-Oracle}, we apply an oracle operation $x[j]\splus 1$ to $x[j]$, the first position of the locus $\kappa_2$ (\ie{$x[j]\sqcupplus x[0,j\sminus 1)$}), which transforms the first basis bit to $\ket{1}$.
Before the application, we rewrite the pre-state containing \(\kappa_1\) below the line \rulelab{EQ} in \Cref{fig:ghz-proof}, to the form corresponding to the locus $\kappa_2$ above the line.

\noindent\textbf{\textit{Quantum Conditionals.}}\label{sec:quantum-conditional} 
As in rule \rulelab{S-If}, for a conditional $\sifq{b}{e}$, we first evaluate the Boolean guard $b$ on $\kappa$'s position bases ($FV(\Omega,b)=\kappa$) of the quantum value state
$q$ to $\denote{b}^{\slen{\kappa}}q$\footnote{This is defined formally as an oracle, same as $\mu$ above, in \Cref{sec:exp-mu}.
} because $b$'s computation might have side-effects in changing $\kappa$'s position bases, as the example in \Cref{sec:quantumwalk}.
The quantum value referred by \(\kappa \sqcupplus \kappa'\) is further partitioned into \(\qfun{q}{\kappa,b} + \qfun{q}{\kappa, \neg b}\) where $\qfun{q}{\kappa,b}$ is a set of basis-kets whose $\kappa$'s position bases satisfying $b$ and $\qfun{q}{\kappa,\neg b}$ is the rest.
  Since the body \({e}\) only affects the basis-kets (\(\qfun{q}{\kappa,b}\)) satisfying the guard \({b}\), we rule out the basis-kets $\qfun{q}{\kappa,\neg b}$ (unsatisfying the guard) in $e$'s computation.
We also need to push $\kappa$'s position bases in $\qfun{q}{\kappa,b}$ to the frozen stacks through the $S^n$ operation to maintain the locus-state simultaneity in \Cref{sec:monitortypes}.

We describe rule \rulelab{S-If} along with an example in \Cref{fig:shorqafny} line 3-5. Here, the $j$-th iteration is unrolled to a quantum conditional $\qif{x[j]}{\qass{y\srange{0}{n}\;}{\;a^{2^j} \cdot y\srange{0}{n}\,\mmod\,N}}$.
The loci involved in the computation are $x[j]$ and $x\srange{0}{j}\sqcupplus y\srange{0}{n}$, and their state transitions are given as:

\vspace*{-0.5em}
{\footnotesize
\begin{center}
$
\begin{array}{ll}
&\textcolor{spec}{
\big{\{}x[j] : \frac{1}{\sqrt{2}}(\ket{0}+\ket{1})\big{\}} \uplus \big{\{}x\srange{0}{j}\sqcupplus y\srange{0}{n}: \schai{2^j \sminus 1}{\frac{1}{\sqrt{2^j}}}{i}{\ket{a^{i}\;\mmod\;N}} \big{\}}
}\\[0.3em]
\equiv
&\textcolor{spec}{
\big{\{}x[j]\sqcupplus x\srange{0}{j}\sqcupplus y\srange{0}{n} :
\schia{2^j\sminus 1}{\frac{1}{\sqrt{2^{j\texttt{+}1}}}}{\ket{1}\ket{i}}{\ket{a^{i}\;\mmod\;N}}
+\schia{2^j \sminus 1}{\frac{1}{\sqrt{2^{j\texttt{+}1}}}}{\ket{0}\ket{i}}{\ket{a^{i}\;\mmod\;N}}
\big{\}}
}
\\[0.3em]
\xrightarrow{\rulelab{S-If}}
&\textcolor{spec}{
\big{\{}x[j]\sqcupplus x\srange{0}{j}\sqcupplus y\srange{0}{n} :
\schia{2^j \sminus 1}{\frac{1}{\sqrt{2^{j\texttt{+}1}}}}{\ket{1}\ket{i}}{\ket{(a^{i} \cdot a^{2^j})\;\mmod\;N}}
+\schia{2^j \sminus 1}{\frac{1}{\sqrt{2^{j\texttt{+}1}}}}{\ket{0}\ket{i}}{\ket{a^{i}\;\mmod\;N}}
\big{\}}
}
\\[0.3em]
\equiv
&\textcolor{spec}{
\big{\{}
x\srange{0}{j\splus 1}\sqcupplus y\srange{0}{n} : \schai{2^{j\splus 1} \sminus 1}{\frac{1}{\sqrt{2^{j\splus 1}}}}{i}{\ket{a^{i}\;\mmod\;N}}
\big{\}}
}
\end{array}
$
\end{center}
}
\vspace*{-0.5em}

The first equation transition ($\equiv$) merges the two locus states and turns the merged state into two sets (separated by $+$), respectively representing basis-kets where $x[j]$'s position bases are $1$ and $0$.
Since the Boolean guard $x[j]$ has no side-effects, the application $\denote{b}^{\slen{\kappa}}$ is an identity.
The \rulelab{S-If} application performs a modulo multiplication oracle application on the basis-ket set where $x[j]$'s position bases being $1$, while the last equation merges the two sets back to one summation formula.
The \rulelab{S-If} application above can be further decomposed into two additional transitions in between:

\vspace*{-0.5em}
{\footnotesize
\begin{center}
$
\begin{array}{ll}
\longrightarrow
&\textcolor{spec}{
\big{\{}
x\srange{0}{j}\sqcupplus y\srange{0}{n} :
\schia{2^j \sminus 1}{\frac{1}{\sqrt{2^{j\texttt{+}1}}}}{\ket{i}}{\ket{a^{i}\;\mmod\;N}}
}\textcolor{stack}{\cmsg{\ket{1}}}
\textcolor{spec}{\big{\}}}
\\[0.4em]
\xrightarrow{\rulelab{S-OP}}
&\textcolor{spec}{
\big{\{}
x\srange{0}{j} \sqcupplus  y\srange{0}{n} :
\schia{2^j \sminus 1}{\frac{1}{\sqrt{2^{j\texttt{+}1}}}}{\ket{i}}{\ket{(a^{i} \cdot a^{2^j})\;\mmod\;N}}
}\textcolor{stack}{\cmsg{\ket{1}}}
\textcolor{spec}{\big{\}}}
\\
\end{array}
$
\end{center}
}
\vspace*{-0.5em}

The first transition removes the $\qfun{q}{\kappa,\neg b}$ part, e.g., the basis-ket set where $x[j]$'s position bases are $0$.
Additionally, for every basis-ket in the $\qfun{q}{\kappa,b}$ set, e.g., the basis-ket set where $x[j]$'s position bases being $1$,
we freeze $\kappa$'s position bases by pushing the bases into the basis-ket's stacks through the function application $S^{\slen{\kappa}}(\qfun{q}{\kappa,b})$, which finds the first $\slen{\kappa}$ bits in every basis-ket and push them into the basis-ket's stack so that $e$'s application targets locus $\kappa'$ instead of $\kappa \sqcupplus \kappa'$.
As the first transition above, for each basis-ket, we push $x[j]$'s position basis ($\ket{1}$) to the basis-ket's stack, as the $\textcolor{stack}{\cmsg{\ket{1}}}$ part and the pointed-to locus is rewritten to $x\srange{0}{j}\sqcupplus y\srange{0}{n}$.
After applying the body $e$ to the state, for every basis-ket, we pop $\kappa$'s position bases ($P(q')$) from the basis-ket's stack and relabel the locus of the state to be $\kappa \sqcupplus \kappa'$; in doing so, we also need to add the unmodified basis-kets $\qfun{q}{\kappa,\neg b}$ back into the whole state.
After applying $\rulelab{S-OP}$ on locus $x\srange{0}{j}\sqcupplus y\srange{0}{n}$ above,
we pop $\textcolor{stack}{\cmsg{\ket{1}}}$ from every basis-ket's stack and assemble the unchanged part ($\textcolor{spec}{\schia{2^j \sminus 1}{\frac{1}{\sqrt{2^{j\texttt{+}1}}}}{\ket{0}\ket{i}}{\ket{a^{i}\;\mmod\;N}}}$) back to the state of locus $x[j]\sqcupplus x\srange{0}{j}\sqcupplus y\srange{0}{n}$;
the result is shown as the state after the $\xrightarrow{\rulelab{S-If}}$ application above.

\noindent\textbf{\textit{Quantum Measurement.}}\label{sec:sem-measurement}
A measurement ($\sexp{x}{\smea{y}}{e}$) collapses a qubit array $y$, binds a $\mmode$-kind outcome to $x$, and restricts its usage in \(e\).
Rule \rulelab{S-Mea} shows the partial measurement behavior \footnote{A complete measurement is a special case of a partial measurement when $\kappa'$ is empty in \rulelab{S-Mea}}.
Assume that the locus containing the qubit array $y$ is $y\srange{0}{n}\sqcupplus\kappa'$, the measurement is essentially a two-step array filter:
(1) the basis-kets of the $\tcht$ typed value is partitioned into two sets (separated by $+$): $(\scha{m}{z_j}{c}{\ket{c_{j}}})+\qfun{q}{\kappa,c\neq \kappa}$ with $\kappa=y\srange{0}{n}$, by randomly picking a $\slen{\kappa}$-length basis $c$ where every basis-ket in the first set have $\kappa$'s position basis $c$;
and (2) we create a new array value by removing all the basis-kets not having $c$ as prefixes (the $\qfun{q}{\kappa,c\neq \kappa}$ part) and also removing the $\kappa$'s position basis in every remaining basis-ket; thus, the quantum value becomes
$\sch{m}{\frac{z_j}{\sqrt{r}}}{\eta_{j}}$. Notice that the element size of the post-state $m\splus 1$ is smaller than the size of the pre-state before the measurement. Since the amplitudes of basis-kets must satisfy $\Msum_i\slen{z_i}^2=1$, we need to normalize the amplitude of each element in the post-state by multiplying a factor $\frac{1}{\sqrt{r}}$, with $r=\Msum_{j=0}^m \slen{z_j}^2$ as the sum of the amplitude squares appearing in the post-state. 
When proving quantum program properties, the amplitudes appearing in basis-kets usually follow a periodic pattern that users can provide, so computing $r$ will be relatively simple, see \Cref{sec:semantics-proof}.
In \Cref{fig:shorqafny}, the measurement (line 8) transitions the state from lines 7 to 9.
Locus $y\srange{0}{n}$'s position basis is $\ket{a^{i}\;\mmod\;N}$ for each basis-ket in $\schai{2^{n}\sminus 1}{\frac{1}{\sqrt{2^{n}}}}{i}{\ket{a^{i}\;\mmod\;N}}$. We then randomly pick the basis value $a^{t}\;\mmod\;N$ as a measurement result, stored in $u$, and the probability of the pick is $\frac{p}{2^n}$ where $p$ is the order of $a$ and $N$. The probability is computed solely based on $p$ because it represents the period of the factorization in Shor's algorithm.
The number ($r$) of remaining basis-kets in range $x\srange{0}{n}$ is computed by rounding $\frac{2^n}{p}$.

\vspace{-0.5em}
\subsection{Qafny Locus Type System}\label{sec:type-checking}

\begin{figure}[t]
\vspace{-0.5em}
  {
    {\scriptsize
      \begin{mathpar}
      \hspace*{-0.3em}
        \inferrule[T-Par]{\sigma \preceq \sigma' \;\quad \Omega;\sigma' \vdash_g e \triangleright \sigma''}{\Omega;\sigma \vdash_g e \triangleright \sigma'' }

        \inferrule[T-ExpC]{x\not\in \dom{\Omega} \;\quad 
          \Omega;\sigma\vdash_g e[n/x] \triangleright \sigma'}{\Omega;\sigma \vdash_g \sexp{x}{n}{e} \triangleright \sigma' }

    \inferrule[T-ExpM]{x\not\in \dom{\Omega} \;\quad  \Omega \vdash a : \mmode \;\quad 
\Omega[x\mapsto \mmode];\sigma\vdash_g e \triangleright \sigma'}{\Omega;\sigma \vdash_g \sexp{x}{a}{e} \triangleright \sigma' }
      \end{mathpar}
     \vspace*{-0.5em}
     \begin{mathpar}
    \inferrule[T-OP]{\circ=op\vee\circ=\mu}{\Omega;\sigma\uplus\{\kappa\sqcupplus\kappa': \tcht\} \vdash_g \ssassign{\kappa}{}{\circ} \triangleright \sigma\uplus\{\kappa\sqcupplus\kappa': \tcht\} }

        \inferrule[T-Mea]{\Omega(y)=\qmode{n} \;\;\quad x\not\in\dom{\Omega} \\\\ \Omega[x\mapsto \mmode];\sigma\uplus\{\kappa : \tcht\}\vdash_{\cmode} e \triangleright  \sigma'}{\Omega;\sigma\uplus\{y\srange{0}{n}\sqcupplus\kappa: \tau\} \vdash_{\cmode} \sexp{x}{\smea{y}}{e} \triangleright \sigma' }

        \inferrule[T-If]{ 
          FV(\Omega,b)=\kappa \;\quad
          FV(\Omega,e) \subseteq \kappa'
          \;\quad
          \Omega;\sigma\uplus\{\kappa' : \tcht\} \vdash_{\mmode} e \triangleright \sigma\uplus\{\kappa' : \tcht\}}
        {\Omega;\sigma\uplus\{\kappa \sqcupplus \kappa' : \tcht\} \vdash_g \sifq{b}{e} \triangleright \sigma\uplus\{\kappa \sqcupplus \kappa' : \tcht\} }
\qquad
        \inferrule[T-Seq]{\Omega;\sigma\vdash_g e_1 \triangleright\sigma_1
          \;\quad\Omega;\sigma_1\vdash_g e_2 \triangleright\sigma_2}
        {\Omega;\sigma \vdash_g \sseq{e_1}{e_2}\triangleright \sigma_2}

        \inferrule[T-Loop]{x\not\in \dom{\Omega} \\
\forall j\in[n_1,n_2)\,.\,\Omega;\sigma[j/x]\vdash_g \sifq{b[j/x]}{e[j/x]} \triangleright \sigma[j\splus 1/x] }
        {\Omega;\sigma[n_1/x] \vdash_g \sqwhile{x}{n_1}{n_2}{b}{\{e\}} \triangleright \sigma[n_2/x]}
      \end{mathpar}
    }
  }
\vspace*{-1.5em}
  \caption{\qafny type system. $FV(\Omega, -)$ gets a locus containing qubits in $-$ w.r.t. $\Omega$ (Appx. \ref{sec:session-gen}). }
  \label{fig:exp-sessiontype}
  \vspace*{-1.4em}
\end{figure}

The \qafny typing judgment
$\Omega; \sigma\vdash_{g} e \triangleright \sigma'$ states that $e$ is well-typed under the context mode $g$ (the syntax of kind $g$ is reused as context modes) and environments
$\Omega$ and $\sigma$.
The kind environment $\Omega$ is populated through \texttt{let} and \texttt{for} loops that introduce $\cmode$ and $\mmode$ kind variables, while $\qmodename$-kind variable mappings in $\Omega$ are given as a global environment.
Selected type rules are in \Cref{fig:exp-sessiontype}; the rules not mentioned are similar and given in \Cref{sec:state-eq}.
For every type rule, well-formed domains \((\Omega \vdash \dom{\sigma})\) are required but hidden from the rules, such that every variable used in all loci of \(\sigma\) must appear in \(\Omega\), while $\Omega \vdash a : \mmode$ judges that the expression $a$ is well-formed and returns an $\mmode$ kind; see \Cref{sec:session-gen,appx:well-formeda}.
The type system enforces three properties below.  

\noindent\textbf{\textit{No Cloning and Observer Breakdown.}}
We enforce no cloning by disjointing qubits mentioned in a quantum conditional Boolean guard and its body.
In rule \textsc{T-If}, $\kappa$ and $\kappa'$ are disjoint unioned, and the two $FV$ side-conditions ensure that the qubits mentioned in the Boolean guard and conditional body are respectively within $\kappa$ and $\kappa'$; thus, they do not overlap.
\qafny is a \emph{flow-sensitive} language, as we enforce no observer breakdown by ensuring no classical variable assignments through the \qafny syntax and no measurements inside a quantum conditional through context restrictions.
Each program begins with the context mode $\cmode$, which permits all \qafny operations.
Once a type rule switches the mode to $\mmode$, as in \textsc{T-If}, measurement operations are suspended in this scope, as \textsc{T-Mea} is valid only if the context mode is $\cmode$.
For instance, let's imagine that the measurement in \Cref{fig:shorqafny} line 8 lives inside the for-loop in line 5, which our type system would forbid because type checking through \rulelab{T-Loop} calls rule \textsc{T-If} that marks the context mode to $\mmode$, while the application of rule \rulelab{T-Mea} requires a $\cmode$ mode context to begin with.

\begin{figure}[t]
\captionsetup[subfigure]{justification=centering}
{\scriptsize
{\hspace*{-2.3em}
\begin{minipage}[t]{0.49\textwidth}
\subcaption{Environment Equivalence}
\vspace*{-2.8em}
\begin{center}
 \[\hspace*{-0.5em}
  \begin{array}{r@{~}c@{~}l}
    \sigma & \preceq & \sigma \\[0.2em]
  \{\emptyset:\tau\} \uplus \sigma &\preceq& \sigma\\[0.2em]
   \{\kappa:\tau\} \uplus \sigma &\preceq& \{\kappa:\tau'\} \uplus \sigma\\
&&\texttt{where}\;\;\tau\sqsubseteq\tau' \\[0.2em]
  \{\kappa_1\sqcupplus s_1 \sqcupplus s_2 \sqcupplus \kappa_2 :\tau\} \uplus \sigma &\preceq& \{\kappa_1\sqcupplus s_2 \sqcupplus s_1 \sqcupplus \kappa_2 : \tau\} \uplus \sigma\\\\[0.2em]
  \{\kappa_1 :\tau\} \uplus \{\kappa_2 :\tau\} \uplus \sigma &\preceq& \{\kappa_1 \sqcupplus \kappa_2 :\tau\} \uplus \sigma \\\\[0.2em]
  \{\kappa_1 \sqcupplus \kappa_2 :\tau\} \uplus \sigma &\preceq& \{\kappa_1 :\tau\} \uplus \{\kappa_2 :\tau\} \uplus \sigma
\\
&&
    \end{array}
  \]
\end{center}
  \label{fig:env-equiv}
\end{minipage}
\vline height -8ex
\begin{minipage}[t]{0.49\textwidth}
\subcaption{State Equivalence}
\vspace*{-2.8em}
\begin{center}
   \[\hspace*{-0.5em}
   \begin{array}{r@{~}c@{~}l}
    \varphi & \equiv & \varphi \\[0.2em]
  \{\emptyset:q\} \uplus \varphi &\equiv& \varphi\\[0.2em]
   \{\kappa:q\} \uplus \varphi &\equiv& \{\kappa:q'\} \uplus \varphi\\
   &&\texttt{where}\;\;q\equiv_{\slen{\kappa}}q' 
   \\[0.2em]
  \{\kappa_1\sqcupplus s_1 \sqcupplus s_2 \sqcupplus \kappa_2 :q\} \uplus \varphi &\equiv& \{\kappa_1\sqcupplus s_2 \sqcupplus s_1 \sqcupplus \kappa_2 : q'\} \uplus \varphi
\\
&&\texttt{where}\;\;q'=q^{\slen{\kappa_1}}\langle \slen{s_1}\asymp \slen{s_2} \rangle
\\[0.2em]
  \{\kappa_1 :q_1\} \uplus \{\kappa_2 :q_2\} \uplus \varphi &\equiv& \{\kappa_1 \sqcupplus \kappa_2 :q'\} \uplus \varphi 
\\
&&\texttt{where}\;\;q'= q_1\bowtie q_2
   \\[0.2em]
  \{\kappa_1 \sqcupplus \kappa_2 :\varphi\} \uplus \sigma &\equiv& \{\kappa_1 :\varphi_1\} \uplus \{\kappa_2 :\varphi_2\} \uplus \sigma
\\
&&\texttt{where}\;\;\varphi_1 \bowtie \varphi_2=\varphi\wedge \slen{\varphi_1}=\slen{\kappa_1}
    \end{array}
 \]
\end{center}
  \label{fig:qafny-stateequiv}
\end{minipage}
\vspace*{-1.5em}
{\scriptsize
\[\hspace*{-1em}
\begin{array}{l}
\textcolor{blue}{\text{Permutation:}}\\[0.5em]
{
\hspace*{-0.2em}
\begin{array}{r@{~}c@{~}l@{~}c@{~}l}
(q_1 \bigotimes q_2 \bigotimes q_3 \bigotimes q_4)^{n}\langle i\asymp k\rangle
&\triangleq&
q_1 \bigotimes q_3 \bigotimes q_2 \bigotimes q_4
&\;\;\;\texttt{where}&\;\;\slen{q_1}=n\wedge \slen{q_2}=i\wedge \slen{q_3}=k\\[0.5em]
(\Msum_{j}{z_j\ket{c_j}\ket{c'_j}\ket{c''_j}\eta_j})^{n}\langle i\asymp k\rangle
&\triangleq&
\Msum_{j}{z_j\ket{c_j}\ket{c''_j}\ket{c'_j}\eta_j}
&\;\;\;\texttt{where}&\;\;\slen{c_j}=n\wedge \slen{c'_j}=i\wedge \slen{c''_j}=k
\end{array}
}
\\[1.4em]
\textcolor{blue}{\text{Join Product:}}\\
{
\setlength\arraycolsep{4pt}
\begin{array}{r@{~}c@{~}l@{~}c@{~}l r@{~}c@{~}l@{~}c@{~}l}
z_1\ket{c_1}&\bowtie& z_2\ket{c_2} &\triangleq& (z_1 \cdot z_2)\ket{c_1}\ket{c_2}
&
\sch{n}{z_j}{\ket{c_j}}&\bowtie&\schk{m}{z_k}{c_k} &\triangleq& \schi{n\cdot m}{z_j\cdot z_k}{c_j}\ket{c_k}
\\[0.5em]
\ket{c_1}&\bowtie&\Msum_{j}{z_j \eta_j}&\triangleq&\Msum_{j}{z_j\ket{c_1}\eta_j}
&
(\ket{0}+\alpha(r)\ket{1})&\bowtie&\Msum_{j}{z_j \eta_j}
&\triangleq&
\Msum_{j}{z_j\ket{0}\eta_j}+\Msum_{j}{(\alpha(r)\cdot z_j)\ket{1}\eta_j}
\end{array}
}
\end{array}
\]
}
\vspace*{-1.4em}
  \caption{\qafny type/state relations. $\cdot$ is math mult. Term $\Msum^{n\cdot m}P$ is a summation omitting the indexing details. $\bigotimes$ expands a $\thadt$ array, as $\frac{1}{\sqrt{2^{n+m}}}\Motimes_{j=0}^{n\splus m\sminus 2}q_j=(\frac{1}{\sqrt{2^{n}}}\Motimes_{j=0}^{n\sminus 1}q_j)\bigotimes(\frac{1}{\sqrt{2^{m}}}\Motimes_{j=0}^{m\sminus 1}q_j)$.}
  \label{fig:qafny-eq}
}
}
\vspace*{-1.5em}
\end{figure}

\noindent\textbf{\textit{Guiding Locus Equivalence and Rewriting.}}
The semantics in \Cref{sec:semantics} assumes that the loci in quantum states can be in ideal forms, e.g., rule \rulelab{S-OP} assumes that the target locus $\kappa$ are always prefixed.
This step is valid if we can rewrite (type environment partial order $\preceq$) the locus to the ideal form through rule $\textsc{T-Par}$, which interconnectively rewrites the locus appearing in the state, through our state equivalence relation ($\equiv$), as the locus state simultaneity enforcement (\Cref{sec:monitortypes}).
The state equivalence rewrites have two components. 

First, the type and quantum value forms have simultaneity, i.e., given a type $\tau_1$ for a locus $\kappa$ in a type environment ($\sigma$), if it is a subtype ($\sqsubseteq$) of another type $\tau_2$, $\kappa$'s value $q_1$ in a state ($\varphi$) can be rewritten to $q_2$ that has the type $\tau_2$ through state equivalence rewrites ($\equiv_n$) where $n$ is the number of qubits in $q_1$ and $q_2$. Both $\sqsubseteq$ and $\equiv_n$ are reflexive and types $\tnort$ and $\thadt$ are subtypes of $\tcht$, which means that a $\tnort$ typed value ($\ket{c}$) and a $\thadt$ typed value ($\shad{2^n}{n\sminus 1}{\alpha(r_j)}$) can be rewritten to an $\tcht$ typed value (\Cref{sec:state-eq}). For example, range $x[0,n)$'s $\thadt$ typed value $\shad{2^n}{n\sminus 1}{}$ in \Cref{fig:shorqafny} line 1 can be rewritten to an $\tcht$ type as $\schii{2^{n}\sminus 1}{\frac{1}{\sqrt{2^{n}}}}{i}$.
If such a rewrite happens, we correspondingly transform $x[0,n)$'s type to $\tcht$ in the type environment. 

Second, type environment partial order ($\preceq$) and state equivalence ($\equiv$) also have simultaneity ---
in a proof judgment, we associate the state predicate, representing a state $\varphi$, with the type environment $\sigma$ by sharing the same domain, i.e., $\dom{\varphi}=\dom{\sigma}$.
Thus, the environment rewrites ($\preceq$) happening in $\sigma$ gear the state rewrites in $\varphi$, e.g., the bottom proof step of \Cref{fig:ghz-proof} transforms locus $x\srange{0}{j}$ in $\sigma$ to locus $\kappa$ ($x[j\sminus 1]\sqcupplus x\srange{0}{j\sminus 1}\sqcupplus x[j]$) above it, and the state rewrites in the pre-condition predicate happen accordingly as (left to right):

\vspace*{-0.4em}
{\footnotesize
\begin{center}
$
\begin{array}{l@{~}c@{~}lclcl}
\textcolor{spec}{
\{x\srange{0}{j} : \schd{d}{1}{\frac{1}{\sqrt{2}}}{\overline{d}}}\}
&
\textcolor{spec}{\uplus}& 
\textcolor{spec}{ \{x[j] : \ket{0}\}
}
&\equiv&
\textcolor{spec}{
\{x\srange{0}{j\,\splus\, 1} : \schd{d}{1}{\frac{1}{\sqrt{2}}}{\overline{d}}\ket{0}\}
}
&\equiv&
\textcolor{spec}{
\{
\kappa : \schd{d}{1}{\frac{1}{\sqrt{2}}}{\overline{d}}\ket{0}\}
}
\\
\textcolor{spec}{
\{x\srange{0}{j} : \tcht \}}
&\textcolor{spec}{\uplus}& 
\textcolor{spec}{\{x[j] : \tnort\}
}
&\preceq&
\textcolor{spec}{
\{x\srange{0}{j\,\splus\,1} : \tcht\}
}
&\preceq&
\textcolor{spec}{
\{\kappa : \tcht\}
}
\end{array}
$
\end{center}
}
\vspace*{-0.4em}

Here, we add qubit $x[j]$ ($\ket{0}$) to the end of locus $x\srange{0}{j}$ and transform locus $x\srange{0}{j\,\splus\,1}$ to $\kappa$, so the upper proof step (\rulelab{P-If}) in \Cref{fig:ghz-proof} can proceed.  The above rewrites are derived by the rules in \Cref{fig:qafny-eq},
where the rules in environment partial order and state equivalence are one-to-one corresponding. 
The first three lines describe the properties of reflective, identity, and subtyping equivalence.
The fourth line enforces that the environment and state are close under locus permutation.
After the equivalence rewrite, the position bases of ranges $s_1$ and $s_2$ are mutated by applying the function $q^{\slen{\kappa_1}}\langle \slen{s_1}\asymp \slen{s_2} \rangle$.
One example is the locus rewrite in \Cref{fig:shorqafny} line 7 from left to right, as:

\vspace*{-0.6em}
{\footnotesize
\begin{center}
$\begin{array}{lcl}
\big{\{}x\srange{0}{n}\sqcupplus y\srange{0}{n} : \tcht \big{\}} &\preceq& \big{\{}y\srange{0}{n}\sqcupplus x\srange{0}{n} : \tcht \big{\}} 
\\[0.2em]
\big{\{}x\srange{0}{n}\sqcupplus y\srange{0}{n} : \textcolor{spec}{\schai{2^{n}\sminus 1}{\frac{1}{\sqrt{2^{n}}}}{i}{\ket{a^{i}\;\mmod\;N}}} \big{\}}
&\equiv& 
\big{\{} y\srange{0}{n}\sqcupplus x\srange{0}{n} : \textcolor{spec}{\schai{2^{n}\sminus 1}{\frac{1}{\sqrt{2^{n}}}}{a^{i}\;\mmod\;N}{\ket{i}}}\big{\}}
\end{array}
$
\end{center}
}
\vspace*{-0.6em}

\ignore{

Here, we add qubit $x[j]$ ($\ket{0}$) to the end of locus $x\srange{0}{j}$ and
transform locus $x\srange{0}{j\,\splus\,1}$ to $\kappa$, so the upper proof step
(\rulelab{P-If}) in \Cref{fig:ghz-proof} can proceed.

The type system tracks the qubit arrangement of loci and permits the locus type environment partial order rewrites ($\preceq$) through rule $\textsc{T-Par}$.
In $\qafny$, we introduce equivalence relations for type environments (partial order $\preceq$) and states ($\equiv$) to permit automated state rewrites, as discussed in \Cref{sec:monitortypes}.
We develop \emph{equivalence relations} over quantum loci, quantum values, 
type environments, and states (\Cref{sec:state-eq}).
The common equivalence relations are state-type rewrites, permutations, split
and join.
Recall in \Cref{sec:monitortypes} that state rewrites are guided by type environment rewrites.
Similarly, type rewrites would also transform quantum state value forms. 
An example is to rewrite an $\tcht$-typed value $\ssum{j=0}{1}{z_j\beta_j}$ to
$\tnort$-typed ${z_0\beta_0}$ in \Cref{sec:quantumwalk}.
}
The last two lines in \Cref{fig:env-equiv,fig:qafny-stateequiv} describe locus joins and splits, where the latter is an inverse of the former but much harder to perform practically.
In the most general form, joining two $\tcht$-type states computes the Cartesian product of their basis-kets, shown in the bottom of \Cref{fig:qafny-eq},
which is practically hard for proof automation.
Fortunately, the join operations in most quantum algorithms are between a $\tnort$/$\thadt$ typed and an $\tcht$-typed state, 
Joining a $\tnort$-typed and $\tcht$-typed state puts extra qubits in the
right location in every basis-ket of the $\tcht$-typed state as discussed in \Cref{sec:monitortypes1}.  
Joining a $\thadt$-typed qubit (single qubit state) and $\tcht$-typed state duplicates the
$\tcht$-typed basis-kets.  
In every loop step in \Cref{fig:shorqafny} line 3-5, we add a $\thadt$-typed qubit
$x[j]$ to the middle of an $\tcht$-typed locus $x\srange{0}{j}\sqcupplus y\srange{0}{n}$, transform the state to:

\vspace*{-0.7em}
{\footnotesize
\begin{center}
$
\big{\{}x\srange{0}{j\splus 1}\sqcupplus  y\srange{0}{n} : \textcolor{spec}{\schai{2^j\sminus 1}{\frac{1}{\sqrt{2^j}}}{i}{\ket{0}\ket{a^{i}\;\mmod\;N}}+\scha{2^j}{\frac{1}{\sqrt{2^j\sminus 1}}}{i}{\ket{1}\ket{a^{i}\;\mmod\;N}}} \big{\}}
$
\end{center}
}
\vspace*{-0.7em}

The state can be further rewritten to the one in \Cref{fig:shorqafny} by merging
the above two parts (separated by $+$).
Notice that the basis-kets are still all distinct because the two parts
are distinguished by $x[j]$'s position basis, i.e., $\ket{0}$ and $\ket{1}$.
\Cref{sec:newtype} shows practical ways to perform additional state joins and splits, including an upgraded dependent type
system to permit a few cases of splitting $\tcht$ typed values.

\noindent\textbf{\textit{Approximating Locus Scope.}}
The type system approximates locus scopes.
In rule \textsc{T-If}, we use $\kappa \sqcupplus \kappa'$ as the approximate locus large enough to describe all possible qubits directly and indirectly mentioned in $b$ and $e$.
Such scope approximation might be over-approximated, which does not cause incorrectness in our proof system, while under-approximation is forbidden.
For example, if we combine two $\thadt$-typed qubits in our system and transform the value to $\tcht$-type as $\frac{1}{2}(\ket{00}+\ket{01}+\ket{10}+\ket{11})$, this is an over-approximation since the two qubits are not entangled.
Partially measuring the first qubit leaves the second qubit's value unchanged as $\frac{1}{\sqrt{2}}(\ket{0}+\ket{1})$.

In addition to the above properties, we allow $\cmode$-kind classical variables introduced by \texttt{let} to be evaluated in the type checking stage \footnote{We consider all computation that only needs classical computers is done in the compilation time.}, while tracks $\mmode$ variables in $\Omega$.
Rule \textsc{T-ExpC} enforces that a classical variable $x$ is replaced with its assigned value $n$ in $e$, and classical expressions in $e$ containing $x$ are evaluated, so the proof system can avoid handling constants.

\vspace*{-0.5em}
\subsection{The Qafny Proof System}\label{sec:semantics-proof}

\begin{figure*}[t]
{\footnotesize
  \begin{mathpar}
\hspace*{-1em}
    \inferrule[P-Frame]{\dom{\sigma}\cap FV(\Omega,R)=\emptyset \\\\
       FV(\Omega,e)\subseteq \dom{\sigma} \\ \fivepule{\Omega}{\sigma}{g}{P}{e}{Q}}
                {\fivepule{\Omega}{\sigma\uplus\sigma'}{g}{P * R}{e}{Q * R}}

    \inferrule[P-Con]{\sigma \preceq \sigma' \\ P \Rightarrow P'\\  \fivepule{\Omega}{\sigma'}{g}{P'}{e}{Q'} \\ Q' \Rightarrow Q}
                {\fivepule{\Omega}{\sigma}{g}{P}{e}{Q}}
      \end{mathpar}
      \begin{mathpar}
      \inferrule[P-OP]{\circ = op \vee \circ = \mu}
      {\fivepule{\Omega}{\{\kappa \sqcupplus \kappa' : \tcht\}}{g}{\kappa \sqcupplus \kappa' \mapsto q}
        {\ssassign{\kappa}{}{\circ}}{\kappa \sqcupplus \kappa' \mapsto \denote{\circ}^{\slen{\kappa}}q}}

    \inferrule[P-ExpC]{\fivepule{\Omega}{\sigma}{g}{P}{e[n/x]}{Q}}
                {\fivepule{\Omega}{\sigma}{g}{P}{\sexp{x}{n}{e}}{Q}}

    \inferrule[P-Mea]{x\not\in\dom{\Omega}\\\fivepule{\Omega[x\mapsto \mmode]}{\sigma\uplus\{\kappa: \tcht\}}{\cmode}{P[\mea(x,n,\kappa)/y\srange{0}{n}\sqcupplus\kappa]}{e}{Q} }
     {\fivepule{\Omega}{\sigma\uplus\{y\srange{0}{n}\sqcupplus\kappa: \tcht\}}{\cmode}{P}{\sexp{x}{\smea{y}}{e} }{Q} }

      \inferrule[P-If]{ FV(\Omega,b) = \kappa  \\ \fivepule{\Omega}{\{\kappa' : \tcht\}}{\mmode}{P[\frz(b,\kappa,\kappa')/ \kappa \sqcupplus \kappa']}{e}{Q}}
      {\fivepule{\Omega}{\{\kappa \sqcupplus \kappa' : \tcht\}}{g}{P}{\sifq{b}{e}}{P[\ufrz(\neg b,\kappa,\kappa \sqcupplus \kappa')/\kappa \sqcupplus \kappa'] * Q[\ufrz(b,\kappa,\kappa \sqcupplus \kappa')/\kappa']}
      }
      \end{mathpar}
      \begin{mathpar}
      \hspace*{-0.5em}
    \inferrule[P-Loop]{n < n' \quad \fivepule{\Omega}{\sigma}{g}{P(j)\wedge j \,\slt\, n'}{\sifq{b}{e}}{P(j\splus 1)} }
     {\fivepule{\Omega}{\sigma[n/j]}{g}{P(n)}{ \sqwhile{j}{n}{n'}{b}{\{e\}} }{P(n')} }
\qquad
    \inferrule[P-Seq]{\Omega;\sigma\vdash_g e_1 \triangleright\sigma_1\\\\\fivepule{\Omega}{\sigma}{g}{P}{e_1 }{P'} \;\quad \fivepule{\Omega}{\sigma_1}{g}{P'}{e_2 }{Q}}
     {\fivepule{\Omega}{\sigma}{g}{P}{\sseq{e_1}{e_2} }{Q} }
  \end{mathpar}
}
\vspace*{-0.8em}
\caption{Select proof rules.}
\label{fig:exp-proofsystem-1}
\vspace*{-1.4em}
\end{figure*}

Every valid proof judgment $\fivepule{\Omega}{\sigma}{g}{P}{e}{Q}$, shown in \Cref{fig:exp-proofsystem-1}, contains a pre- and post-condition predicates $P$ and $Q$ (syntax in \Cref{fig:vqimp}) for the statement $e$, satisfying the type restriction that $\Omega;\sigma\vdash_g e \triangleright \sigma'$; we also enforce that the predicate well-typed restrictions $\Omega;\sigma\vdash P$ and $\Omega;\sigma'\vdash Q$, meaning that all loci mentioned in $P$ must be in the right forms and as elements in $\dom{\sigma}$, introduced in \Cref{appx:well-formeda}. We state the restrictions as the typechecking constraint ($\mathit{TC}$)  below:

\vspace*{-0.5em}
{\small
\begin{center}
$
TC(\sigma,P,Q)\triangleq\Omega;\sigma\vdash_g e \triangleright \sigma_1
\wedge
\Omega;\sigma\vdash P 
\wedge
\Omega;\sigma_1\vdash Q
$
\end{center}
}
\vspace*{-0.5em}

Rule \rulelab{P-Con} describes the consequence rule where a well-formed
pre- and post-conditions \(P\) and $Q$ under \(\sigma\) is replaced by \(P'\) and $Q'$, well-formed under \(\sigma'\). 
Under the new conditions, we enforce a new type constraint $\mathit{TC}(\sigma', P', Q')$.
Rule \rulelab{P-Frame} is a specialized separation logic frame rule that separates the locus type environment and the quantum value to support local reasoning on quantum states.
Rule \rulelab{M-Frame} in \Cref{fig:qafny-model-rules} describes the predicate semantics of the separating conjunction $*$, where $\psi$ is a local store mapping from $\mmode$-kind variables to $\mmode$-kind values $(r,n)$; we require $\dom{\psi}\subseteq\dom{\Omega}$ and $\mmode$-kind variables modeled by \rulelab{M-Local}.
Besides predicate well-formedness, the predicate semantic judgment (\(\Omega; \psi;\sigma; \varphi \models_g P\)) also ensures the states (\(\varphi\)) being well-formed ($\Omega;\sigma\vdash_g \varphi$), defined as follows:

\begin{definition}[Well-formed \qafny state]\label{def:well-formed}\rm 
  A state $\varphi$ is \emph{well-formed}, written as
  $\Omega;\sigma \vdash_g \varphi$, iff $\dom{\sigma} = \dom{\varphi}$,
  $\Omega\vdash\dom{\sigma}$ (all variables in $\varphi$ are in $\Omega$), and:
  \begin{itemize}
  \item For every $\kappa \in \dom{\sigma}$, s.t. $\sigma(\kappa) = \tnort$,
    $\varphi(\kappa)=z\ket{c}\tos{\beta}$ and $\slen{\kappa} = \slen{c}$
    and $\slen{z}\le 1$; specifically, if $g=\cmode$,
    $\beta=\emptyset$ and $\slen{z} = 1$. \footnote{$\slen{\kappa}$ and $\slen{c}$ are the lengths of $\kappa$ and
      $c$, and $\slen{z}$ is the norm. } 
\item For every $\kappa \in \dom{\sigma}$, s.t. $\sigma(\kappa) = \thadt$,
    $\varphi(\kappa)=\shad{2^n}{n \sminus 1}{\alpha(r_j)}$ and $\slen{\kappa}=n$.
\item For every $\kappa \in \dom{\sigma}$, s.t. $\sigma(\kappa) = \tcht$,
    $\varphi(\kappa)=\sch{m}{z_j}{\ket{c_j}}\tos{\beta_j}$, and for all $j$, $\slen{\kappa}= \slen{c_j}$ and
    $\sum_{j=0}^m\slen{z}^2\le 1$; specifically, if $g=\cmode$,
    for all $j$, $\beta_j=\emptyset$ and $\sum_{j=0}^m\slen{z}^2 = 1$.
  \end{itemize}
\end{definition}

Here, an $\mmode$ mode state, representing a computation living in an $\mmode$ mode context, has a relaxed well-formedness, where $\sum_{j=0}^m\slen{z}^2\le 1$ and $\beta_j\neq \emptyset$. 
This is needed for describing the state inside the execution of a conditional body in rule \rulelab{S-If} in \Cref{sec:quantum-conditional}, where unmodified basis-kets are removed before the execution. 
There is a trick to utilizing the frozen stacks for promoting proof automation, as the modeling rule \rulelab{M-Map} equates two quantum values by discarding the frozen stack qubits, and we will see an example in \Cref{sec:controled-ghz}.

The predicate syntax (\Cref{fig:vqimp}) introduces three locus predicate transformers $\frz$, $\ufrz$, and $\mea$ in the locus syntax category, but their semantics (\Cref{fig:qafny-model-rules}) essentially transform quantum states in the predicates, as we define them in equational style, explained below.

\noindent\textbf{\textit{Assignment and Heap Mutation Operations.}}
Rule \rulelab{P-ExpC} describes $\cmode$-kind variable substitutions.
Rule \rulelab{P-OP} is a classical separation logic style heap mutation rule for state preparations $\ssassign{\kappa}{}{op}$ and oracles $\ssassign{\kappa}{}{\mu}$, which analogize such operations as classical array map operations mentioned in \Cref{fig:intros-example}.
Here, we discuss the cases when the state of the target loci $\kappa\sqcupplus\kappa'$ is of type $\tcht$, while some other cases are introduced in \Cref{appx:proof-rules}.
Each element in the array style pre-state $q$, for locus \(\kappa\sqcupplus\kappa'\), represents a basis-ket $z_j\ket{c_j}\eta_j$, with $\slen{\kappa}=\slen{c_j}$. Here, we first locate $\kappa$'s position basis $\ket{c_j}$ in each basis-ket of $q$, and then apply the operations $op$ or $\mu$ to $\ket{c_j}$.

\begin{figure*}[t]
{\scriptsize
$\hspace*{-0.5em}
\begin{array}{ll}
\textcolor{blue}{\footnotesize \text{Predicate Model Rules:}}\\[0.3em]
\begin{array}{ll@{\;}c@{\;}l}
\rulelab{M-Map}&
\Omega;\psi;\sigma;\varphi\uplus\{\kappa : \Msum_{j}{z_j}{\ket{c_j}}{\tos{\beta_j}}\} &\models_g& \kappa \mapsto \Msum_{j}{z_j}{\ket{c_j}}{\tos{\beta'_j}}
\\[0.3em]
\rulelab{M-Local}&
\Omega[x\mapsto \mmode];\psi[x\mapsto (r,v)];\sigma; \varphi &\models_g& P 

\qquad\qquad
\;\;\,\quad\text{if} \quad
\Omega;\psi;\sigma; \varphi \models_g P[(r,v)/x]
\\[0.3em]
\rulelab{M-Frame}&
\Omega;\psi;\sigma \uplus \sigma'; \varphi \uplus \varphi' &\models_g& P * Q

\qquad\qquad\text{if} \quad
\Omega;\psi;\sigma; \varphi \models_g P \text{  \;and\;  } \Omega;\psi;\sigma'; \varphi' \models_g Q
\\[0.3em]
\end{array}
\\
\textcolor{blue}{\footnotesize \text{Transformation Rules:}}
\\[0.1em]
\begin{array}{cl}
\rulelab{M-\(\frz\)}&
 \frz(b,\kappa,\kappa') \mapsto q
\qquad\qquad\qquad\qquad\qquad\;\;\;
=\quad \kappa' \mapsto \Msum_{j}{z_j\beta_j\tos{\ket{c_j}\beta'_j}}
\\[0.3em]
&
\qquad\qquad
\qquad\qquad
\text{ where }\;
\denote{b}^{\slen{\kappa}}q=\qfun{q}{\kappa,b}+\qfun{q}{\kappa,\neg b}
\;\wedge\;
\qfun{q}{\kappa,b}=\Msum_{j}{z_j}{\ket{c_{j}}}\beta_j{\tos{\beta'_j}}
\;\wedge\;
\forall j.\; \slen{c_{j}}=\slen{\kappa}
\\[0.2em]
\rulelab{M-\(\ufrz\)}&
{\hspace*{-1em}
\begin{array}{l}
\;\ufrz(\neg b,\kappa,\kappa') \mapsto \Msum_{j}{z_j}{\ket{c_{j}}}{\eta_j}+\qfun{q}{\kappa,\neg b}
\\[0.3em]
\;\;\;  * \ufrz(b,\kappa,\kappa') \mapsto \Msum_{j}{z'_j \beta_j\tos{\ket{c_{j}}\beta'_j}}
\end{array}
}
\,
=\quad\kappa' \mapsto \Msum_{j}{z'_j}{\ket{c_{j}}}{\beta_j\tos{\beta'_j}}+\qfun{q}{\kappa,\neg b}
\quad\quad\,
\text{ where } \forall j.\,\slen{\kappa}=\slen{c_j}
\\[0.3em]
\rulelab{M-\(\mea\)}&
\mea(x,n,\kappa) \mapsto \Msum_{j}{z_j}{\ket{c}}{{\eta_{j}}}+\qfun{q}{\kappa,c \neq \kappa}
\;\;=\quad \kappa \;\mapsto \Msum_{j}{\frac{z_j}{\sqrt{r}}}{\eta_{j}} * x = (r,\tov{c})

\qquad\quad\quad
\text{ where } n=\slen{c}
\end{array}
\end{array}
$
}
\vspace{-0.8em}
\caption{Predicate semantics.}
\label{fig:qafny-model-rules}
\vspace*{-1.7em}
\end{figure*}

\noindent\textbf{\textit{Quantum Conditionals.}}
As in \Cref{sec:guidedtypeexp,fig:ghz-proof}, the key in designing a proof rule for a quantum conditional \(\small\qif{b}{e}\) with its locus scope \(\kappa \sqcupplus \kappa'\), is to encode two transformers: $\frz$ and \(\ufrz\).
In rule \rulelab{P-If} (\Cref{fig:exp-proofsystem-1}), we require the $\sigma$ only contains locus \(\kappa \sqcupplus \kappa'\), which can be done through \rulelab{P-Frame}.
We then utilize \(\frz{(b, \kappa,\kappa')}\) to finish two tasks: (1) it computes $b$'s side-effects on the $\kappa$'s position bases ($\denote{b}^{\slen{\kappa}}q$), and (2) it freezes all basis-kets that are irrelevant when reasoning about the body \(e\).
This freezing mechanism modeled by the equation \rulelab{M-\(\frz\)} (\Cref{fig:qafny-model-rules}) is accomplished at two levels: stashing all kets unsatisfying \(b\) ($\qfun{q}{\kappa, \neg{b}}$) and moving \(\kappa\)'s position bases to basis stacks for the rest of basis-kets.
After substituting \(\kappa\sqcupplus\kappa'\) for \(\frz{(b, \kappa,\kappa')}\), besides expelling the parts not satisfying \(b\), we also shrink the locus \(\kappa\sqcupplus\kappa'\) to \(\kappa'\), which in turn marked the \(\kappa\)'s position basis in every basis-ket inaccessible as \(\kappa\) is now invisible in the locus type environment.

After the body $e$'s proof steps, the post-state $Q$ describes the computation result for $\kappa'$ without the frozen parts.
To reinstate the state for \(\kappa\sqcupplus\kappa'\) by retrieving the frozen parts, we first substitute locus \(\kappa\sqcupplus\kappa'\) for \(\ufrz(\neg{b},\kappa,\kappa\sqcupplus\kappa')\) in \(P\), which represents the unmodified part, unsatisfying \(b\), in the pre-state. We also substitute $\kappa'$ for \(\ufrz({b},\kappa, \kappa\sqcupplus\kappa')\) in \(Q\), which represents the part satisfying \(b\), evolved due to the execution of \(e\).
Rule \rulelab{M-\(\ufrz\)} (\Cref{fig:qafny-model-rules}) describes the predicate transformation, empowered by the locus construct $\ufrz$, that utilizes the innate relation of separating conjunction and logical complement to assemble the previously unmodified and the evolved parts.
Rule \textsc{P-Loop} proves a \texttt{for} loop where \(P(j)\) is the loop invariant parameterized over the loop counter \(j\). Other rules are introduced in \Cref{appx:proof-rules}.

\ignore{
\begin{figure*}[t]
{\footnotesize
  \begin{mathpar}
    \inferrule[M-\(\mathpzc{F}\)]{\slen{c}=n \\
           \Omega;\sigma;\varphi\models \kappa \mapsto 
                      \sch{m}{\frac{z_j}{\sqrt{r}}}{c_{j}} * x = (r,\tov{c})}
          {\Omega;\sigma,\varphi\models \mathpzc{F}(x,n,\kappa) \mapsto \scha{m}{z_j}{c}{\ket{c_{j}}}+q(n,\neq c) }

  \end{mathpar}
\begin{mathpar}
    \inferrule[P-Mea]{\fivepule{\Omega[x\mapsto \mmode]}{\sigma[\kappa\mapsto \tcht]}{\cmode}{P[\mathpzc{F}(x,n,\kappa)/y\srange{0}{n}\sqcupplus\kappa]}{e }{Q} }
     {\fivepule{\Omega[y\mapsto \qmode{n}]}{\sigma[y\srange{0}{n}\sqcupplus\kappa\mapsto \tcht]}{\cmode}{P}{\sexp{x}{\smea{y}}{s} }{Q} }

  \end{mathpar}
}
{\footnotesize
$r= \Msum_{k=0}^{m} \slen{z_k}^2
\qquad\quad
q(n,\neq c) = \schk{m'}{z'_k}{c'.c'_{k}} \;\;\texttt{where}\;c'\neq c
$
}
\caption{Semantic and Proof Rules for Measurement. $\tov{c}$ turns basis $c$ to an integer, and $r$ is the likelihood that the bitstring $c$ appears in a basis-ket. }
\label{fig:qafny-mea-rules}
\end{figure*}
}

\noindent\textbf{\textit{Measurement.}}
A measurement ($\sexp{x}{\smea{y}}{e}$) collapses a qubit array $y$, binds an $\mmode$ kind outcome to $x$ and restricts its usage in \(e\).
These statements usually appear in periodical patterns in many quantum algorithms, which users formalize as predicates to help verify algorithm properties.
In rule \textsc{P-Mea}, we first select an $n$-length prefix bitstring $c$ from one of range $y\srange{0}{n}$'s position bases; it then computes the probability $r$ and assigns $(r,\tov{c})$ to variable $x$.
We then replace the locus $y\srange{0}{n}\sqcupplus\kappa$ in $P$ with a locus predicate transformer $\mea(x,n,\kappa)$ and update  the type state $\Omega$ and $\sigma$ by replacing $y\srange{0}{n}\sqcupplus \kappa$ with $\kappa$. 
The construct $\mea(x,n,\kappa)$, with its transformation rule \rulelab{M-$\mea$} (\Cref{fig:qafny-model-rules}), is introduced to do exactly the two steps in \Cref{sec:sem-measurement} for describing measurement operations, i.e., we remove basis-kets not having $c$ as $y\srange{0}{n}$'s position bases ($\qfun{q}{\kappa,c \neq \kappa}$) and truncate $y\srange{0}{n}$'s position bases in the rest basis-kets. 

\vspace*{-1em}
{\footnotesize
  \begin{mathpar}
   \inferrule[]
   { \fivepule{\Omega[u\mapsto \mmode]}{\{x\srange{0}{n} : \tcht\}}{\cmode}
  {\textcolor{spec}{\mea(u,n,x\srange{0}{n})\mapsto C}}{ \{\} }{\textcolor{spec}{x\srange{0}{n} \mapsto D * E}} }
   { \fivepule{\Omega}{\{y\srange{0}{n}\sqcupplus x\srange{0}{n} : \tcht\}}{\cmode}
      {\textcolor{spec}{y\srange{0}{n}\sqcupplus x\srange{0}{n}\mapsto C}}{ \sexp{u}{\smea{y}}{\{\}} }{\textcolor{spec}{x\srange{0}{n} \mapsto D * E}}
     }
  \end{mathpar}
{
\begin{center}
\vspace*{-0.5em}
$\begin{array}{l}
C \triangleq \sch{2^{n}\sminus 1}{\frac{1}{\sqrt{2^{n}}}}{\ket{a^{j}\;\mmod\;N}\ket{j}}
\;\;\;
D \triangleq \smch{\frac{1}{\sqrt{r}}}{r\sminus 1}{t\,\splus\,k p}
\;\;\;
E \triangleq p = \texttt{ord}(a,N)
*
u=(\frac{p}{2^n},a^{t}\;\mmod\;N)
*
r=\texttt{rnd}(\frac{2^n}{p})
\end{array}$
\end{center}
}
}
\vspace*{-0.5em}

We show a proof fragment above for the partial measurement in \Cref{fig:shorqafny} line 8.
The proof applies rule $\textsc{P-Mea}$ by replacing locus $y\srange{0}{n}\sqcupplus x\srange{0}{n}$ with $\mea(u,n,x\srange{0}{n})$.
On the top, the pre- and post-conditions are equivalent, as explained below.
In locus $y\srange{0}{n}\sqcupplus x\srange{0}{n}$'s state, for every basis-ket, range $y\srange{0}{n}$'s position basis is $\ket{a^{j}\;\mmod\;N}$; the value $j$ is range $x\srange{0}{n}$'s position basis for the same basis-ket. Randomly picking a basis value $a^{t}\;\mmod\;N$ also filters a specific $j$ in range $x\srange{0}{n}$, i.e., we collect any $j$ having the relation $a^{j}\;\mmod\;N=a^{t}\;\mmod\;N$. Notice that modulo multiplication is a periodic function, which means that the relation can be rewritten $a^{t+kp}\;\mmod\;N=a^{t}\;\mmod\;N$, and $p$ is the period order. Thus, the post-measurement state for range $x\srange{0}{n}$ can be rewritten as a summation of $k$: $\smch{\frac{1}{\sqrt{r}}}{r\sminus 1}{t\,\splus\,k p}$. The probability of selecting $\ket{a^{j}\;\mmod\;N}$ is $\frac{r}{2^n}$.
In the \qafny implementation, we include additional axioms for these periodical theorems to grant this pre- and post-condition equivalence so that we can utilize \qafny to verify Shor's algorithm.

\vspace*{-0.5em}
\subsection{Qafny Metatheory}\label{sec:theorems}

We now present \qafny's type soundness and its proof system's soundness and relative completeness. These results have all been verified in Coq.
We prove our type system's soundness with respect to the semantics, assuming well-formedness (\Cref{def:well-formed-ses} and \Cref{def:well-formed}).
The type soundness shows that our type system ensures the three properties in \Cref{sec:type-checking} and that the "in-place" style \qafny semantics can describe all different quantum operations without losing generality because we can always use the equivalence rewrites to rewrite the locus state in ideal forms.

\begin{theorem}[\qafny type soundness]\label{thm:type-progress-oqasm}\rm 
If $\Omega;\sigma \vdash_g e \triangleright \sigma'$ and
  $\Omega;\sigma \vdash_g \varphi$, then there exists $\varphi'$ such that $(\varphi,e)\Downarrow \varphi'$ and
  $\Omega;\sigma' \vdash_g \varphi'$.
\end{theorem}

Our proof system is sound and relatively complete w.r.t. its semantics for well-typed \qafny programs.
Our system utilizes a subset of separation logic admitting completeness by excluding qubit array allocation and pointer aliasing.
Since every quantum program in \qafny converges, the soundness and completeness refer to the total correctness of the \qafny proof system. 
$\psi(e)$ refers to that we substitute every variable $x \in \dom{\psi}$ in $e$ with $\psi(x)$.

\begin{theorem}[proof system soundness]\label{thm:proof-soundness}\rm 
  For any program $e$, such that $\Omega;\sigma\vdash_g e \triangleright \sigma'$ and $\fivepule{\Omega}{\sigma}{g}{P}{e}{Q}$, and for every $\psi$ and $\varphi$, such that $\Omega;\sigma\vdash_g \varphi$ and $\Omega;\psi;\sigma;\varphi\models_g P$, there exists a state $\varphi'$, such that $(\varphi,\psi(e))\Downarrow \varphi'$ and $\Omega;\sigma'\vdash_g \varphi'$ and $\Omega;\psi;\sigma';\varphi'\models_g Q$.
\end{theorem}

\begin{theorem}[proof system relative completeness]\label{thm:proof-completeness}\rm 
  For a well-typed program $e$, such that $\Omega;\sigma\vdash_g e \triangleright \sigma'$, $(\varphi,e)\Downarrow \varphi'$ and $\Omega;\sigma\vdash_g \varphi$, and for all predicates $P$ and $Q$ such that $\Omega;\emptyset;\sigma;\varphi\models_g P$ and $\Omega;\emptyset;\sigma';\varphi' \models Q$, we have $\fivepule{\Omega}{\sigma}{g}{P}{e}{Q}$.
\end{theorem}
 \vspace*{-0.5em}
\section{Qafny Compilation and Implementation Evaluation}

Here, we focus on the \qafny proof system compilation to a subset of separation logic.

\subsection{Translation from Qafny to Separation Logic}\label{sec:dafny-compilation}

The \qafny types and loci are extra annotations associated with \qafny programs and predicates for proof automation.
In the \qafny implementation in Dafny, these annotations are not present---qubits are arranged as simple array structures without extra metadata such as locus types.
This section shows how \qafny annotations can be safely erased with no loss of expressiveness, e.g., loci are represented as virtual-level dynamic sequences without types, and equational rewrites are compiled with extra operations in the compiled language. We present a compilation algorithm that converts from \qafny to \sep, a C-like language admitting a subset of an array separation logic proof system. 

\begin{figure}[t]
\begin{minipage}[b]{0.27\textwidth}
{\footnotesize
\begin{center}
  $\begin{array}{l@{~}c@{~}l} 
       A,B &::=& \overline{n}\\
       \widetilde{op}&::=&\mu\mid op\\
       \widetilde{e} &::=& x=\texttt{alloc}{(A)}\\
               &\mid& \texttt{mutate}{(n,\widetilde{op},x)}\\
               &\textcolor{red}{\mid}& \textcolor{red}{(r,n)=\texttt{pick}{(x,m)}}\\
               &\textcolor{red}{\mid}& \textcolor{red}{\texttt{filter}{(x,b)}}\\
               &\textcolor{red}{\mid}& \textcolor{red}{\texttt{amp}{(x,r)}}\\
               &\mid& ...
    \end{array}
 $
\end{center}
}
\vspace*{-1em}
  \caption{\sep Syntax}
\label{fig:sep-syntax}
\end{minipage}
\begin{minipage}[b]{0.32\textwidth}
\center
  \includegraphics[width=0.4\textwidth]{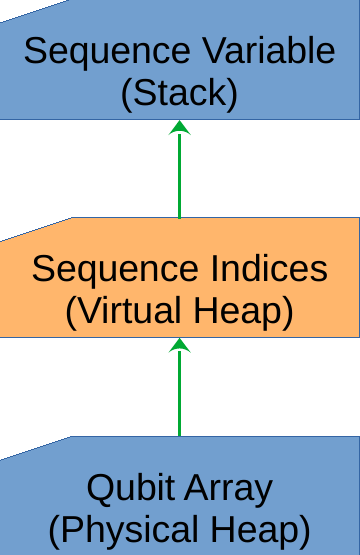}
\caption{Heap Layout}
\label{fig:compile-diagram}
\end{minipage}
\begin{minipage}[b]{0.41\textwidth}
\center
  \includegraphics[width=0.68\textwidth]{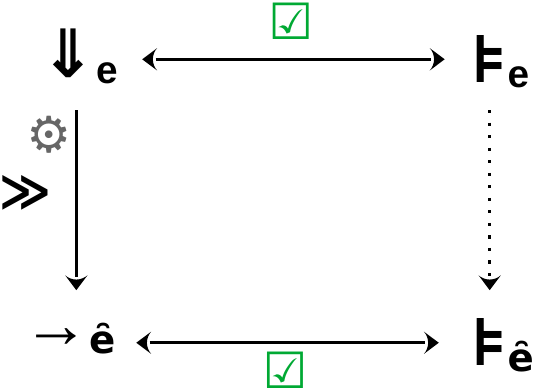}
  \caption{Compilation Proof Diagram}
\label{fig:compile-diagrama}
\end{minipage}
\vspace*{-2.8em}
\end{figure}

\noindent\textbf{\textit{Target Compilation Language.}}
\sep is based on a variant of the separation logic introduced by Yang and O'Hearn \cite{10.1007/3-540-45931-6_28}, which is sound and complete.
Mainly, we utilize the allocation (\cn{alloc}), heap lookup and mutation operations (\cn{mutate}) in the work, with three additional operations (marked red in \Cref{fig:sep-syntax}).
In \sep, program states are divided into virtual and physical levels in \Cref{fig:compile-diagram}.
Every \sep program starts with a physical qubit array, analogous to physical heap structures. The program operations are applied to qubit sequences of array indices ($A$), representing a collection of physical qubit locations that live at the virtual level. \sep permits immutable program variables ($x$) representing these sequences. 

An allocation $x=\texttt{alloc}{(A)}$ allocates a new array $x$ that copies $A$'s content and has the same length as $A$, while a heap mutation $\texttt{mutate}{(n,\widetilde{op},x)}$ mutates the first $n$~elements of the array pointed to by $x$, by applying the operation $\widetilde{op}$. Operations $\cn{pick}$, $\cn{filter}$, and $\cn{amp}$ are variations of heap lookups and mutations.
$(r,n)=\texttt{pick}{(x,m)}$ measures the first $m$ qubits in $x$, with the outcome $n=\tov{c}$ and its probability $r$, similar to the computation in \rulelab{S-Mea} (\Cref{fig:exp-semantics-1}). $\texttt{filter}{(x,b)}$ mutates $x$'s pointed-to quantum value by filtering out basis-kets that are not satisfying $b$, and $\texttt{amp}{(x,r)}$ multiples $r$ to every basis-kets in $x$'s quantum value. 
Yang and O'Hearn maintain completeness by carefully designing the proof rules so heap mutations do not modify pointer references. \sep ensures the same property by immutable variables, i.e., if a sequence changes, we allocate a new array and a new variable pointing to the array. For example, to join two loci represented by two sequences $A$ and $B$, respectively pointed to by variables $x$ and $y$, we allocate a new space for the two sequences' concatenation ($A@B$); so we compile the join to $u=\cn{alloc}(A@B)$. We must abandon using $x$ and $y$ after the join and only refer to $u$ in the following computation.

\noindent\textbf{\textit{Compilation Procedure.}}
As shown in \Cref{fig:compile-diagrama}, we compile the \qafny language to \sep and achieve the proof system compilation through the proof system completeness in \qafny and \sep.
For every \qafny program, we translate the program and states to \sep.
Then, every provable triple in \qafny can be translated to a provable \sep triple through the language translation from \qafny to \sep as the diagram (\Cref{fig:compile-diagrama}).
The compilation is defined by extending \qafny's typing judgment thusly: $\Omega;\sigma\vdash_g (\theta,\varphi,e) \gg (\theta,\widetilde{\psi},\widetilde{\varphi},\widetilde{e})$.
We include an initial \qafny state $\varphi$, the output local store ($\widetilde{\psi}$), mapping variables to qubit location sequences, and state ($\widetilde{\varphi}$), mapping from locations to quantum values.
$\theta$ and $\theta'$ are maps from locus locations in $\varphi$ to \sep qubit locations in $\widetilde{\varphi}$.

Here, we explain the rules for compilation by examples of compiling the \qafny operations to \sep. 
The locus rewrites (\Cref{sec:type-checking}) are compiled to the array allocations, such as the join operation above.
Additionally, the split of a locus is compiled to two consecutive allocations of two sequences, respectively representing the two split result loci.
In compiling a measurement statement ($\sexp{x}{\smea{y}}{...}$), where $y$ locates in the locus $y\srange{0}{n}\sqcupplus \kappa$, let's assume that the locus is mapped in $\theta$ by sequence $[0,n+m)$, pointed to by $u$ in $\widetilde{\psi}$, while the range $y\srange{0}{n}$ is mapped by the sequence $[0,n)$; the operation is compiled to:

\vspace*{-0.5em}
{\small
\begin{center}
$\sseq{(r,p)=\cn{pick}{(u,m)}}{\sseq{\cn{filter}(u,u[0,n) = p)}{\sseq{\cn{amp}{(u,\sum_j\frac{z_j}{\sqrt{r}})}}{t=\cn{alloc}({[n,n\splus m))}}}}$
\end{center}
}
\vspace*{-0.5em}

We first pick a key $p$, filter out the basis-kets whose $u[0,n)$'s position bases are not $p$, normalize the amplitudes of the remaining basis-kets (\Cref{sec:semantics-proof}), and allocate a new space $t$ for the quantum residue after the measurement.
We also update $\kappa$ in $\theta$ to map to the newly allocated space of $t$ instead of $[n,n\splus m)$.
We compile an operation $\qass{x\srange{0}{n}\sqcupplus y[0]}{\qbool{x}{<}{5}{y[0]}}$ with its initial state $\varphi$ ($C=\shad{2^n}{n\sminus 1}{}$) to \sep, with $D=\scha{2^n\sminus 1}{}{j}{\ket{{j\,\slt\,5}}}$. Such an operation computes the Boolean comparison of $x\,\slt\,5$ and stores the value to $y[0]$.

\vspace*{-1em}
{\footnotesize
\begin{center}
$\hspace*{-2em}
\begin{array}{l@{~}c@{~}l}
\begin{array}{l}
\quad\textcolor{spec}{\varphi=\big{\{}x\srange{0}{n}: C, y[0]:\ket{0}\big{\}}}\\[0.1em]
\qquad\qass{x\srange{0}{n}\sqcupplus y[0]\;}{\;\qbool{x}{<}{5}{y[0]}}\\[0.1em]
\quad\textcolor{spec}{\varphi'=\big{\{}x\srange{0}{n}\sqcupplus y[0]:D \big{\}}}
\end{array}
&
\text{\scalebox{1.5}{\quad$\gg$}\quad}
&
\begin{array}{l}
\textcolor{spec}{\widetilde{\psi}=\big{\{}p: [0,n), t : \{n\} \big{\}}, \widetilde{\varphi}=\big{\{} [0,n) : C * \{n\}:\ket{0}\big{\}}}
\\[0.2em]
\quad{\sseq{u=\cn{alloc}([0,n\splus 1))}{\cn{mutate}(n\splus 1,u[0,n)\,\slt\,5@ u[n],u)}}
\\[0.2em]
\textcolor{spec}{
\widetilde{\psi'}=
\widetilde{\psi}\cup
\big{\{}
u: [n\,\splus\, 1\,,\;2 n\,\splus\, 2)\big{\}}, \widetilde{\varphi'}=\widetilde{\varphi}\cup\big{\{} [n\,\splus\, 1\,,\;2 n\,\splus\, 2): D \big{\}}}
\end{array}
\end{array}
$
\end{center}
}
\vspace*{-0.8em}

After the compilation, we create two local variables $p$ and $t$ to represent the loci $x$ and $y$, mapping to sequences $[0,n)$ and $\{n\}$. We then add $u=\cn{alloc}([0,n\,\splus\, 1))$ allocating a new space $[n\splus 1,2 n\,\splus\, 2)$ to join the two loci. The post-state contains a new variable $u$, pointing to the concatenated new sequence $[n\,\splus\, 1\,,\,2 n\,\splus\, 2)$.
The old arrays $p$ and $t$ are still in the stores, but we refer to the locus $x\srange{0}{n}\sqcupplus y[0]$ as the newly allocated space in the following computation by mapping the locus to the new space in $\theta'$.
As a future work, we will prove the proof system compilation correctness from \qafny to \sep, proof strategy in \Cref{fig:compile-diagrama}.

\ignore{
The language compilation correctness from \qafny to \sep is stated below,
which is validated by testing the compiler's semantic equivalence on 203 unit test programs.

\begin{theorem}[\qafny to \sep Compilation Correctness]\label{thm:proof-compile-sep}\rm 
For any $e$, $\theta$ and $\theta'$, such that $\Omega;\sigma \vdash_g e \triangleright \sigma'$, $\Omega;\sigma \vdash_g \varphi$, and $\Omega;\sigma\vdash_g (\theta,\varphi,e) \gg (\theta',\widetilde{\psi},\widetilde{\varphi},\widetilde{e})$; thus, if $(\varphi,e)\Downarrow \varphi'$, then
$(\widetilde{\psi},\widetilde{\varphi},\widetilde{e})\Downarrow (\widetilde{\psi'},\widetilde{\varphi'})$, where $\widetilde{\varphi'}$ is the compiled results of $\varphi'$ with the $\theta'$ map.
\end{theorem}
}

\vspace*{-0.5em}
\subsection{Implementation and Comparison to Existing Quantum Verifications}\label{sec:implementation}
\vspace*{-0.5em}

We have implemented a prototype \qafny to Dafny compiler, which faithfully respects the presented \qafny to \sep compilation algorithm.
To validate the soundness of the compiler implementation, we create many test cases for the compiler.
We then insert a number of bugs in these test cases; Qafny has been able to detect all of them.
Dafny's proof engine cannot be used to verify arbitrary separation-logic assertions because it only has an implicit frame rule implementation that allows users to set up variables that can be modified in a function.
However, \qafny only requires a subset of separation logic, and the \qafny loci disjoint property and non-aliasing guarantee ensures that the \qafny separation conjunctions are captured by Dafny's implicit frames when we compile \qafny to Dafny.
We utilize the \qafny to Dafny compiler as an automated verification framework to verify six quantum programs, shown in \Cref{sec:arith-oqasm,fig:circ-evaluation1}.

\begin{figure*}[t]
\begin{minipage}[b]{0.65\textwidth}
\hspace*{-0.3em}
{\scriptsize
\begin{tabular}{|l|@{}c@{}|c|@{}c@{}|c|@{}c@{}|c|}
\hline
\multirow{2}{*}{\texttt{Algorithm}}
&
\multicolumn{2}{c|}{Qafny}
&
\multicolumn{2}{c|}{\qbricks}
&
\multicolumn{2}{c|}{\sqir}
\\
\cline{2-7}
&
  \begin{tabular}{c}\texttt{Runtime}\\ \texttt{(sec)} \end{tabular} & \texttt{LOC}
& \begin{tabular}{c}\texttt{Runtime}\\ \texttt{(sec)} \end{tabular} & \texttt{LOC} 
& \begin{tabular}{c}\texttt{Runtime}\\ \texttt{(sec)} \end{tabular} & \texttt{LOC} \\
\hline
\texttt{GHZ} & 14.2 & 16 & - & - & 141 & 119\\
\texttt{Deutsch–Jozsa} & 8.3 & 13 & 74 & 108 & 163 & 408 \\

\texttt{Grover's search} & 26.7 & 27 & 253 & 233 & 148 & 1018\\
\texttt{Shor's algorithm} & 36.3  & 36 & 1328 & 1163 & 1244 & 8464\\
\hline                           
\end{tabular}
}
\vspace*{-0.5em}
\caption{Running time (include theory loading) $\&$ LOC comparison, in an i7 Ubuntu Mach. 8G RAM; \texttt{-}: no data. }
\label{fig:circ-evaluation}
\end{minipage}
\hfill\hfill
\hspace*{0.5em}
\begin{minipage}[b]{0.32\textwidth}
{\scriptsize
\begin{tabular}{|l|@{}c@{}|c|}
\hline
\texttt{Algorithm}
&
\begin{tabular}{c}\texttt{Runtime}\\ \texttt{(sec)} \end{tabular} & \texttt{LOC}
\\
\hline
\texttt{Controlled GHZ} & 6.4 & 12 \\

\texttt{Quantum Walk} & 43.1 & 49 \\
\hline                           
\end{tabular}
}
\vspace*{-0.5em}
\caption{\qafny data for case studies in Section~\ref{sec:arith-oqasm}.}
\label{fig:circ-evaluation1}
\end{minipage}
\vspace*{-2.8em}
\end{figure*}

There were two main approaches to verifying quantum programs: program and measurement-based.
The former views quantum program transitions as a state machine and verifies the inductive relations among transitions,
while the latter focuses on the relations between quantum program measurements and the post-processing classical components --- they typically view quantum components as black boxes with specifications.

\qafny is program-based and other program-based mechanized frameworks for formally verifying quantum programs, including \qwire~\cite{RandThesis}, \sqir~\cite{VOQC, PQPC} (an upgrade of \qwire), and \qbricks~\cite{qbricks}, provide libraries in an interactive proof assistant to guide users for building inductive proofs for quantum programs; each has verified 7-10 quantum programs.
The core of \sqir and \qbricks, as well as other executable quantum verification platforms, is a circuit-description language.
Verifying a program in these frameworks inductively builds a unitary or density matrix as the program's quantum circuit semantic interpretation.
For example, to verify Shor's algorithm, both \qbricks and \sqir require inductive proofs based on elementary circuit gates to derive the unitary or density matrix semantics of different sub-components, including state preparations, oracles, and $\texttt{QFT}^{-1}$ gates.
Additionally, program verification in these frameworks requires the development of theories and tactics to capture program properties, which usually involves the proof of additional theorems.
This approach is qualitatively different from \qafny, where program verification involves embedding assertions in a program for completing a proof. \qafny identifies a few program structures, such as oracles and quantum conditionals, and formulates inductive patterns involving these quantum components as proof rules.
These rules interact with quantum program operations and states for deriving verification proofs, so proofs are largely automated based on this small set of structures.

\Cref{fig:circ-evaluation} shows a quantitative comparison of \qafny, regarding theory/proof statement running time and numbers of lines (LOC), with respect to \qbricks and \sqir for verifying several quantum programs.
The results show that \qafny has the shortest running time and LOCs for verifying programs with our automated proof engine. The \qbricks verification has better LOCs but a similar running time compared to \sqir.
\sqir provides a complete verification \cite{shorsprove} by proving every mathematical theorem involved in the verification, so its verification proofs are longer than \qbricks; \qbricks performs better by providing some automatic tactics for sequence operations ($\sseq{e}{e}$) and taking many math theorems as assumptions without proof.
In testing the two frameworks, we found that the previous claim \cite{VOQC, PQPC} that rigorous quantum proofs are one-time cost is problematic because inductive theorem provers update constantly. Once an update happens, users might need to fix the proofs (not programs or specifications) in their history code, e.g., our researcher spent three days fixing minor bugs in the proofs in \sqir and \qbricks due to Coq and Why3 version issues. Moreover, a new program verification in \sqir typically required detailed proofs of additional theorems beyond the program specifications. 

\qafny provides fast prototyping, where we apply the automated verification mechanisms in many classical systems \cite{DBLP:conf/pldi/Qiu0SM13,dafnyref} to verify quantum programs and save programmers' effort. There is a subjective survey in \Cref{appx:human} to show the human effort benefits of verifying programs in \qafny compared to other platforms. Verifying programs in many inductive theorem provers takes weeks and months to finish, while the same tasks in \qafny cost researchers a few days due to the \qafny features mentioned above.
The fast prototyping in \qafny can also help users to explore and understand new quantum program patterns such as the two case studies in \Cref{sec:arith-oqasm}, whose running time and LOCs are shown in \Cref{fig:circ-evaluation1}.
Compared to the data of well-known algorithms in \Cref{fig:circ-evaluation}, the data for verifying the new programs do not show a significant difference, showing \qafny's ability to explore new algorithm behaviors without proving many new theories, which usually appears in the above quantum verification frameworks.
 \vspace*{-0.5em}
\section{Case Studies}
\label{sec:arith-oqasm}
\vspace*{-0.5em}

With two examples, we show \qafny as a rapid prototyping tool for
quantum programs.
\vspace*{-0.5em}
\subsection{Controlled GHZ: Composing Quantum Algorithms from Others}\label{sec:controled-ghz}

\begin{figure}[h]
\vspace*{-1em}
{
\begin{center}
{
\begin{tabular}[b]{c @{\qquad} c}
  \raisebox{3\height}{\small
  \Qcircuit @C=0.5em @R=0.5em {
    \lstick{\ket{0}} & \gate{H} & \qw & \ctrl{1} & \qw & \qw  & \\
    \lstick{\ket{0}} & \qw & \qw & \multigate{3}{\texttt{GHZ}} & \qw &\qw &  \\
    & \vdots &   &  &  & & \\
    & \vdots &  & \dots & & &  \\
    \lstick{\ket{0}} & \qw & \qw & \ghost{\texttt{GHZ}} &\qw &\qw &
    }}

&
{
$\small
\begin{array}{r l}
\textcolor{blue}{1}
&
\textcolor{spec}{
\big{\{}x[0]\mapsto \ket{0},y\srange{0}{n}\mapsto \ket{\overline{0}}\big{\}}}
\\[0.2em]
\textcolor{blue}{2}
&
\qass{x[0]\;}{\;\ihadh};\\[0.2em]

\textcolor{blue}{3}
&
\textcolor{auto}{
\big{\{}x[0]\mapsto \frac{1}{\sqrt{2}}(\ket{0}+\ket{1}) * y\srange{0}{n}\mapsto \ket{\overline{0}}\big{\}}
}
\\[0.2em]

\textcolor{blue}{4}
&
\textcolor{auto}{
\Rightarrow
\big{\{}x[0]\sqcupplus y\srange{0}{n}\mapsto \schd{d}{1}{}{d}{\ket{\overline{0}}}\big{\}}
}
\\[0.2em]

\textcolor{blue}{5}
&
\sifq{x[0]}{\;\texttt{ghz(}y\cn{)}};
\\[0.2em]

\textcolor{blue}{6}
&
\textcolor{spec}{
\big{\{}x[0]\sqcupplus y\srange{0}{n}\mapsto \frac{1}{\sqrt{2}}\ket{\overline{0}}+\frac{1}{2}\ket{1}\ket{\overline{0}}+\frac{1}{2}\ket{\overline{1}} \big{\}}
}
\end{array}
$
}
\end{tabular}
}
\end{center}
}
\vspace*{-1.5em}
\caption{Controlled GHZ circuit and proof. $\texttt{ghz}(y)$ is in Fig. \ref{fig:background-circuit-example}. Lines 3-4 automatically inferred.}
\label{fig:background-circuit-example-controlled}
\vspace*{-0.8em}
\end{figure}
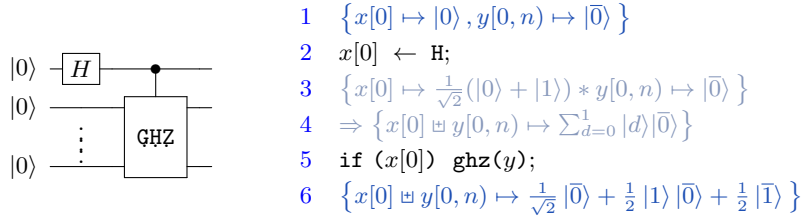

Automated verification frameworks such as Dafny encourage programmers
to build program proofs based on the reuse of subprogram proofs.
However, this perspective is more or less overlooked in previous quantum proof systems.
In \sqir, for example, verifying the correctness of a \emph{controlled GHZ}, a simple circuit
constructed by extending GHZ with an extra control qubit,
requires generalizing the GHZ circuit to any arbitrary inputs.
In \qafny, users do not need to do this, as shown here.

\Cref{fig:background-circuit-example-controlled} provides a proof of the
Controlled GHZ algorithm based on a proven GHZ method in
\Cref{fig:background-circuit-example-proof}. 
The focal point is the quantum conditional on line 5.
For verifying a GHZ circuit, its input is an $n$-qubit $\tnort$-typed value of all $\ket{0}$, 
but the given value, in line 4, is an $\tcht$-typed entanglement
$\schd{d}{1}{}{d}{\ket{\overline{0}}}$.
Here is where \sqir gets stuck.
In \qafny, we automatically verify the proof by rule \rulelab{P-If} and the
equivalence relation to rewrite a singleton $\tcht$ value to a $\tnort$ one, as
$\sch{0}{z_j}{\ket{c_j}}\equiv z_0\ket{c_0}$.
The detailed proof for the conditional
is given below, where $U(b)=\ufrz(b,x[0],\kappa)$ and $\kappa=x[0]\sqcupplus y\srange{0}{n}$.

\vspace*{-1.4em}
{\scriptsize
  \begin{mathpar}
\inferrule[]{
   \inferrule[]
   { 
  { \fivepule{\Omega}{\{y\srange{0}{n}:\tnort\}}{\mmode}
{\textcolor{spec}{
y\srange{0}{n}\mapsto {\ket{\overline{0}}}}
\textcolor{stack}{\cmsg{\ket{1}}}
}{ \texttt{ghz(}y\cn{)} }{
\textcolor{spec}{
y\srange{0}{n}\mapsto \schd{d}{1}{\frac{1}{\sqrt{2}}}{\overline{d}}}
\textcolor{stack}{\cmsg{\ket{1}}}
}
} }
   {
\fivepule{\Omega}{\{y\srange{0}{n}:\tcht\}}{\mmode}
{\textcolor{spec}{
\frz(x[0],y\srange{0}{n})\mapsto \schd{d}{1}{}{d}{\ket{\overline{0}}}
}
}{ \texttt{ghz(}y\cn{)} }{
\textcolor{spec}{
y\srange{0}{n}\mapsto \schd{d}{1}{\frac{1}{\sqrt{2}}}{\overline{d}}}
\textcolor{stack}{\cmsg{\ket{1}}}
}} \;\rulelab{EQ}} {
\inferrule[]
{\fivepule{\Omega}{\{\kappa : \tcht\}}{\cmode}
{\textcolor{spec}{
\kappa\mapsto \schd{d}{1}{}{d}{\ket{\overline{0}}}
}
}{ \sifq{x[0]}{\texttt{ghz(}y\cn{)}} }{
\textcolor{spec}{
U(\neg x[0])\mapsto \schd{d}{1}{}{d}{\ket{\overline{0}}}
*
U(x[0])\mapsto \schd{d}{1}{\frac{1}{\sqrt{2}}}{\overline{d}}}
\textcolor{stack}{\cmsg{\ket{1}} }}}
{\fivepule{\Omega}{\{\kappa : \tcht\}}{\cmode}
{\textcolor{spec}{
\kappa\mapsto \schd{d}{1}{}{d}{\ket{\overline{0}}}
}
}{ \sifq{x[0]}{\texttt{ghz(}y\cn{)}} }{
\textcolor{spec}{
\kappa \mapsto \frac{1}{\sqrt{2}}\ket{\overline{0}}+\frac{1}{2}\ket{1}\ket{\overline{0}}+\frac{1}{2}\ket{\overline{1}}
} }}
}\;\rulelab{P-If}
  \end{mathpar}
}
\vspace*{-1.2em}

After rule \rulelab{P-If} is applied, locus
$y\srange{0}{n}$'s value is rewritten to a $\tnort$ type value on the top as
$\textcolor{spec}{\ket{\overline{0}}}\textcolor{stack}{\cmsg{\ket{1}}}$, where
$\textcolor{spec}{\ket{\overline{0}}}$ is $n$ qubits and
$\textcolor{stack}{\cmsg{\ket{1}}}$ is frozen in the stack. 
Since two values are equivalent as \qafny discards stacks,
$\textcolor{spec}{\ket{\overline{0}}}\textcolor{stack}{\cmsg{\ket{1}}}$ is equivalent
to $\textcolor{spec}{\ket{\overline{0}}}$, which satisfies the input condition
for GHZ, so all proof obligations introduced to invoke the
\texttt{ghz} method are discharged.

\subsection{Case Study: Understanding Quantum Walk}\label{sec:quantumwalk}

\begin{wrapfigure}{r}{3.3cm}
\includegraphics[width=0.22\textwidth]{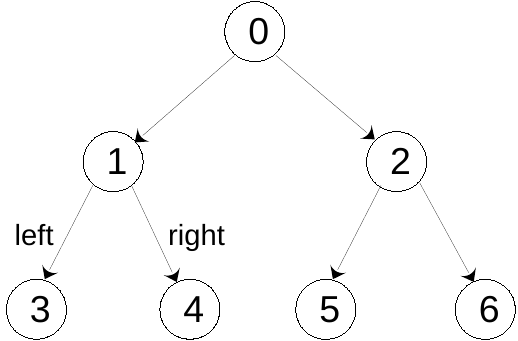}
\caption{Tree Structure}
\label{fig:tree-structure}
\end{wrapfigure} 

Quantum walk \cite{Wong_2022,ChildsNAND,Venegas_Andraca_2012} is a quantum version of the classical random walk \cite{RayleighThePO} and an important framework for developing quantum algorithms.
However, most quantum walk analyses are based on Hamiltonian simulation, which deters many computer programmers from the quantum walk framework.
Here, we show that the discrete-time quantum walk, at its very least, is a quantum version of breadth-first search; thus, many algorithms \cite{ChildsNAND} based on Quantum walk can be understood as performing search algorithms in the quantum setting.

\Cref{fig:background-circuit-example-walk} lists the proof outline for the core loop of a discrete-time quantum walk algorithm to traverse a complete binary tree (structure in \Cref{fig:tree-structure}); each node has a unique key. The $m$-depth nodes in the tree have keys $j\in [2^m\,\sminus\, 1,2^{m\splus 1}\, \sminus\, 2)$, which form a sequence from left to right in depth $m$-th, such as the sequence $3,4,5,6$ in depth $2$ in \Cref{fig:tree-structure}.
Thus, a node (key $j$) has a depth $m=\lfloor\log\;(j\splus 1) \rfloor$, and its $\texttt{left}$ and $\texttt{right}$ children have keys $2\cdot j \,\splus\, 1$ and $2\cdot j \,\splus\, 2$, respectively representing the $\texttt{left}$ and $\texttt{right}$ operation semantics in \Cref{fig:background-circuit-example-walk}, i.e., for any basis $\ket{j}$, with $m$ being the depth and $j$ a node key, the outputs of applying $\texttt{left}$ and $\texttt{right}$ are $\ket{2\cdot j \,\splus\, 1}$ and $\ket{2\cdot j \,\splus\, 2}$, respectively 
\footnote{The tree structure is a simplification; comprehensive implementations use Szegedy walk encoding \cite{10.5555/1070432.1070591}.}.

\begin{figure}[t]
{\centering
{\footnotesize
\[\hspace*{-0.5em}
\begin{array}{l}
q(j)\triangleq\schai{2^{(j\splus 1)}\,\sminus\,3}{z_i}{\lfloor\log\;(i\splus 1) \rfloor}{\ket{\texttt{pat}(i,j)}\ket{i\splus 1}\ket{d_i}}
\;\;\texttt{where}\;\;
\texttt{pat}(i,j) \triangleq \ket{0}^{\otimes \lfloor\log\;(i\splus 1) \rfloor}\ket{1}^{\otimes (j\,\sminus\,\lfloor\log\;(i\splus 1) \rfloor)}\\[0.6em]
\begin{array}{r l}
\textcolor{blue}{1}
&
\textcolor{spec}{
\big{\{} x\srange{0}{t} \mapsto \shadi{2^t}{t\sminus 1}{} *
   y\srange{0}{m} \mapsto \ket{\overline{0}} *
   h\srange{0}{n} \mapsto \ket{\overline{0}} 
   * u[0] \mapsto \ket{0} * m = 2^t \cdot m < n\big{\}}
}
\\[0.3em]
\textcolor{blue}{2}
&
\textcolor{auto}{
\Rightarrow
\big{\{} x\srange{0}{t}\sqcupplus y\srange{0}{0}\sqcupplus h\srange{0}{n}\sqcupplus u[0] \mapsto
 \ssum{k=0}{2^t\sminus 1}{\frac{1}{\sqrt{2^t}}\ket{k}\ket{\overline{0}}\ket{0}}
 * y\srange{0}{m} \mapsto \ket{\overline{0}}
 * m = 2^t \cdot m < n\big{\}}
}
\\[0.3em]
\textcolor{blue}{3}
&
\qfor{j}{0}{m}{\qbool{x\srange{0}{t}}{\,<\,}{j\,\splus\, 1}{\,y[j]}}
\\[0.3em]

\textcolor{blue}{4}
&
\;\;
\textcolor{spec}{\big{\{} x\srange{0}{t}\sqcupplus y\srange{0}{j}\sqcupplus h\srange{0}{n}\sqcupplus u[0] \mapsto q(j) + \ssum{k=j}{2^t\sminus 1}{\frac{1}{\sqrt{2^t}}\ket{k}\ket{\overline{0}}\ket{\overline{0}}\ket{0}}
 * y\srange{j}{m} \mapsto \ket{\overline{0}}
\big{\}}}
\\[0.3em]

\textcolor{blue}{5}
&
\{
\;\;\;
\qass{u[0]\;}{\;\texttt{H}};
\\[0.2em]

\textcolor{blue}{6}
&
\quad
\sifq{u[0]}{\;\texttt{left(}\lfloor\log\;(j\splus 1) \rfloor\,,\,h\srange{0}{n}\cn{)}};
\\[0.2em]

\textcolor{blue}{7}
&
\quad
\sifq{\neg u[0]}{\;\texttt{right(}\lfloor\log\;(j\splus 1) \rfloor\,,\,h\srange{0}{n}\cn{)}}; \;\}
\\[0.4em]

\textcolor{blue}{8}
&
\textcolor{spec}{
\big{\{} x\srange{0}{t}\sqcupplus y\srange{0}{m}\sqcupplus h\srange{0}{n}\sqcupplus u[0] \mapsto q(m)
\big{\}}
}
\end{array}
\end{array}
\]
}
}
\vspace*{-1.2em}
\caption{Quantum walk reachable node verification for a complete binary tree. $\texttt{left}$ and $\texttt{right}$ reach corresponding children. $q(j)$ is a quantum value with var $j$. $i+1$ is a node key in a tree.}
\label{fig:background-circuit-example-walk}
\vspace*{-1.2em}
\end{figure}

The algorithm in \Cref{fig:background-circuit-example-walk} requires four quantum ranges: a $t$-qubit range $x$ in superposition, an $m$-qubit range where \(y[j]\)'s position bases keep the result of evaluating $\qboola{x\srange{0}{t}}{<}{j\,\splus\, 1}$ for $j$-th loop step, an \(n\)-qubit node register $h$ storing the node keys, and a single qubit $u$ acting as the random walk coin and determining the moving direction of the next step ($1$ for the left and $0$ for the right). 
In line 2, we merge the ranges $x$, $u$, and $h$ as the locus $x\srange{0}{t}\sqcupplus y\srange{0}{0}\sqcupplus h\srange{0}{n}\sqcupplus u[0]$ ($y\srange{0}{0}$ is empty); at each loop step (lines 3-7), we entangle a qubit in the range $y$ into the locus.
Finally, at line 8, the loop program entangles all these ranges together as a locus $x\srange{0}{t}\sqcupplus y\srange{0}{m}\sqcupplus h\srange{0}{n}\sqcupplus u[0]$.

In the $j$-th loop step, we abbreviate locus $x\srange{0}{t}\sqcupplus y\srange{0}{j}\sqcupplus h\srange{0}{n}\sqcupplus u[0]$ as $\qfun{\kappa}{j}$, and locus $\qfun{\kappa}{j}$'s state value is split into two basis-ket sets, separated by $+$ in \Cref{fig:background-circuit-example-walk} line 4.
To verify a step, we first split the $y[j]$ qubit, having position basis $\ket{0}$, from range $y\srange{j}{m}$, and merge the qubit into $\qfun{\kappa}{j}$. The split rewrites are given as:

\vspace*{-0.5em}
{\scriptsize
\begin{center}
$
\begin{array}{l}
\textcolor{spec}{
\big{\{}
\qfun{\kappa}{j}\mapsto \schai{2^{(j\splus 1)}\,\sminus\,3}{z_i}{\lfloor\log\;(i\splus 1) \rfloor}{\ket{\texttt{pat}(i,j)}\ket{i\splus 1}\ket{d_i}} 
+
\ssum{k=j}{2^t\sminus 1}{\frac{1}{\sqrt{2^t}}\ket{k}\ket{\overline{0}}\ket{\overline{0}}\ket{0}}
*
y\srange{j}{m} \mapsto \ket{\overline{0}}
\big{\}}
}
\\[0.2em]
\textcolor{spec}{
\equiv\;
\big{\{}
\qfun{\kappa}{j\splus 1} \mapsto \schai{2^{(j\splus 1)}\,\sminus\,3}{z_i}{\lfloor\log\;(i\splus 1) \rfloor}{\ket{\texttt{pat}(i,j)}\ket{0}\ket{i\splus 1}\ket{d_i}} 
+
\ssum{k=j}{2^t\sminus 1}{\frac{1}{\sqrt{2^t}}\ket{k}\ket{\overline{0}}\ket{0}\ket{\overline{0}}\ket{0}}
*
y\srange{j\splus 1}{m} \mapsto \ket{\overline{0}}
\big{\}}
}
\\[0.2em]
\textcolor{spec}{
\equiv\;
\big{\{}
\qfun{\kappa}{j\splus 1} \mapsto {q'}{(\ket{0})}
+
\ssum{k=j}{2^t\sminus 1}{\frac{1}{\sqrt{2^t}}\qfun{\delta}{k,0}}
*
y\srange{j\splus 1}{m} \mapsto \ket{\overline{0}}
\big{\}}
}
\end{array}
$
\end{center}
}
\vspace*{-0.5em}

At the last rewrite above, we abbreviate the first part of $\qfun{\kappa}{j\splus 1}$'s value to be $\textcolor{spec}{{q'}{(\ket{0})}}$, and the second part to be $\textcolor{spec}{\ssum{k=j}{2^t\sminus 1}{\frac{1}{\sqrt{2^t}}\qfun{\delta}{k,0}}}$ where $\textcolor{spec}{\qfun{\delta}{k,c}=\ket{k}\ket{\overline{0}}\ket{c}\ket{\overline{0}}\ket{0}}$.
Below, we show the proof steps (only the pre-condition transitions) for a conditional step in \Cref{fig:background-circuit-example-walk} line 3, which can be divided into two small steps. 
Here, $e$ is the conditional body in lines 5-7, and we apply \rulelab{P-Frame} to frame out locus $y\srange{j\splus 1}{m}$ from the states, so the bottom state only refers to the locus $\qfun{\kappa}{j\splus 1}$.
We further split the second part above ($\textcolor{spec}{\ssum{k=j}{2^t\sminus 1}{\frac{1}{\sqrt{2^t}}\qfun{\delta}{k,0}}}$) into two basis-ket sets: $\textcolor{spec}{\frac{1}{\sqrt{2^t}}\qfun{\delta}{j,0}}$ and $\textcolor{spec}{\ssum{k=j\splus 1}{2^t\sminus 1}{\frac{1}{\sqrt{2^t}}\qfun{\delta}{k,0}}}$.

\vspace*{-0.5em}
{\scriptsize
  \begin{mathpar}
   \inferrule[]
   { 
  \inferrule[]
{
{
\fivepule{\Omega}{\{h\srange{0}{n}\sqcupplus u[0] : \tcht\}}{\mmode}
{\textcolor{spec}{
h\srange{0}{n}\sqcupplus u[0] \mapsto {\textstyle\schia{2^{(j\splus 1)}\,\sminus\,3}{z_i}{\ket{i\splus 1}\ket{d_i}\textcolor{stack}{\cmsg{\beta}}}}
+\frac{1}{\sqrt{2^t}}\ket{\overline{0}}\ket{0}\textcolor{stack}{\cmsg{\beta'}}
}
}{ e }{
\textcolor{spec}{...} }
}
}
   { 
  { \fivepule{\Omega}{\{\qfun{\kappa}{j\splus 1} : \tcht\}}{\mmode}
{\textcolor{spec}{
{\textstyle
\qfun{\kappa}{j\splus 1} \mapsto {q'}{(\ket{1})}+\frac{1}{\sqrt{2^t}}\qfun{\delta}{j,1}+\ssum{k=j\splus 1}{2^t\sminus 1}{\frac{1}{\sqrt{2^t}}\qfun{\delta}{k,0}}
}
}
}{ \sifq{\qbool{x\srange{0}{t}}{\,<\,}{j\,\splus\, 1}{y[j]}}{e} }{
\textcolor{spec}{...} }
} } }
   {
\fivepule{\Omega}{\{\qfun{\kappa}{j\splus 1} : \tcht\}}{\mmode}
{\textcolor{spec}{
{\textstyle
\qfun{\kappa}{j\splus 1} \mapsto {q'}{(\ket{0})}+\frac{1}{\sqrt{2^t}}\qfun{\delta}{j,0}+\ssum{k=j\splus 1}{2^t\sminus 1}{\frac{1}{\sqrt{2^t}}\qfun{\delta}{k,0}}
}
}
}{ \sifq{\qbool{x\srange{0}{t}}{\,<\,}{j\,\splus\, 1}{y[j]}}{e} }{
\textcolor{spec}{...} }
}
  \end{mathpar}
}
\vspace*{-0.5em}

We split the \rulelab{P-If} proof step (Line 3 in \Cref{fig:background-circuit-example-walk}) into two small steps above.
The bottom step represents the first half of the $\frz$ transformer application (\Cref{fig:exp-proofsystem-1}) in $\rulelab{P-If}$. It views the Boolean guard $\qbool{x\srange{0}{t}}{<}{j\,\splus\, 1}{y[j]}$ as an oracle application and for every basis-ket in the locus $\qfun{\kappa}{j\splus 1}$, we compute the Boolean value $\qboola{x\srange{0}{t}}{<}{j\,\splus\, 1}$ and store it to $y[j]$'s position bases.
Unlike the simple Boolean guards appearing in \Cref{fig:background-circuit-example-proof,fig:shorqafny}, the Boolean guard here has side-effects that modify $y[j]$'s position bases. $\qfun{\kappa}{j\splus 1}$'s value is split into three basis-ket sets separated by $+$.
In the set $\textcolor{spec}{\qfun{q}{0}}$, range $x\srange{0}{t}$'s position bases have the form $\textcolor{spec}{\ket{\lfloor\log\;(i\splus 1) \rfloor}}$ (the depth of a node key $i\splus 1$) and the bases' natural number interpretations are smaller than $j\splus 1$; in the sets $\textcolor{spec}{\frac{1}{\sqrt{2^t}}\qfun{\delta}{j,0}}$ and $\textcolor{spec}{\ssum{k=j\splus 1}{2^t\sminus 1}{\frac{1}{\sqrt{2^t}}\qfun{\delta}{k,0}}}$, range $x\srange{0}{t}$'s position bases are $\textcolor{spec}{\ket{j}}$ and $\textcolor{spec}{\ket{k}}$ ($j < k$).
The former's natural number interpretation is less than $j\splus 1$, while the latter is not.
Thus, we flip $y[j]$'s position bases of the two sets after applying the bottom rule above while leaving the third set unchanged.

The middle step in the above \rulelab{P-If} proof step rules out the basis-ket set $\textcolor{spec}{\ssum{k=j\splus 1}{2^t\sminus 1}{\frac{1}{\sqrt{2^t}}\qfun{\delta}{k,0}}}$, because the $y[j]$'s position bases are all $0$; then, we push bases $\textcolor{spec}{\ket{\lfloor\log\;(i\splus 1) \rfloor}\ket{\texttt{pat}(i,j)}\ket{0}}$ and $\textcolor{spec}{\ket{k}\ket{\overline{0}}\ket{0}}$ to the frozen stacks as $\textcolor{stack}{\cmsg{\beta}}$ and $\textcolor{stack}{\cmsg{\beta'}}$, respectively for the remaining two sets.

\vspace*{0.2em}
{\noindent\hspace{-1em}
\scriptsize
$
\begin{array}{l}
{\Omega ;\{u[0]\sqcupplus h\srange{0}{n}: \tcht\}}
\vdash_{\mmode}
\\[0.1em]

{\big{\{}
\textcolor{spec}{
u[0]\sqcupplus h\srange{0}{n} \mapsto {\textstyle\schia{2^{(j\splus 1)}\,\sminus\,3}{z_i}{\ket{d_i}\ket{i\splus 1}\textcolor{stack}{\cmsg{\beta}}}}
+\frac{1}{\sqrt{2^t}}\ket{0}\ket{\overline{0}}\textcolor{stack}{\cmsg{\beta'}}
}
\big{\}}
}
\\[0.1em]
\quad
{ \qass{u[0]\;}{\;\texttt{H}} }
\\[0.1em]

{
\big{\{}
\textcolor{spec}{
u[0]\sqcupplus h\srange{0}{n} \mapsto {\textstyle\schia{2^{(j\splus 1)}\,\sminus\,3}{\frac{1}{\sqrt{2}}z_i}{\ket{d_i}\ket{i\splus 1}\textcolor{stack}{\cmsg{\beta}}}}
+\schia{2^{(j\splus 1)}\,\sminus\,3}{\frac{1}{\sqrt{2}}z_i}{\ket{d_i\splus 1}\ket{i\splus 1}\textcolor{stack}{\cmsg{\beta}}}
+
\frac{1}{\sqrt{2^{t\splus 1}}}\ket{0}\ket{\overline{0}}\textcolor{stack}{\cmsg{\beta'}}
+\frac{1}{\sqrt{2^{t\splus 1}}}\ket{1}\ket{\overline{0}}\textcolor{stack}{\cmsg{\beta'}}
}
\big{\}}
} 
\end{array}
$
}

For the \texttt{H} application in line 5 (\Cref{fig:background-circuit-example-walk}), we first rewrite the locus, in the pre- and post-states, from $h\srange{0}{n}\sqcupplus u[0]$ to $u[0]\sqcupplus h\srange{0}{n}$; shown as the proof triple above.
There is a hidden uniqueness assumption \footnote{In the Dafny implementation, this needs to be an explicit assumption given by the users.} for all basis-kets in $q(j)$ (\Cref{fig:background-circuit-example-walk}): $\forall z\beta\ket{d_i} \in q(j) \Rightarrow \forall z'\,.\,z'\beta\ket{d_i\splus 1} \not\in q(j)$, i.e., if we truncate the $u[0]$ qubit, every basis is still unique in $q(j)$.
For each basis-ket, the \texttt{H} application duplicates the non-$u[0]$ part, with the flip of $u[0]$'s position bases.
During the process, the amplitude of each basis-ket is reduced by $\frac{1}{\sqrt{2}}$.
The \cn{left} and \cn{right} in lines 6 and 7 then move the node key (range $h[0,n)$'s position basis) of each basis-ket to its \cn{left} and \cn{right} child, depending on the coin bit stored as $u[0]$'s position basis; thus, the uniqueness property is preserved (left and right children always have different keys).
Remember that the number of basis-kets is doubled in the \texttt{H} gate application; after the $j$-th loop step, all $(j\sminus 1)$-th depth nodes become $j$-th depth and the two root node basis-kets ($\textcolor{spec}{\frac{1}{\sqrt{2^{t\splus 1}}}\ket{0}\ket{\overline{0}}}\textcolor{stack}{\cmsg{\beta'}}$ and $\textcolor{spec}{\frac{1}{\sqrt{2^{t\splus 1}}}\ket{1}\ket{\overline{0}}}\textcolor{stack}{\cmsg{\beta'}}$) become $1$-st depth nodes; thus, the state, after $j$-th loop step, is in superposition containing all nodes up to $j$-th depth, except the root node.

The above applications also show the necessity of frozen basis stacks.
When applying the conditional, we hide $x\srange{0}{t}$'s position bases to frozen basis stacks, and there are $2^{(j\splus 1)}\sminus 3$ different such stacks.
We need to record the position bases in the frozen basis stacks; so, 1) when we apply the $\cn{H}$ gate, we know which basis-kets are associated with a specific position basis; 2) once the conditional is over, the position bases can be retrieved.

The verification in \Cref{fig:background-circuit-example-walk} describes the basic property of the quantum walk algorithm framework.
The biggest advantage of the framework is to permit the manipulation of different quantum applications on different tree nodes in each loop step, which is why many algorithms \cite{Childs_2009,ChildsNAND,Low_2017,1366221} have been developed based on it.
 \section{Related Work}
\label{sec:related}

This section gives related work beyond the discussion in \Cref{sec:implementation}.

\noindent\textbf{\textit{Measurement-based Quantum Proof Systems.}}
Except for the works in \Cref{sec:implementation}, previous quantum proof systems are measurement-based,
including quantum Hoare logic \cite{qhoare,qhoreusage,10.1145/3456877,10.1007/s00165-018-0465-3}, quantum separation logic (QSL) \cite{qseplocal,qsepa}, quantum relational logic
\cite{relationlogic,10.1145/3290346}, and probabilistic Hoare logic for quantum programs \cite{10.1007/978-3-642-10622-4_7}, informed the \qafny development.
They differ from \qafny in three main respects, however:
1) their conditionals are solely classical, while \qafny has quantum conditionals;
2) they mainly focus on the probabilistic relations between the quantum measurement results and classical components and view quantum program components as black boxes specified by Hilbert spaces or density matrices;
and 3) most of them have no mechanized implementations, and they do not have a quantum program compiler.
The verification procedure in these frameworks, to some extent, shows the possibility of verifying mainly hybrid classical-quantum (HCQ) programs by requiring the input of black-boxed and verified quantum program components.

QSL \cite{qseplocal} develops a new separation logic theory (with no executable proof examples, however) for Hilbert spaces and classical controls, mainly for verifying HCQ programs by black-boxing quantum components.
This differs from the \qafny system, based on classical separation logic for classical array operations.
QSL is based on a notion of frame rules that split a tensor product state into two parts, similar to
our $\thadt$ typed state split equation.
However, they do not specify when and how a quantum state separation may happen.
As in \Cref{sec:overview}, in many cases, quantum state separation is not trivial and might not be automatically inferred by a proof engine.

Liu \textsf{et al.} \cite{qhoreusage} contains an example verification for Grover's search algorithm based on the \sqir inductive verification style, albeit with worse proof automation (3184 LOC vs.\/ 1018 LOC in \Cref{fig:circ-evaluation}).
\coqq \cite{10.1145/3571222} provides a mechanized automated verification framework for HCQ programs.
However, their proof automation is to connect quantum and classical components, i.e., they view quantum circuit components as black-boxes.
By giving pre- and post-conditions, they perform proof automation on verifying HCQ programs that view the quantum components as sub-procedures. There are some examples in \coqq to verify quantum components, but they are handled in the same style as \sqir and \qbricks above. See \Cref{appx:human}.

\noindent\textbf{\textit{Classical Proof Systems.}}
We are informed by separation logic, as articulated in the classic paper \cite{separationlogic}, and other papers as well
\cite{10.1145/3453483.3454087,arxiv.1609.00919,10.1007/978-3-319-89960-2_13,10.1007/978-3-319-89960-2_2,nat-proof-fun,10.1145/3453483.3454036}.
Primarily, we show a compilation from \qafny to \sep, representing a subset of separation logic admitting sound and completeness \cite{10.1007/3-540-45931-6_28}, which was also studied by \cite{TATSUTA20191,FAISALALAMEEN201673}.
The \qafny implementation is compiled to Dafny
\cite{10.1007/978-3-642-17511-4_20}, a language designed to simplify writing correct code.
The natural proof methodology \cite{nat-proof-fun,nat-proof-frame,10.1145/2103621.2103673} informs the \qafny development,
where it embeds the proofs of data-structures to a recursive search problem.
\ignore{
The \qafny implementation of the natural proof methodology identifies a subclass
of proofs $\mathpzc{N}$ such that (1) a large class of valid verification
specifications of near-term classical-quantum hybrid programs have proof in
$\mathpzc{N}$, and (2) searching for a proof in $\mathpzc{N}$ is efficiently
decidable. In the original natural proof, the subclass identification is based
on mapping heap data structures, such as trees and link lists, to logical
invariant properties. 
In QNP, the identification is through the \qafny type system in classifying
quantum loci and state types so that quantum operation applications on a specific locus can be compiled into classical aggregate operations that have
rich-proof infrastructures in Dafny.
}

 \section{Conclusion and Future Work}\label{sec:limit}

We present \qafny, a system for expressing and automatically verifying quantum programs that can be compiled into quantum circuits.
We develop a proof system that views quantum operations as classical array aggregate operations, e.g., viewing quantum measurements as array filters, so that we can map the proof system, which is sound and complete with respect to the \qafny semantics for well-typed programs, to classical verification infrastructure. 
We implement a prototype compiler in Dafny and use it to verify several quantum programs.
We believe that programmers can utilize \qafny to develop quantum algorithms and verify them through our automated verification engine with a significant saving of human effort, as demonstrated in \Cref{sec:arith-oqasm}.
The \qafny language is universal in terms of the power of expressing quantum programs since all gates in the universal RzQ gate set $\{\texttt{H}, \texttt{X}, \texttt{RZ}, \texttt{CNOT}\}$ \cite{Nam2018} are definable. However, being able to define all possible quantum programs does not mean that we can utilize \qafny to verify all quantum programs, especially HCQ programs, automatically.
Verifying HCQ programs requires the full power of quantum mixed states, i.e., users might want to know the probabilistic output of executing a quantum program with a quantum input state being associated with a probability.
Verifying all such programs requires a powerful classical probability distribution library beyond the current scope of \qafny, although the existing \qafny implementation does include a restricted library for reasoning about probability distributions~\cite{zeprop} that can verify some HCQ programs.

In future work, we plan to build and verify a complete \qafny implementation in Dafny; especially,
we intend to enhance the probability distribution libraries to automatically verify more HCQ programs. 
We also want to show the soundness proof of the implementation as well as the circuit compilation correctness from \qafny to \sqir (\Cref{sec:vqir-compilation}).
We will further investigate integrating \qafny with other tools, such as \coqq \cite{10.1145/3571222}, to verify HCQ programs automatically.

\bibliography{reference}

\newpage
\appendix
\section{Qafny Kind Checking, $FV$ function, and Locus Generation}\label{sec:session-gen}

\begin{figure}[h]
{\small
  \begin{mathpar}
    \inferrule[ ]{\Omega(x)=\cmode \vee \Omega(x)=\mmode }{\Omega \vdash x : \Omega(x)}

    \inferrule[ ]{\Omega(x)=\qmode{n} \\0 \le i \le j \le n}{\Omega \vdash x : x\srange{i}{j}}

    \inferrule[ ]{ \Omega \vdash a_1 : q_1 \\ \Omega \vdash a_2 : q_2 }{\Omega \vdash a_1 \,\splus\, a_2 : q_1 \sqcup q_2}   

    \inferrule[ ]{ \Omega \vdash a_1 : q_1 \\ \Omega \vdash a_2 : q_2 }{\Omega \vdash a_1 \cdot a_2 : q_1 \sqcup q_2}   
 
    \inferrule[ ]{ \Omega \vdash a_1 : q_1 \\ \Omega \vdash a_2 : q_2 \\ \Omega \vdash x[n] : q_3  }{\Omega \vdash \qbool{a_1}{=}{a_2}{x[n]} : q_1 \sqcup q_2\sqcup q_3}   

    \inferrule[ ]{ \Omega \vdash a_1 : q_1 \\ \Omega \vdash a_2 : q_2 \\ \Omega \vdash x[n] : q_3  }{\Omega \vdash \qbool{a_1}{<}{a_2}{x[n]} : q_1 \sqcup q_2\sqcup q_3}

    \inferrule[ ]{ \Omega \vdash b : q}{\Omega \vdash \neg b : q}  

    \inferrule[ ]{ \Omega \vdash e : \zeta_2 \sqcup \zeta_1 }{\Omega \vdash e : \zeta_1 \sqcup \zeta_2}   

  \end{mathpar}
}
{\footnotesize
\[
\begin{array}{l}
g \sqcup \kappa = \kappa
\qquad
\cmode \sqcup \mmode = \mmode
\qquad
g \sqcup g = g
\qquad
\kappa \sqcup \kappa' = \kappa\sqcupplus\kappa'
\end{array}
\]
}
  \caption{Arith and Bool Kind Checking}
  \label{fig:exp-well-typeda}
\end{figure}

The $FV(\Omega,-)$ function calculation is based on the kind checking procedure $\Omega\vdash - : (g_k,\kappa)$.
When the term $-$ is a $\cmode$ or $\mmode$ kind term, the output is $(g,\emptyset)$, while if the term $-$ is a quantum value with kind $\qmode{n}$, the output is $(\qmode{n},\kappa)$, where $\kappa$ contains all qubits mentioned in the term $-$ and its length $\slen{\kappa}=n$. As an abbreviation, we can think of the kind checking procedure has the form $\Omega\vdash - : g_k$ when the term $-$ is $\cmode$ or $\mmode$, while it has the form $\Omega\vdash - : \kappa$ when the term $-$ is of kind $\qmodename$.

\Cref{fig:exp-well-typeda} provides the kind check rules for arithmetic and Boolean expressions.
Here, we enforce a partial order over variable kinds as $\cmode \le \mmode \le \qmodename$,
such that the kind of an expression depends on the most general variable kind in the expression.
When two $\qmodename$ variables meet, their locus $\kappa$ and $\kappa'$ cannot be overlapped, as we enforce the disjoint union property $\kappa\sqcupplus\kappa'$, because quantum expressions are reversible. For any arithmetic operation, if the two parts of an expression, such as $q_1$ and $q_2$, is an addition $q_1 \splus q_2$, have overlapped locus fragments, the reversibility fails.
In the Boolean guards $\qbool{a_1}{=}{a_2}{x[n]}$ and $\qbool{a_1}{<}{a_2}{x[n]}$, the index appearing in quantum array $x$ must be a $\cmode$-kind value,
so the \qafny type system simplifies the expression to a natural number $n$ in the compilation time, which is why we use the value $n$ instead of an expression $a_3$ in the guards. 

The $FV(\Omega,-)$ function produces a locus in the term $-$, similar to computing free variables in a term.
Based on the kind checking procedure, the $FV$ calculation is defined as follows:

{
\begin{center}
$\Omega\vdash - :  \kappa \Rightarrow FV(\Omega,-) = \kappa$
\end{center}
}

The construct $-$ here refers to arithmetic, Boolean equations, or a statement.

\section{Qafny Well-formed and Well-typed Definitions}\label{appx:well-formeda}

The kind checking procedure $\Omega\vdash - : (g_k,\kappa)$ in \Cref{fig:exp-well-typeda} is essentially a well-formedness restriction that we place on the term $-$. There are two additional well-formed and well-typed restrictions in our type and proof systems.

\begin{definition}[Well-formed locus domain]\label{def:well-formed-ses}\rm 
  The domain of a environment $\sigma$ (or state $\varphi$) is \emph{well-formed}, written as
  $\Omega \vdash \dom{\sigma}$ (or $\dom{\varphi}$), iff for every locus $\kappa\in \dom{\sigma}$ (or $\dom{\varphi}$):
\begin{itemize}
\item $\kappa$ is disjoint unioned, i.e., for every two ranges $x\srange{i}{j}$ and $y\srange{i'}{j'}$, $x\srange{i}{j}\cap y\srange{i'}{j'}=\emptyset$.

\item For every range $x\srange{i}{j}\in\kappa$, $\Omega(x)=\qmode{n}$ and $[i,j) \subseteq [0,n)$.
\end{itemize}
\end{definition}

\begin{definition}[Well-formed state predicate]\label{def:well-formed-pred}\rm 
  A predicate $P$ is well-formed, written as $\Omega;\sigma \vdash P$, iff every variable and locus appearing in $P$ is defined in $\Omega$ and $\sigma$, respectively; particularly, if $P=\kappa \mapsto q * P'$, $\kappa \in \dom{\sigma}$.
\end{definition}

\section{Qafny Full Definition and More Examples}
\label{sec:qafny-app}

Here, we provide the details of the \qafny language and proof system, with more examples.

\subsection {Full Syntax}\label{appx:syntax-types}

\begin{figure}[h]
{
  \small
  \[\begin{array}{llcl} 
      \text{\oqasm Expr} & \mu\\
      \text{Arith Expr} & a & ::= & x \mid v \mid a \,\splus\, a \mid a \cdot a \mid ... \\
      \text{Bool Expr} & b & ::= & x[a] \mid \qbool{a_1}{=}{a_2}{\textcolor{red}{x[a]}} \mid \qbool{a_1}{<}{a_2}{\textcolor{red}{x[a]}} \mid ... \\
      \text{Predicate Locus} & \Kappa & ::= & \kappa \mid \mea(x,n,\kappa) \mid \frz(b,\kappa,\kappa) \mid \ufrz(b,\kappa,\kappa)\\
      \text{Predicate} & P,Q,R & ::= & a_1 = a_2 \mid a_1 < a_2 \mid \Kappa \mapsto q \mid P \wedge P \mid P * P \mid ... \\
      \text{Gate Expr} & op & ::= & \texttt{H} \mid \iqft[\lbrack -1 \rbrack]{}{} \\
      \text{C/M Kind Expr}& am & ::= & a \mid \smea{y} \\
      \text{Statement} & e & ::= & \sskip \mid \sexp{x}{am}{e} \mid  \ssassign{\kappa}{}{op} \mid \ssassign{\kappa}{}{\mu} 
                                 \\ & & \mid & \sseq{e}{e} \mid \sifq{b}{e} \mid
                                     \sqwhile{j}{a_1}{a_2}{b}{\{e\}}
                                 \\ & & \textcolor{red}{\mid} & \textcolor{red}{\sifb{b}{e}{e}\mid\ssassign{\kappa}{}{\sdis} \mid \ssassign{\kappa}{}{\sreduce{c}{n}}}
    \end{array}
  \]
}
\caption{\qafny Full Syntax }
  \label{fig:vqimp-full}
\end{figure}

\Cref{fig:vqimp-full} provides the \qafny syntax definition with extended constructs. 
In the Boolean guard syntax, the marked red parts can be omitted if the expressions do not contain any quantum variables, e.g., if variables $x$ and $y$ in expression $x=y$ are classical, we do not need an additional field after it to store the Boolean result.
The marked red part in a statement syntax contains three additional operations.
$\sifb{b}{e}{e}$ is a classical conditional operation, where $b$ has kind $\cmode$ or $\mmode$.
A quantum diffusion operation $\ssassign{\kappa}{}{\sdis}$ manipulates the amplitude of basis-kets of a locus containing \(\kappa\).
$\ssassign{\kappa}{}{\sreduce{c}{n}}$ is a quantum amplification operation that multiplies the basis-ket $\ket{c}$ in $\kappa$ with an amplitude $\frac{1}{\sqrt[4]{2^n}}$ and any other basis-kets with an amplitude $\sqrt{1-\frac{1}{\sqrt{2^n}}}$. Essentially, it reduces the amplitude volume in the basis-ket $\ket{c}$ while keeping the amplitudes of other basis-kets.

\subsection{Additional Semantic Rules}\label{sec:add-sem-rules}

\begin{figure*}[t]
{\scriptsize
  \begin{mathpar}
    \ignore{\inferrule[S-Par]{\varphi \equiv \varphi' \\ (\varphi',e)\Downarrow \varphi''}{(\varphi,e) \Downarrow \varphi''}}

    \inferrule[SH-Nor]{}{ (\varphi\uplus \{\kappa : \ket{c}\},\ssassign{\kappa}{}{\texttt{H}}) \Downarrow \varphi\uplus\{\kappa: \shad{2^{\slen{\kappa}}}{\slen{\kappa}\sminus 1}{\alpha(\frac{1}{2^{c[j]}})}\} }

    \inferrule[SH-Had]{\slen{c}=n\\ (r_j\ge\frac{1}{2}\Rightarrow c[j]=1) \vee (r_j<\frac{1}{2}\Rightarrow c[j]=0)}{ (\varphi\uplus\{\kappa : \shad{2^n}{n\sminus 1}{\alpha(r_j)}\},\ssassign{\kappa}{}{\texttt{H}}) \Downarrow \varphi\uplus\{\kappa : \ket{c}\} }

      \inferrule[S-IfT]{  (\varphi,e_1)\Downarrow \varphi' } 
      {(\varphi,\sifb{\cn{true}}{e_1}{e_2}) \Downarrow \varphi' }

      \inferrule[S-IfF]{  (\varphi,e_2)\Downarrow \varphi' } 
      {(\varphi,\sifb{\cn{false}}{e_1}{e_2}) \Downarrow \varphi' }

    \inferrule[S-Dis]{}
      {(\varphi\uplus \{\kappa\sqcupplus\kappa': q\},\ssassign{\kappa}{}{\sdis}) \Downarrow \varphi\uplus\{\kappa \sqcupplus\kappa' : \mathpzc{D}(\slen{\kappa},q)\} }

    \inferrule[S-Red]{}
      {(\varphi\uplus \{\kappa\sqcupplus\kappa': q\},\ssassign{\kappa}{}{\sreduce{c}{n}}) \Downarrow \varphi\uplus\{\kappa\sqcupplus\kappa' : \mathpzc{A}(\slen{\kappa},q)\} }
  \end{mathpar}
}
{\scriptsize
$
\begin{array}{l}
\forall u\in [0, 2^t).\, \tob{u}=c\Rightarrow z_{cu}=\frac{1}{\sqrt[4]{2^n}} \wedge \tob{u}\neq c\Rightarrow z_{cu}=\sqrt{1-\frac{1}{\sqrt{2^n}}}
\\[0.2em]
\begin{array}{l@{~}c@{~}l}
\mathpzc{D}(n,\Msum_{i=0}^{m}\scha{2^{n}\sminus 1}{z_{ij}}{j}{\ket{c_{ij}}}) &\triangleq&
\Msum_{i=0}^{m}\scha{2^n}{(\frac{1}{2^{n-1}}\Msum_{u=0}^{2^n\sminus 1}z_{iu} - z_{ij})}{j}{\ket{c_{ij}}}
\\[0.2em]
\mathpzc{A}(n,\Msum_{i=0}^{m}\scha{2^{t}\sminus 1}{z_{ij}}{j}{\ket{c_{ij}}})&\triangleq& \Msum_{i=0}^{m}\scha{2^t\sminus 1}{(\frac{1}{2^{t-1}}\Msum_{u=0}^{2^t\sminus 1}z_{cu}\cdot z_{iu} - z_{cj}\cdot z_{ij})}{j}{\ket{c_{ij}}}
\end{array}
\end{array}
$
}
\caption{Additional big-step semantic rules. $c[j]$ gets the $j$-th bit in the bitstring $c$. $\tob{u}$ turns an integer $u$ to a bitstring as basis.}
\label{fig:exp-semantics-3}
\end{figure*}

Here, we describe additional big-step semantics of the ones in \Cref{fig:exp-semantics-1}.
The rules are given in \Cref{fig:exp-semantics-3}.
Rule \rulelab{SH-Nor} describes the semantics for an $\cn{H}$ gate when its applicant state value is of $\tnort$ type,
while rule \rulelab{SH-Had} deals with the case when the applicant state value is of $\thadt$ type.
These two rules show any example semantic rules for defining the behaviors of state preparation and oracle operations when their applicant state types are not $\tcht$ type.
The rules for the other state preparation and oracle operations are defined in the same manner.
One thing worth noting is that any $\tnort$ or $\thadt$ type value can be rewritten to an $\tcht$ type value,
so these rules for $\tnort$ or $\thadt$ type values are equivalent shortcuts of the rules for $\tcht$ type values.

Rules \rulelab{S-IfT} and \rulelab{S-IfF} describe the semantics for classical conditional. Since the \qafny big-step semantics utilize a substitution-based approach for dealing with classical variables, the Boolean guards should have been freed of variables at the time when they are being evaluated; thus, they can be simplified as either $\cn{true}$ or $\cn{false}$.
Rules \rulelab{S-Dis} and \rulelab{S-Red} describe the semantics of amplitude diffusion and amplification operations, which are explained in \Cref{appx:proof-rules-amp}.

\subsection{Additional Typing Rules and Equivalence Relations}\label{sec:state-eq}

\begin{figure}
{
{\scriptsize
  \begin{mathpar}
    \inferrule[T-Skip]{}{\Omega;\sigma \vdash_g \sskip \triangleright \sigma}

    \inferrule[TH-Nor]{}{\Omega;\sigma\uplus\{\kappa:\tnort\} \vdash_g \ssassign{\kappa}{}{op} \triangleright \sigma\uplus\{\kappa:\thadt\}}

    \inferrule[TH-Had]{}{\Omega;\sigma\uplus\{\kappa:\thadt\} \vdash_g \ssassign{\kappa}{}{\ihadh} \triangleright \sigma\uplus\{\kappa:\tnort\} }

     \inferrule[T-IfC]{\Omega \vdash b : g\\
          \Omega;\sigma \vdash_{g'} e_1 \triangleright \sigma'\\\Omega;\sigma \vdash_{g'} e_2 \triangleright \sigma'}
        {\Omega;\sigma \vdash_g \sifb{b}{e_1}{e_2} \triangleright \sigma' }

        \inferrule[T-Dis]{}{\Omega,\sigma\uplus\{\kappa\sqcupplus\kappa':\tcht\} \vdash_g \ssassign{\kappa}{}{\sdis} \triangleright \sigma\uplus\{\kappa\sqcupplus\kappa':\tcht\} }

    \inferrule[T-Red]{}{\Omega,\sigma\uplus\{\kappa\sqcupplus\kappa':\tcht\} \vdash_g \ssassign{\kappa}{}{\sreduce{c}{n}} \triangleright \sigma\uplus\{\kappa\sqcupplus\kappa':\tcht\} }

  \end{mathpar}
}
}
  \caption{Additional \qafny type rules}
  \label{fig:exp-sessiontypea}
\end{figure}

\begin{figure}
\captionsetup[subfigure]{justification=centering}
{\scriptsize
{
\begin{minipage}[b]{0.35\textwidth}
\begin{center}
 \[
  \begin{array}{l@{~}cl}
  \tau & \sqsubseteq & \tau \\[0.3em]
  \tnort &\sqsubseteq& \tcht\\[1.3em]
  \thadt &\sqsubseteq& \tcht
    \end{array}
  \]
\end{center}
\subcaption{Subtyping}
  \label{fig:qafny-subtype}
\end{minipage}
\begin{minipage}[b]{0.5\textwidth}
\begin{center}
   \[\hspace*{-3em}
   \begin{array}{l@{~}cl}
   q & \equiv_{\slen{q}} & q \\[0.3em]
  \ket{c} &\equiv_n& \sch{0}{ }{\ket{c}}\\[0.5em]
  \shad{2^n}{n\sminus 1}{\alpha(r_j)} &\equiv_n& \sch{2^n\sminus 1}{\frac{\alpha(\Msum_{k=0}^{n\sminus 1} r_k \cdot \tob{j}[k])}{\sqrt{2^n}}}{\ket{j}}
    \end{array}
 \]
\end{center}
\subcaption{Quantum Value Equivalence}
  \label{fig:qafny-sequiv}
\end{minipage}
\hfill{}
\begin{minipage}[t]{0.8\textwidth}
\begin{center}
 \[
  \begin{array}{l @{~} c @{~} l l@{~}c @{~} l l @{~}c @{~} l l @ {~} c @{~} l}
\kappa &\equiv & \kappa
&
\qquad
x[n,n) & \equiv & \emptyset
&
\qquad
\emptyset \sqcupplus \kappa & \equiv & \kappa
&
\qquad
x[n,m)\sqcupplus\kappa & \equiv & x[n,j)\sqcupplus x[j,m) \sqcupplus\kappa
\\
&&&&&&&&&&&
\texttt{where}\;\;n \le j < m

    \end{array}
  \]
\end{center}
\subcaption{locus Equivalence}
  \label{fig:qafny-ses-equal}
\end{minipage}

  \caption{\qafny type/state relations.}
  \label{fig:qafny-eq1}
}
}
\end{figure}

\Cref{fig:exp-sessiontypea} provides additional \qafny type rules.
Rule \rulelab{T-Skip} defines the identity type rule for $\texttt{SKIP}$ operations.
Rules \rulelab{TH-Nor} and \rulelab{TH-Had} to track the types for Hadamard $\cn{H}$ state preparation operations with the applicant types to be $\tnort$ and $\thadt$ type values.
Rule \rulelab{T-IfC} defines the behaviors for classical conditionals with $b$ being of kind $\cmode$ or $\mmode$.
We require that the two different branches $e_1$ and $e_2$ output the same type of judgment ($\sigma'$).
Rules \rulelab{T-Dis} and \rulelab{T-Red} types the amplitude diffusion and amplification operations.
We require the input and output types for $\kappa$, as a locus fragment of $\kappa\sqcupplus\kappa'$, have the type $\tcht$.

As we suggested in \Cref{sec:monitortypes}, \qafny utilizes type environment and quantum state equations to rewrite loci and states in predicates to a canonical form, where the automated verification tool can apply proof rules by pattern matching.
Essentially, quantum computation is implemented as circuits. In \Cref{fig:background-circuit-examplea}, if the first and second circuit lines and qubits are permuted, it is intuitive that the two circuit results are equivalency up to the permutation. Additionally, as mentioned in \Cref{sec:semantics-proof}, we sometimes need quantum loci to be split and regrouped. All these properties are formulated in \qafny as equational properties in \Cref{fig:qafny-ses-equal} that rely on locus rewrites, which can then be used as built-in libraries in the proof system. As one can imagine, the equational properties might bring non-determinism in the \qafny implementation, such that the automated system does not know which equations to apply in a step. In dealing with the non-determinism, we design a type system for \qafny to track the uses, split, and join of loci, as well as the three value types in every transition step so that the system knows how to apply an equation.

\Cref{fig:qafny-eq1} shows the equivalence relations on types and values.
\Cref{fig:qafny-subtype} shows the subtyping relation such that $\tnort$ and $\thadt$ subtype to $\tcht$.
Correspondingly, the subtype of $\tnort$ to $\tcht$ represents the first line equation in \Cref{fig:qafny-sequiv}, where a $\tnort$-typed value is converted to an $\tcht$ typed value. Similarly, a $\thadt$-typed value can be converted to an $\tcht$ value in the second line.
Additionally, \Cref{fig:qafny-ses-equal} defines the equivalence relations for the locus concatenation operation $\sqcupplus$: associative, identified with the identity empty locus element $\emptyset$.
We also view a range $x[n,n)$ to be empty ($\emptyset$), and a range $x[n,m]$ can be split into two ranges in the locus as $x[n,j) \uplus x[j,m)$. 

\subsection{Additional Proof Rules}\label{appx:proof-rules}

\begin{figure*}[h]
{\scriptsize
  \begin{mathpar}
    \inferrule[P-Skip]{}{\fivepule{\Omega}{\varphi}{g}{P}{\sskip}{P}}

    \inferrule[PH-Nor]{}{\fivepule{\Omega}{\{\kappa : \tnort\}}{g}{\kappa \mapsto \ket{c}}{\ssassign{\kappa}{}{\texttt{H}}}{\kappa\mapsto \shad{2^{\slen{\kappa}}}{\slen{\kappa}\sminus 1}{\alpha(\frac{1}{2^{c[j]}})}}} 

    \inferrule[PH-Had]{\slen{c}=n\\ (r_j\ge\frac{1}{2}\Rightarrow c[j]=1) \vee (r_j<\frac{1}{2}\Rightarrow c[j]=0)}{
\fivepule{\Omega}{\{\kappa : \thadt\}}{g}{\kappa \mapsto \shad{2^n}{n\sminus 1}{\alpha(r_j)}}{\ssassign{\kappa}{}{\texttt{H}}}{\kappa\mapsto \ket{c}}}

    \inferrule[P-ExpM]{\Omega \vdash a : \mmode\\x\not\in \dom{\Omega}\\\\ \fivepule{\Omega[x\mapsto \mmode]}{\sigma}{g}{P*x=a}{e}{Q*x=a}}
                {\fivepule{\Omega}{\sigma}{g}{P}{\sexp{x}{a}{e}}{Q}}

      \inferrule[P-IfC]{\fivepule{\Omega}{\sigma}{g}{b\wedge P}{e_1}{Q}\\\fivepule{\Omega}{\sigma}{g}{\neg b\wedge P}{e_2}{Q}}{\fivepule{\Omega}{\sigma}{g}{P}{\sifb{b}{e_1}{e_2}}{Q} }

    \inferrule[P-Dis]{}
     {\fivepule{\Omega}{\{\kappa\sqcupplus\kappa':\tcht\}}{g}{\kappa\sqcupplus\kappa'\mapsto q}{\ssassign{\kappa}{}{\sdis} }{\kappa\sqcupplus\kappa'\mapsto \mathpzc{D}(\slen{\kappa},q)} }

    \inferrule[P-Loop1]{ n \ge n' }
                  {\fivepule{\Omega}{\sigma}{g}{P}{\sqwhile{j}{n}{n'}{b}{\{e\}}}{P}}

    \inferrule[P-Red]{}
     {\fivepule{\Omega}{\{\kappa\sqcupplus\kappa':\tcht\}}{g}{\kappa\sqcupplus\kappa'\mapsto q}{\ssassign{\kappa}{}{\sreduce{c}{n}} }{\kappa\sqcupplus\kappa'\mapsto \mathpzc{A}(\slen{\kappa},q)} }
  \end{mathpar}
}
\caption{Additional Proof Rules}
\label{fig:exp-proofsystem-seqb}
\end{figure*}

\Cref{fig:exp-proofsystem-seqb} provides additional \qafny proof rules.
Rule \rulelab{P-Skip} shows the proof rule for a \texttt{SKIP} operation.
Rules \rulelab{PH-Nor} and \rulelab{PH-Had} are proof rules for the $\texttt{H}$ state preparation operation with its value being of $\tnort$ and $\thadt$ typed values. As mentioned above, rules for state preparation operations with states other than $\tcht$ type are shortcuts for our automated verification tool.

Rule \rulelab{P-ExpM} describes the proof rule for a let-binding with $\mmode$-kind variable, which introduces an invariant property of the statement, where $x$ is assigned to an invariant value ($x=a$) along $e$'s evaluation, and the invariant is removed after $x$'s scope is ended. 
Rule \rulelab{P-IfC} shows a classical conditional rule, which is the same as a conditional rule in Hoare Logic.
The other rules (\rulelab{P-Dis} and \rulelab{P-Red}) are amplitude manipulation operations, which will be introduced in \Cref{appx:proof-rules-amp}.
Rule \rulelab{P-Loop1} defines the loop case when the upper bound $n'$ is greater than the lower bound $n$.

\begin{figure}[t]
{\footnotesize
\[\hspace*{-0.5em}
\begin{array}{r l}
\textcolor{blue}{1}
&
\textcolor{spec}{
\{A(x) * A(y) * B \}}
\;\;\text{where}\;\;
A(\beta) = \beta\srange{0}{n} \mapsto \ket{\overline{0}} 
\\
&
\qquad\qquad\qquad
\qquad\qquad\quad\;\;
B = 1 < a < N * n > 0 \;* N < 2^n \cdot \texttt{gcd}(a,N)=1
\\[0.4em]
\textcolor{blue}{2}
& \qass{x\srange{0}{n}\;}{\;\ihadh};\\[0.2em]

\textcolor{blue}{3}
&
\textcolor{auto}{
\{x\srange{0}{n} \mapsto \dabs{\ihadh}(\ket{\overline{0}}) * A(y) * B \}
}
\\[0.4em]
\textcolor{blue}{4}
&
\textcolor{auto}{
\Rightarrow
\{x\srange{0}{n} \mapsto C * A(y) * B \}
}
\;\;\text{where}\;\;
C = \shad{2^n}{n\sminus 1}{}
\\[0.4em]
\textcolor{blue}{5}
& \qass{y\srange{0}{n}\;}{\;y\srange{0}{n}\,\splus\, 1};
\\[0.4em]

\textcolor{blue}{6}
&
\textcolor{auto}{
\{x\srange{0}{n} \mapsto C * y\srange{0}{n} \mapsto \dabs{y\splus 1}(\ket{\overline{0}}) * B \}
}
\\[0.4em]
\textcolor{blue}{7}
&
\textcolor{auto}{
\Rightarrow
\{x\srange{0}{n} \mapsto C * y\srange{0}{n} \mapsto \ket{\overline{0}}\ket{1} * B\}
}
\\[0.4em]
\textcolor{blue}{8}
& 
\textcolor{auto}{
\Rightarrow
\{ E(0) * B\}
}
\;\;
\texttt{where}\;\;
E(t) =
\\
&
\qquad\qquad\qquad\quad
\begin{array}{l}
x\srange{t}{n} \mapsto \shadi{2^{n \,\sminus\,  t}}{n \sminus t\sminus 1}{}\;*
\\
x\srange{0}{t}\sqcupplus y\srange{0}{n} \mapsto \schai{2^t\sminus 1}{\frac{1}{\sqrt{2^t}}}{i}{\ket{a^{i}\;\mmod\;N}}
\end{array}
\\[0.4em]
\textcolor{blue}{9}
&
\qfor{j}{0}{n\,\sminus\,1}{x[j]}
\\[0.4em]
\textcolor{blue}{10}
&
\quad
\textcolor{spec}{\{E(j) * B\}}
\\[0.4em]
\textcolor{blue}{11}
&
\quad\qass{y\srange{0}{n}\;}{\;a^{2^j}\cdot y\srange{0}{n}\;\mmod\; N};
\\[0.4em]
\textcolor{blue}{12}
&
\textcolor{auto}{
\{E(n) * B\}
}
\\[0.4em]
\textcolor{blue}{13}
&
\textcolor{auto}{
\Rightarrow
\{x\srange{0}{n}\sqcupplus y\srange{0}{n} \mapsto \schai{2^{n}\sminus 1}{\frac{1}{\sqrt{2^{n}}}}{i}{\ket{a^{i}\;\mmod\;N}} * B\}
}
\\[0.4em]
\textcolor{blue}{14}
& \sexp{u}{\smea{y}}{}
\\[0.2em]
\textcolor{blue}{15}
&
\textcolor{auto}{
\{
x\srange{0}{n} \mapsto \smch{\frac{1}{\sqrt{s}}}{s\sminus 1}{t\,\splus\,k p} 
*
p = \texttt{ord}(a,N)
*
u=(\frac{p}{2^n},a^{t}\;\mmod\;N)
*
s=\texttt{rnd}(\frac{2^n}{p}) * B
\}
}
\\[0.4em]
\textcolor{blue}{16}
&
\quad\qass{x\srange{0}{n}\;}{\;\iqft[-1]{}{}}; 
\\[0.4em]
\textcolor{blue}{17}
&
\textcolor{auto}{
\{
x\srange{0}{n} \mapsto \smich{\frac{1}{\sqrt{s 2^n}}}{2^n\sminus 1}{(\omega^{tk}\Msum_{j=0}^{s\sminus 1}\omega^{tkj})}{k} 
*
p = \texttt{ord}(a,N)
*
u=(\frac{p}{2^n},a^{t}\;\mmod\;N)
*
s=\texttt{rnd}(\frac{2^n}{p}) * B
\}
}
\\[0.4em]
\textcolor{blue}{18}
&
\quad\sexp{w}{\smea{x}}{}
\\[0.4em]
\textcolor{blue}{19}
&
\textcolor{auto}{
\{
w=(\frac{4}{\pi ^ {2 p}},p)
*
p = \texttt{ord}(a,N)
*
u=(\frac{p}{2^n},a^{t}\;\mmod\;N)
*
s=\texttt{rnd}(\frac{2^n}{p}) * B
\}
}
\\[0.4em]
\textcolor{blue}{20}
&
\quad\quad\texttt{post}(w)
\\[0.4em]
\textcolor{blue}{21}
&
\textcolor{spec}{
\{
\texttt{post}(w)=(\frac{4e^{-2}}{\pi^2 \texttt{log}_2^4 N},p)
*
p = \texttt{ord}(a,N)
*
u=(\frac{p}{2^n},a^{t}\;\mmod\;N)
*
s=\texttt{rnd}(\frac{2^n}{p}) * B
\}
}
\end{array}
\]
}
\caption{The order finding of Shor's algorithm as the quantum portion. Type environments are on the right. $\sord{a,N}$ gets the order of $a$ and $N$. $\srnd{r}$ rounds $r$ to the nearest integer. The right-hand side contains the types of loci involved. $\ket{i}$ is an abbreviation of $\ket{\tob{i}}$.
$\tob{i}$ turns a number $i$ to a bitstring basis. 
$\omega=e^{\frac{2\pi i}{2^n}}$}
\label{fig:shorqafnya}
\end{figure}

\subsection{The Proof of The Order Finding Portion in Shor's Algorithm}\label{appx:shors}

Here, we show the complete proof scheme of Shor's algorithm~\cite{Shor94} in \qafny. 
Given an integer $N$, Shor's algorithm finds its nontrivial prime factors, which has the following step: (1) randomly pick a number $1 < a < N$ and compute $k=\texttt{gcd}(a,N)$;
(2) if $k \neq 1$, $k$ is the factor; (3) otherwise, $a$ and $N$ are coprime and we find the order $p$ of $a$ mode $N$;
(4) if $p$ is even and $a^{\frac{p}{2}} \neq -1\; \mmod\; N$, $\texttt{gcd}(a^{\frac{p}{2}}\pm 1,N)$ are the factors; otherwise, we repeat the process. Step (2) is Shor's algorithm's quantum component, an order-finding algorithm.
\Cref{fig:shorqafny} discusses its pre-measurement part. Without the type environments, we show the full \qafny proof in \Cref{fig:shorqafnya}.
In the Qafny implementation, verifying a quantum program requires only the user input of the pre-, post-, and loop invariant. For example, the proof of \Cref{fig:shorqafnya} in \qafny requires no states grayed out; only the conditions marked blue. 
In \Cref{fig:shorqafnya}, The steps before line 5 prepare the $\thadt$ type value for the range $x[0,n)$ and add 1 to the range $y\srange{0}{n}$, which flips $y$'s lowest bit. The two results result in the state as the beginning of \Cref{fig:shorqafny}.

After the two steps, \Cref{fig:shorqafny} discusses all the steps up to line 15 in \Cref{fig:shorqafnya}.
Here, we focus on the later parts. Line 16-18 is the quantum phase estimation part of the order finding algorithm.
In line 16, we apply an inversed QFT gate. The main functionality is to push the $k$ values to the basis-ket's amplitudes as $\omega^{tk}\Msum_{j=0}^{s}\omega^{tkj}$. Then, line 18 measures variable $x$. In this procedure, we will get the period $p$ as part of the measurement result $w$ as $(\frac{4}{\pi ^ {2 p}},p)$. However, $w$ cannot be directly used as an order number in the actual order-finding implementation. Instead, we need a post step, the continued fraction method, to retrieve the order, which is stated as $\texttt{post}(w)$ in line 20. The post step takes an $\mmode$-kind value and returns a $\mmode$-kind value $\texttt{post}(w)=(\frac{4e^{-2}}{\pi^2 \texttt{log}_2^4 N},p)$, where $p$ is the order number that we are looking for, and $\frac{4e^{-2}}{\pi^2 \texttt{log}_2^4 N}$ is its probability value.
In the \qafny implementation in Dafny, we implement the conditioned fraction post steps and all these probability theorems by writing them as Dafny library functions and providing mechanized proofs by mimicking the proofs in \cite{shorsprove}.

{\scriptsize
  \begin{mathpar}
\mprset{flushleft}
\inferrule[]{
   \inferrule[]
   { \inferrule[] {
    \fivepule{\Omega}{\sigma_2}{\mmode}{X(\snext{j}) * x\srange{0}{j}\sqcupplus y\srange{0}{n}\mapsto C(j).1}{ e }{X(\snext{j}) * x\srange{0}{j}\sqcupplus y\srange{0}{n}\mapsto C'(j).1} }
  { \fivepule{\Omega}{\sigma_2}{\mmode}{X(\snext{j}) * \frz(b,x[j\splus 1],x\srange{0}{j}\sqcupplus y\srange{0}{n})\mapsto 0.C(j)+1.C(j)}{ e }{X(\snext{j}) * x\srange{0}{j}\sqcupplus y\srange{0}{n}\mapsto C'(j).1} } }
   {  \Omega,\sigma_1\vdash_{\mmode} \{X(\snext{j}) * x\srange{0}{j\splus 1}\sqcupplus y\srange{0}{n}\mapsto 0.C(j)+1.C(j)\} \sifq{x[j]}{e}
     \\\\\qquad\qquad \{ X(\snext{j}) * \ufrz(\neg x[j],x[j],x\srange{0}{j\splus 1}\sqcupplus y\srange{0}{n})\mapsto 0.C(j)
          * \ufrz(x[j],x[j],x\srange{0}{j\splus 1}\sqcupplus y\srange{0}{n})\mapsto C'(j).1\}
     } } {
\fivepule{\Omega}{\sigma}{\mmode}{X(j) * x\srange{0}{j}\sqcupplus y\srange{0}{n}\mapsto C(j)}{ \sifq{x[j]}{e} }{X(j\,\sminus\,1) * x\srange{0}{j\splus 1}\sqcupplus y\srange{0}{n} \mapsto 0.C(j)+1.C'(j)} }
  \end{mathpar}
{
$\begin{array}{l}
X(j)=x\srange{j}{n}\mapsto \shad{2^{n \,\sminus\,  j}}{n \sminus j\sminus 1}{}
\qquad
C(j)=\sch{2^k\sminus 1}{\frac{1}{\sqrt{2^k}}}{\ket{\tob{j}^{k}}\ket{a^{\tob{j}^k}\;\mmod\;N}}
\\[0.2em]
d.C(j)=\sch{2^{\snext{k}}\sminus 1}{\frac{1}{\sqrt{2^{\snext{k}}}}}{\ket{\tob{j}^k}\ket{d}\ket{a^{\tob{j}^k}\;\mmod\;N}}
\qquad
C'(j).d=\sch{2^{\snext{k}}\sminus 1}{\frac{1}{\sqrt{2^{\snext{k}}}}}{\ket{a^{\tob{j}^k.1}\;\mmod\;N}\ket{\tob{j}^k}\ket{d}}
\\[0.2em]
d.C'(j)=\sch{2^{\snext{k}} \sminus 1}{\frac{1}{\sqrt{2^{\snext{k}}}}}{\ket{\tob{j}^k}\ket{d}\ket{{a^{\tob{j}^k.1}\;\mmod\;N}}}
\qquad
\sigma_1=\{x\srange{0}{n\sminus (j\splus 1)} : \thadt, \{x\srange{0}{j\splus 1},y\srange{0}{n}\} : \tcht \}
\\[0.2em]
\sigma=\{x\srange{0}{n\sminus (j\splus 1)} : \thadt, y\srange{0}{n\sminus (j\splus 1)} : \tcht \}
\qquad
e=\qass{y\srange{0}{n}\;}{\;a^{2^j}\cdot y\srange{0}{n}\;\mmod\; N}
\end{array}$
}
}

As an additional example for the frozen and unfrozen functions in the proof rule of quantum conditionals in \Cref{sec:semantics-proof}, we provide an additional proof tree for the quantum conditional in a loop step in lines 9-11 in \Cref{fig:shorqafnya} above.
Here, $\tob{j}$ means to turn natural number $j$ to a bitstring and $c.d$ means concatenating a single bit $d$ to the end of bitstring $c$.
The proof is built from the bottom up. We first cut the $\thadt$ type value into two loci ($x\srange{0}{n\sminus (j\splus 1)}$ and $x[j]$), join $x[j]$ with locus $x\srange{0}{j}\sqcupplus y\srange{0}{n}$, and duplicate the value elements to be $0.C(j)+1.C(j)$, which is proved by applying the consequence rule (\rulelab{P-Con}) in \Cref{fig:exp-proofsystem-1}. Notice that the type environment also transitions from $\sigma$ to $\sigma_1$.
By the same strategy of the $\ufrz$ transformer in \Cref{sec:semantics-proof} , we combine the two $\ufrz$ terms into the final result. 
The second step applies rule \textsc{P-IF} to substitute locus $x\srange{0}{j\splus 1}\sqcupplus y\srange{0}{n}$ with the frozen construct $\frz(b,x[j\splus 1],x\srange{0}{j}\sqcupplus y\srange{0}{n})$ in the pre-condition and create two $\ufrz$ terms in the post-condition. The step on the top applies the modulo multiplication on every element in the frozen value $\mapsto C(j).1$ by rule \textsc{P-OP}.

\subsection{Translation from Qafny to SQIR}\label{sec:vqir-compilation}

\newcommand{\tget}{\texttt{get}}
\newcommand{\tstart}{\texttt{start}}
\newcommand{\tfst}{\texttt{fst}}
\newcommand{\tsnd}{\texttt{snd}}
\newcommand{\tucom}[1]{\texttt{ucom}~{#1}}
\newcommand{\tif}{\texttt{if}}
\newcommand{\tthen}{\texttt{then}}
\newcommand{\telse}{\texttt{else}}
\newcommand{\tlet}{\texttt{let}}
\newcommand{\tin}{\texttt{in}}

We translate \qafny to \sqir by mapping loci---representing qubit arrays---to \sqir concrete qubit indices and expanding the \qafny instructions to sequences of \sqir gates.
The translation is expressed as the judgment
$\Omega;\Upsilon;n\vdash e \gg \epsilon$, where: $\Omega$ is the kind environment, mainly used for recording the range qubit size represented by a \texttt{Q}~kind variable; 
$\epsilon$ is the output \sqir circuit; and $\Upsilon$ maps a \qafny range variable
position $x[i]$, in the range $x\srange{j}{k}$ where $i \in [j,k)$, to a \sqir concrete qubit index (i.e., offset into a 
global qubit register), and $n$ is the current index bound.  
At the start of translation, for every
variable $x$ and $i < \texttt{nat}(\Omega(x))$ ($\Omega(x)=\qmode{m}$ and $\texttt{nat}(\qmode{m})=m$), $\Upsilon$ maps $x[i]$ to a unique concrete index, chosen from 0 to $n$.

\begin{figure}[h]
{\footnotesize

  \begin{mathpar}

    \inferrule{\Omega;\Upsilon;n\vdash b\,\cn{@}\,x[i] \gg \textcolor{blue}{\epsilon}\\
                \Omega;\Upsilon;n\vdash e \gg \textcolor{blue}{\epsilon'}}
        {\Omega;\Upsilon;n\vdash \sifq{b\,\cn{@}\,x[i]}{e} \gg \textcolor{blue}{\sseq{\epsilon}{\texttt{ctrl}(\Upsilon(x[i]),\epsilon')}}}    

    \inferrule{\forall t\in[i,j).\,\Omega;\Upsilon;n\vdash \sifq{b[t/x]}{e[t/x]} \gg \textcolor{blue}{\epsilon_{t}}}
        {\Omega;\Upsilon;n\vdash \sqwhile{x}{i}{j}{b}{e} \gg \textcolor{blue}{\epsilon_{i}\,;...;\,\epsilon_{j\,\sminus\,1}}}  
        
    \inferrule{}{\Omega;\Upsilon;n\vdash \sskip \to \textcolor{blue}{\texttt{SKIP}}}

    \inferrule{x\not\in \dom{\Omega} \\\\ \Omega[x\mapsto \cmode];\Upsilon;n\vdash s[m/x] \to\textcolor{blue}{\epsilon}}
           {\Omega;\Upsilon;n\vdash \sexp{x}{m}{s} \to \textcolor{blue}{\epsilon}}

    \inferrule{x\not\in \dom{\Omega} \\\\ \Omega[x\mapsto \mmode];\Upsilon;n\vdash s[(r,n)/x] \to\textcolor{blue}{\epsilon}}
           {\Omega;\Upsilon;n\vdash \sexp{x}{(r,n)}{s} \to \textcolor{blue}{\epsilon}}

    \inferrule{}
         {\Omega;\Upsilon;n\vdash \ssassign{\kappa}{}{op} \to \textcolor{blue}{op(\Upsilon(\kappa))}}

    \inferrule{\Omega;\Upsilon;n\vdash \mu \to \textcolor{blue}{\epsilon} }
         {\Omega;\Upsilon;n\vdash \ssassign{\kappa}{}{\mu} \to \textcolor{blue}{\epsilon}}
 
      \inferrule{}
         {\Omega;\Upsilon;n\vdash \ssassign{\kappa}{}{\sdis} \to \textcolor{blue}{\sdis(\Upsilon(\kappa))}}

      \inferrule{\Omega;\Upsilon;n\vdash s_1:\textcolor{blue}{\epsilon_1} \\
                        \Omega;\Upsilon;n\vdash s_2:\textcolor{blue}{\epsilon_2} }
         {\Omega;\Upsilon;n\vdash \sseq{s_1}{s_2} \to \textcolor{blue}{\sseq{{\epsilon}_1}{{\epsilon}_2}}}
        
  \end{mathpar}
}
\caption{\qafny to \sqir translation rules (\sqir circuits are marked blue)}
\label{appxfig:compile-vqira}
\end{figure}

\Cref{appxfig:compile-vqira} depicts two translation rules with the \textcolor{blue}{blue} part being the \sqir circuits.
The first rule describes the translation of a quantum conditional to a controlled operation in \sqir.
Here, we first translate the Boolean guard $b\,\cn{@}\,x[i]$ \footnote{Here, $b$ is $a_1 \,\cn{<}\, a_2$, $a_1\,\cn{=}\,a_2$, or $\texttt{false}$ referring to the Boolean equation parts of $\qbool{a_1}{<}{a_2}{x[i]}$, $\qbool{a_1}{=}{a_2}{x[i]}$, or $x[i]$. }, and sequentially we translate the conditional body as an \sqir expression controlled on the $x[i]$ position. \texttt{ctrl} generates the controlled version of an arbitrary \sqir program using standard decompositions \cite[Chapter 4.3]{mike-and-ike}.
The last rule translates \qafny for-loops. A for-loop is compiled into a series of conditionals with different loop step values. 
For the compiler correctness, we validate \Cref{thm:proof-compile-sqir} by testing, where $\llbracket \epsilon \rrbracket(\llbracket \varphi \rrbracket_{\Omega})$ is the density matrix semantic interpretation of $\epsilon$ on the density matrix state $\llbracket \varphi \rrbracket_{\Omega}$ in \sqir \cite{VOQC}.

\begin{theorem}[\qafny to \sqir Compilation Correctness]\label{thm:proof-compile-sqir}\rm 
For any $e$, such that $\Omega;\sigma \vdash_g e \triangleright \sigma'$, $\Omega;\sigma \vdash_g \varphi$, and $\Omega;\Upsilon;n\vdash e \gg \epsilon$; thus, if $(\varphi,e)\Downarrow \varphi'$, then
$\llbracket \epsilon \rrbracket(\llbracket \varphi \rrbracket_{\Omega})=\llbracket \varphi' \rrbracket_{\Omega}$.
\end{theorem}

Here, we extract the \qafny semantic interpreter and the \qafny-to-\sqir compiler to OCaml and translate the compiled \sqir programs to OpenQASM \cite{cross2021openqasm} via the \sqir compiler. Then, we run the test programs in the \qafny OCaml interpreter and compiler to see if the results match. Overall, we have run 135 unit test programs to check individual operations and small, composed programs, and all the results from the \qafny interpreter and compiler are matched.

\Cref{appxfig:compile-vqira} also provides the other compilation rules for the operations in \Cref{fig:vqimp}. In compiling the value and diffusion operations, we utilize the verified \sqir infrastructure \cite{VOQC} for the $\texttt{H}$, $\texttt{QFT}$, and diffusion circuits. The function $\Upsilon(\kappa)$ translates the locus $\kappa$ to a qubit array with the physical indices required in \sqir.
The last two rules in the first line show how a $\texttt{let}$-binding is handled for a $\cmode$ and $\mmode$ kind variable.
Similar to the type system, we assume that \texttt{let}-binding introduces new variables, probably with proper variable renaming.
The rule for $\ssassign{\kappa}{}{\mu}$ describes an oracle application compilation, which is handled by the \oqasm compiler Li \textsf{et al.} \cite{oracleoopsla}. The \qafny type system ensures that the qubits mentioned in $\mu$ are the same as qubits in locus $\kappa$ so that the translation does not rely on $\kappa$ itself.

\section{Human Effort Comparison}\label{appx:human}

\begin{figure*}[h]
\begin{center}
{\footnotesize
\begin{tabular}{|l|l|l|l|l|l|}
\hline
Algorithm
&
\qafny
&
\qbricks
&
\sqir
&
QHL
&
\coqq
\\
\hline
GHZ & < 1 & - & 10 & - & - \\
Deutsch–Jozsa & < 1 & 3 & 12 & - & - \\

Grover's search & 2 & 3 & 7 & 30 & 14 \\
Shor's algorithm & 5 & 28 & 1040 & - & - \\
Controlled GHZ & < 1 & - & - & - & - \\
Quantum Walk & 3 & - & - & - & -\\
\hline                           
\end{tabular}
}
\end{center}
\caption{Human effort comparison, which measures the time for a single person to finish programming and verifying an algorithm; \texttt{-} means no data. }
\label{fig:circ-evaluation2}
\end{figure*}

We did a small survey regarding the human efforts of different groups of researchers who work on quantum program verification.
We sent out survey questionnaires to the different research groups listed in \Cref{fig:circ-evaluation2} through emails regarding two major questions: 1) the human efforts in verifying the list of quantum programs and 2) the quantum research experience the researchers had before verifying the programs.

The data are summarized in \Cref{fig:circ-evaluation2}.
We show the total number of days for each algorithm by summing the time for individuals finishing the algorithm verification, each column per group.
We show the human efforts for \qafny, \qbricks~\cite{qbricks}, \sqir~\cite{VOQC, PQPC}, Quantum Hoare Logic Implementation (QHL) \cite{qhoreusage}, and \coqq \cite{10.1145/3571222}.
We show the results for verifying four different algorithms, including GHZ, Deutsch–Jozsa, Grover's search, and Shor's algorithm, because these are the common quantum algorithms that many research groups verified.
The verification of Controlled GHZ and Quantum Walk is uniquely done in \qafny, as the examples show that \qafny is suitable for verifying new quantum algorithms beyond the standard ones, such as the four algorithms above.
QHL and \coqq mainly focus on verifying HCQ programs, and the only quantum program among the four algorithms they verified is Grover's search algorithm. 

Since we are measuring human efforts, the survey subjects to the personal efforts of individuals, such as the hours that these researchers spent on the project per day and their familiarity with the systems.
The verification of quantum programs in \qafny was finished by a student who was not the \qafny developer and had one year of experience in the study of quantum computing.
For \sqir, the verification of the GHZ and Deutsch–Jozsa algorithms was done by a non-developer student having two-year of experience in quantum computing, while Grover's search verification was finished by the \sqir developer, having studied quantum computing for three years at the time of performing the verification.
Shor's algorithm was verified by four experienced researchers, including the developer.
The algorithm verification, done by \qbricks, QHL, and \coqq, was finished by the developer of the frameworks, who had at least three years of quantum computing experience, respectively.

Some of the algorithm verification has different content. As we mentioned in \Cref{sec:related}, verifying Shor's algorithm from \qbricks is finished by assuming without proving many mathematical theorems. In contrast, \sqir verified every mathematical theorem in verifying their version of Shor's algorithm. 

Unlike \qafny, for both \qbricks and \sqir, we can see that the framework verification is not scalable to large program verification, as verifying a large algorithm, such as Shor's algorithm, requires significantly more time than verifying small and well-known programs.
The GHZ, Deutsch–Jozsa, and Grover's search algorithms are relatively small, and their step-by-step behaviors are well-known in much literature, while some parts of Shor's algorithm, such as the probability bounds, are not known until the verification is done by \sqir.

The only algorithm verification done in QHL is Grover's search algorithm, which has 3184 lines of code.
It took the QHL developer one month to finish the verification.
\coqq inherits the style of verification from QHL, with a proof system to verify hybrid classical-quantum (HCQ) programs. 
The style of the quantum component verification in QHL and \coqq is similar to \sqir.
Other than Grover's search, \coqq finishes several other kinds of HCQ program verification, 
which is beyond the scope of the paper. These algorithms typically assume that a large portion of quantum components is given as black boxes and verify the behaviors of utilizing the black-boxed quantum components as procedures in a classical program.
For example, the definition of the quantum phase estimation (QPE) program in \coqq differs from the \sqir one.
In \sqir, the QPE's oracle part is defined as a loop structure, each loop step contains a controlled-$U$ gate, while the oracle part in the \coqq version is defined as a big black box, with the given assumptions of the pre- and post-conditions of the black box.

\section{Amplitude Manipulation Operations}\label{appx:proof-rules-amp}

\begin{wrapfigure}{r}{5.5cm}
  \includegraphics[width=0.37\textwidth]{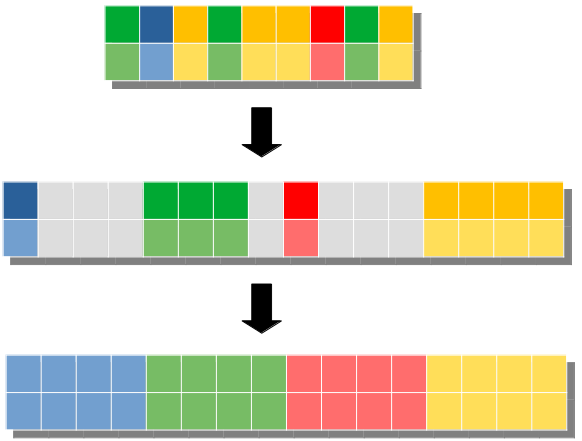}
  \caption{Diffusion Analogy}
\label{fig:qafny-dis-analog}
\end{wrapfigure} 

Quantum amplitude manipulation operations are to manipulate the amplitudes of basis-kets in a quantum state.
As shown in \Cref{sec:quantumwalk}, the quantum walk framework is a framework for manipulating specific basis-kets in a state if certain conditions are met, such as the controlled Hadamard operation in \Cref{fig:background-circuit-example-walk} line 5.
In fact, the controlled Hadamard operation is a simple case for using amplitude manipulation operations in a quantum walk framework. Here, its utility duplicates basis-kets so that we can allow two children's tree nodes to appear simultaneously.
In \Cref{fig:exp-semantics-3,fig:exp-proofsystem-seqb},
rules \rulelab{S-Dis} and \rulelab{P-Dis} as well as \rulelab{S-Red}  and \rulelab{P-Red}
show the semantic and proof rules for quantum amplitude manipulation operations.
The former two show the rules for amplitude diffusion, and the latter two show the rules for amplification operations.
The difference between a diffusion and an amplification is that the different functionality, i.e.,
the former is used to duplicate basis-kets or average the amplitudes in different basis-kets, while the latter is to amplify the amplitudes of some basis-kets so they can have a better chance to measure out.
As we know, amplitudes associated with basis-kets represent the likelihood the basis-ket is present when a measurement happens.
Amplifying the amplitudes certainly increases the chance of measuring the basis-kets the amplitudes are attached to.

Quantum diffusion operations ($\ssassign{\kappa}{}{\sdis}$) reorient the amplitudes of basis-kets. They are analogized to an aggregate operation of reshape and mean computation; both appeared in some programming languages, such as Python.
The aggregate operation first applies a reshape, where elements are regrouped into a normal form, as the first arrow of \Cref{fig:qafny-dis-analog}. More specifically, the diffusion function $\mathpzc{D}(n,q)$ (Rule \textsc{SDis} and \textsc{PDis} in \Cref{fig:exp-semantics-3,fig:exp-proofsystem-seqb}) first takes an $n'$-element $\tcht$ typed value $\schi{t=0}{z_t}{c_t}$, where $n$ corresponds to the number of qubits in $\kappa$.
Then, we rearrange the value by extending the element number from $n'$ to $m*n$ by probably adding new elements that originally have zero amplitude (the white elements in \Cref{fig:exp-proofsystem-seqb}).
Here, let us view a basis $c_{t}$ as a little-endian (LSB) number
$\tov{c_{t}}$. The rearrangement of changing bases $c_{t}$ (for all $t$) to
$\tob{j}.c_{ij}$ is analogized to rewrite a number $\tov{c_{t}}$ to be the form
$2^n i+j$, with $j\in [0,2^n)$, i.e., the reshaping step rearranges the
basis-kets to be a periodic counting sequence, with $2^n$ being the order.
The mean computation analogy (the second arrow in \Cref{fig:qafny-dis-analog}) takes every period in the reshaped value. For each basis-ket in a period, we redistribute its amplitude by the formula $(\frac{1}{2^{n-1}}\sum_{u=0}^{2^n\sminus 1}z_{iu} - z_{ij})$.
The diffusion function $\mathpzc{D}(n,q)$ is implemented as a locus predicate transformer in \qafny, similar to $\frz$ and $\ufrz$ in \Cref{sec:semantics-proof}.
An example of the quantum walk algorithm that uses diffusion operations is given in \Cref{sec:groverexample}.

Rules \rulelab{SRed} and \rulelab{PRed} are the semantic and proof rules for the quantum amplification ($\texttt{reduce}$). It is similar to the diffusion operation, except we only need to multiply an additional amplitude factor $z_{cu}$ for each amplitude term.
For each $u \in [0,2^t)$, if $\tob{u}=c$, we multiply the factor $\frac{1}{\sqrt[4]{2^n}}$, if $\tob{u}\neq c$, we multiply the factor 
$\sqrt{1-\frac{1}{\sqrt{2^n}}}$. One example usage is to rewrite the quantum walk algorithm in \Cref{fig:background-circuit-example-walk} to a new one by replacing the diffusion operation with $\ssassign{u}{}{\sreduce{\overline{0}}{n}}$, where $n$ is an arbitrary number. If we do so, whenever a new basis-ket is introduced to the active set, its amplitude will be reduced compared to other basis-kets in the active set.
The amplitudes representing the basis-ket likelihood are low for each basis-ket here. In quantum computing, amplification operations serve the opposite of quantum diffusion to increase certain basis-ket amplitudes.
Notice that every new basis-ket that becomes active in each loop step starts with the root node, which means that basis-kets having $\snext{j}$-th depth nodes have higher amplitudes than basis-kets having $j$-th depth nodes so that those leaf nodes having the biggest amplitudes, which is ideal because most tree algorithms are likely to work on the leaves rather than the middle transition nodes. 

\subsection{The Grover's Search Proof in Qafny with a New Amplitude Representation}\label{sec:groverexample}

\begin{figure}[t]
{\footnotesize
\[\hspace*{-0.5em}
\begin{array}{r l}
\textcolor{blue}{1}
&
\textcolor{spec}{
\big{\{} x\srange{0}{n}\mapsto \ket{\overline{0}} * N = \lfloor \frac{\xi}{\chi} \rfloor \big{\}}}
\;\;
\texttt{where}\;\;
\xi=\cos^{-1}(\frac{1}{\sqrt{2^n}})
\wedge \chi=\pi - 2\xi
\\[0.4em]
\textcolor{blue}{2}
& \qass{x\;}{\;\ihadh};
\\[0.2em]
\textcolor{blue}{3}
&
\textcolor{auto}{
\big{\{}x\srange{0}{n} \mapsto \shad{2^n}{n\sminus 1}{} * N = \lfloor \frac{\xi}{\chi} \rfloor \big{\}}
}
\\[0.4em]
\textcolor{blue}{4}
&
\textcolor{auto}{
\Rightarrow
\big{\{}x\srange{0}{n} \mapsto \frac{1}{\sqrt{2^n}}\ket{k} + \frac{1}{\sqrt{2^n}}\ssum{j=0\wedge j\neq k}{2^n\sminus 1}{\ket{j}} * N = \lfloor \frac{\xi}{\chi} \rfloor \big{\}}
}
\\[0.4em]
\textcolor{blue}{5}
&
\textcolor{auto}{
\Rightarrow
\big{\{}x\srange{0}{n} \mapsto \cos\varrho_0 \ket{k} + \sin\varrho_0 \cdot q * N = \lfloor \frac{\xi}{\chi} \rfloor \big{\}}
\;\;\texttt{where}\;\;q=\frac{1}{\sqrt{2^n-1}}\ssum{j=0\wedge j\neq k}{2^n\sminus 1}{\ket{j}}
\wedge \varrho_i=\xi-j\cdot \chi
}
\\[0.4em]
\textcolor{blue}{6}
&
\texttt{for}~\;{j \in [0,N)}
\\[0.4em]
\textcolor{blue}{7}
&
\;\;
\textcolor{spec}{\big{\{} x\srange{0}{n} \mapsto \cos\varrho_j \ket{k} + \sin\varrho_j \cdot q * N = \lfloor \frac{\xi}{\chi} \rfloor \big{\}}}
\\[0.4em]
\textcolor{blue}{8}
& 
\{
\\[0.4em]
\textcolor{blue}{9}
& 
\quad
\qass{x\;}{\;U()};
\\[0.4em]

\textcolor{blue}{10}
& \quad\qass{x\;}{\;\sdis};
\\[0.4em]
\textcolor{blue}{11}
&
\}
\\[0.4em]
\textcolor{blue}{12}
&
\textcolor{auto}{
\big{\{} x\srange{0}{n} \mapsto \cos\varrho_N \ket{k} + \sin\varrho_N \cdot q * N = \lfloor \frac{\xi}{\chi} \rfloor \big{\}}
}
\\[0.4em]
\textcolor{blue}{13}
&
\sexp{y}{\smea{x}}{\sskip}
\\[0.4em]
\textcolor{blue}{14}
&
\textcolor{spec}{
\big{\{} y=(p,k) * p > \cos^2(\chi) * N = \lfloor \frac{\xi}{\chi} \rfloor \big{\}}
}
\end{array}
\]
}
\caption{Grover's Search in \qafny.}
\label{fig:grovesearch}
\end{figure}

Amplitudes $z$ in \Cref{fig:qafny-state} are complex numbers. In general, they are hard to admit proof automation.
In the \qafny implementation in Dafny, we observe the common patterns of many quantum algorithms,
and implement amplitudes $z$ as $r\cdot e^{2\pi i \cdot \frac{1}{n}}$, where $r$ is a real number in the range $[0,1)$ and $e^{2\pi i \cdot \frac{1}{n}}$ is a phase. Since a phase always has the pattern $e^{2\pi i \cdot \frac{1}{n}}$, we can represent it as a natural number $n$. In addition, $r$ can also be approximated by rational numbers. Thus, in Dafny, we represent an amplitude by a pair of a rational number, representing the $r$ amplitude, and a natural number $n$, representing the phase part $e^{2\pi i \cdot \frac{1}{n}}$.
With this setting, we enable many rewrite theories of rational and natural numbers for proof automation on capturing the changes of amplitudes, which successfully help us verify the programs in \Cref{fig:circ-evaluation2}, except Grover's search.

A quantum program involving amplitude manipulation operations usually requires a comprehensive representation with a new theory library to promote proof automation.
Here, we utilize Grover's search proof \cite{grover1996} as an example to highlight the use of a comprehensive representation: the trigonometric amplitude representations.
\Cref{fig:grovesearch} provides the proof steps of the algorithm in \qafny.
The \qafny proof system can automatically infer the gray parts in the proof with some understanding of trigonometric theorems defined by our researcher in the Dafny framework.
The function $U()$ is a predefined oracle function. There are two properties for it.
First, we know that $k$ is the key in the oracle function. $U()\ket{j}=-1\cdot{j}$, if $j=k$; otherwise, $U()\ket{j}=-1\cdot{j}$.
Second, we also require that such a key is unique for all $j \in [0,2^n)$. 
Thus, if we input an $\tcht$ typed value $\sch{2^n}{}{\ket{j}}$ to the oracle function, the result value can be divided into two parts:
$-\frac{1}{\sqrt{2^n}}\ket{k} + \frac{1}{\sqrt{2^n}}\ssum{j=0\wedge j\neq k}{2^n\sminus 1}{\ket{j}}$.

Notice that for any quantum value $\sch{m}{z_j}{\ket{c_j}}$, we have $\sum_{j=0}^{m}\slen{z_j}^2 = 1$.
For the above two parts, if we view $\frac{1}{\sqrt{2^n}}$ as $\cos\xi$, the second part can be rewritten as $\sin\xi \cdot q$ with $q=\frac{1}{\sqrt{2^n-1}}\ssum{j=0\wedge j\neq k}{2^n\sminus 1}{\ket{j}}$.
Thus, the loop invariant in line 7 is given as 
$x\srange{0}{n} \mapsto \cos\varrho_j \ket{k} + \sin\varrho_j \cdot q * N = \lfloor \frac{\xi}{\chi} \rfloor$.
Here, we define $\varrho_j = \xi-j\cdot \chi$, so $\varrho_0=\xi=\cos^{-1}(\frac{1}{\sqrt{2^n}})$.
When we apply the $U()$ function inside the loop, we see that $x$'s value is transformed to $-\cos\varrho_j \ket{k} + \sin\varrho_j\cdot q$. The second step is to apply the quantum diffusion operation, which increases the amplitude values.
The result gives us the post-loop step state for $x$ as $\cos\varrho_{j\splus 1} \ket{k} + \sin\varrho_{j\splus 1} \cdot q$.
The reason is set up as an additional theorem in Dafny, and the transitions are as follows:

{\footnotesize
\[
\begin{array}{l}
		2\paren{-\cos\xi\cos\varrho_j+\sin\xi\sin\varrho_j}\paren{\cos\xi\ket{k}+\sin\xi \cdot q}
		 -\paren{-\cos\varrho_j\ket{k}+\sin\varrho_j\cdot q}\\
		= \paren{-2\cos^2\xi\cos\varrho_j+2\sin\xi\sin\varrho_j\cos\xi +\cos\varrho_j}\ket{k}
		 + \paren{-2\sin\xi\cos\xi\cos\varrho_j+2\sin^2\xi\sin\varrho_j -\sin\varrho_j}\cdot q\\
		= \paren{\paren{1-2\cos^2\xi}\cos\varrho_j+2\sin\xi\sin\varrho_j\cos\xi}\ket{k} 
		 + \paren{-2\sin\xi\cos\xi\cos\varrho_j+2\sin\varrho_j\paren{\sin^2\xi - 1}}\cdot q\\
		=\paren{-\cos\paren{2\xi}\cos\varrho_j+\sin\paren{2\xi}\sin\varrho_j}\ket{k} 
		 + \paren{-\sin\paren{2\xi}\cos\varrho_j-\sin\varrho_j\cos\paren{2\xi}}\cdot q\\
		= \paren{\cos\paren{\pi - 2\xi}\cos\varrho_j+\sin\paren{\pi - 2\xi}\sin\varrho_j}\ket{k} 
		 + \paren{-\sin\paren{\pi - 2\xi}\cos\varrho_j+\sin\varrho_j\cos\paren{\pi - 2\xi}}\cdot q\\
		= \paren{\cos\paren{\chi}\cos\varrho_j+\sin\paren{\chi}\sin\varrho_j}\ket{k}
		 + \paren{-\sin\paren{\chi}\cos\varrho_j+\sin\varrho_j\cos\paren{\chi}}\cdot q\\
		= \cos\paren{\varrho_j-\chi}\ket{k} + \sin\paren{\varrho_j-\chi}\cdot q\\
		= \cos\paren{\varrho_{j+1}}\ket{k} + \sin\paren{\varrho_{j+1}}\cdot q
\end{array}
\]
}

A quantum diffusion operation can also be viewed as a trigonometric rotation, and the above calculation shows that the rotation eventually turns the $\varrho_j$ degree to $\varrho_{j+1}$.
At the end of the program in \Cref{fig:grovesearch}, before the measurement, 
the $x$'s value is $\cos\varrho_N \ket{k} + \sin\varrho_N$ with $\varrho_N = \xi-\lfloor \frac{\xi}{\chi} \rfloor\cdot \chi$.
Notice that $0 \le \varrho_N < \chi$, so we will have the probability of $p$ reaching the measurement result $k$ to be greater than $\cos^2(\chi)$.
The estimation can be better, but any other analysis will follow the same pattern introduced here.

In the \qafny implementation, we set trigger rules to rewrite the term in line 4 in \Cref{fig:grovesearch} to the trigonometric format in line 5. Then, we also provide theorems to support the rewrite of $x$'s value from $-\cos\varrho_j \ket{k} + \sin\varrho_j\cdot q$ to 
$\cos\paren{\varrho_{j+1}}\ket{k} + \sin\paren{\varrho_{j+1}}\cdot q$ in line 10. The Dafny proof system automatically derives the other proof.

\section{An Upgraded Type System and Equivalence Relations}\label{sec:newtype}

Here, we describe an upgraded type system based on the one in \Cref{sec:type-checking}, which upgrades it to a dependent type system by including bases in every basis-kets. The type system is under construction in the \qafny implementation.
 We first show some additional syntax in \Cref{fig:qafny-typesyntax}.

Each construct type is only associated with additional flags depending on the quantum states. The flag $\topt{c}$ for $\tnort$ means we can track the basis-ket bitstring of the $\tnort$ typed value $z\ket{c}$ in the type checking stage. Sometimes, we might be unable to track the basis so we will classify it as $\infty$. Similar to the $\tnort$ typed situation, we use $\bigcirc$ to represent that a $\thadt$ typed value is in uniformed superposition. If not, we can classify the $\thadt$ typed basis-ket as $\infty$. In dealing with $\tcht$ typed values, we will use a set of bases $\overline{c}(m)$ to refer to the basis-ket bitstrings appearing in an $\tcht$ typed value $\sch{m}{z_j}{c_j}$.
In $\overline{c}(m)$, $\overline{c}$ is a basis set, and $m$ is the element number, and it can be abbreviated as $\overline{c}$. 

\begin{figure}
\captionsetup[subfigure]{justification=centering}
{\small
{
\begin{minipage}[t]{0.8\textwidth}
\begin{center}
{\small
\[
\begin{array}{llcl}
      \text{Indexed basis set} & \overline{c}(m) & ::= & \{c_0,c_1,...,c_{m-1}\}       \\
      \text{Option} & p \in\; \topt{'a} & ::= & 'a \mid \infty \\
      \text{Uniform Distribution} & \bigcirc \\
      \text{Type} & \tau & ::= & \tnor{(\topt{c})}\mid \thad{(\topt{\bigcirc})}
                    \mid \tch{(\topt{\overline{c}(m)})} 
\end{array}
\]
}
\end{center}
\subcaption{Upgraded Type Syntax}
  \label{fig:qafny-typesyntax}
\end{minipage}
}
{
\begin{minipage}[t]{0.4\textwidth}
\begin{center}
 \[
  \begin{array}{l@{~}cl}
  \tnor{\infty} &\sqsubseteq_n& \tch{\infty}\\
  \tnor{c} &\sqsubseteq_n& \tch{\{c\}}\\
  \tch{\overline{c}(1)} &\sqsubseteq_n& \tnor{\overline{c}[0]}\\
  \thad{p} &\sqsubseteq_n& \tch{\{\tob{j}|j\in[0,2^n)\}(2^n)}\\
  \tch{\{0,1\}} &\sqsubseteq_1& \thad{\infty}\\
  \tch{p} &\sqsubseteq_n& \tch{\infty}
    \end{array}
  \]
\end{center}
\subcaption{Upgraded Subtyping}
  \label{fig:qafny-subtypea}
\end{minipage}
\begin{minipage}[t]{0.45\textwidth}
\begin{center}
   \[
   \begin{array}{l@{~}cl}
  \ket{c} &\equiv_n& \sch{0}{ }{\ket{c}}\\
  \sch{0}{z_j}{\ket{c_j}} &\equiv_n& \ket{c_0}\\
  \shad{2^n}{n\sminus 1}{\alpha(r_j)} &\equiv_n& \sch{2^n\sminus 1}{\frac{\alpha(\sum_{k=0}^{n\sminus 1} r_k \cdot \tob{j}[k])}{\sqrt{2^n}}}{\ket{j}}\\
  \sch{1}{z_j}{c_j} &\equiv_1& \shad{2}{0}{\frac{\sqrt{2}z_1}{z_0}}\\
   &&\qquad\texttt{when }c_0=0\;\;c_1=1
    \end{array}
 \]
\end{center}
\subcaption{Upgraded Quantum Value Equivalence}
  \label{fig:qafny-sequiva}
\end{minipage}
  \caption{\qafny type/value state relations. $\overline{c}[n]$ produces the $n$-th element in set $\overline{c}$. $\{\tob{j}|j\in[0,2^n)\}(2^n)$ defines a set $\{\tob{j}|j\in[0,2^n)\}$ with the emphasis that it has $2^n$ elements. $\{0,1\}$ is a set of two single element bitstrings $0$ and $1$. $\cdot$ is the multiplication operation, $\tob{j}$ turns a number $j$ to a bitstring basis, $\tob{j}[k]$ takes the $k$-th element in the bitstring $\tob{j}$, and $\ket{j}$ is an abbreviation of $\ket{\tob{j}}$.}
  \label{fig:qafny-eqa}
}
}
\end{figure}

The most significant advantage of the upgraded type system is to permit additional quantum state equational rewrites based directly on type predicates without developing additional built-in predicates in the \qafny to Dafny compilation. We have new subtyping and quantum value equivalence relations in \Cref{fig:qafny-eqa}.
The subtyping relation $\sqsubseteq_n$ and the value equivalence $\equiv_n$ is not associated with a qubit number (locus length number) $n$, such that they establish relations between two quantum values describing a locus of length $n$.
$\sqsubseteq_n$ in \Cref{fig:qafny-subtypea} describes a type term on the left that can be used as a type on the right. For example, a \texttt{Nor} type qubit array $\tnor{c}$ can be used as a single element entanglement type term $\tch{\{c\}}$ \footnote{If a qubit array only consists of $0$ and $1$, it can be viewed as an entanglement of unique possibility. }. 
Correspondingly, value equivalence relation $\equiv_n$, in \Cref{fig:qafny-sequiva}, describes the two values to be equivalent; specifically, the left state term can be used as the right one, e.g., a single element entanglement value $\sch{0}{z_j}{\ket{c_j}}$ can be used as a \texttt{Nor} typed value $\ket{c_0}$ with the fact that $z_0$ is now a global phase that can be neglected.

In addition to the rewrites in \Cref{fig:qafny-eq}, we now permit the rewrites from an $\tcht$ typed value to a $\tnort$ typed one if we find that the $\tcht$ typed value contains a singleton basis-ket, while an $\tcht$ typed value can also be rewritten to a $\thadt$ typed one if we find that the length of the $\tcht$ typed value is $1$, and it contains exactly two basis-kets.

\begin{figure}
{\footnotesize
\begin{center}
 \[
  \begin{array}{r@{~}c@{~}l@{~}cl}
  &&\{\emptyset:\tau\} \cup \sigma &\preceq& \sigma\\
  \tau\sqsubseteq_{\slen{\kappa}}\tau' &\Rightarrow& \{\kappa:\tau\} \cup \sigma &\preceq& \{\kappa:\tau'\} \cup \sigma\\
  &&\{\kappa_1\sqcupplus s_1 \sqcupplus s_2 \sqcupplus \kappa_2 :\tau\} \cup \sigma &\preceq& \{\kappa_1\sqcupplus s_2 \sqcupplus s_1 \sqcupplus \kappa_2 : \texttt{mut}(\tau,\slen{\kappa_1})\} \cup \sigma\\
  &&\{\kappa_1 :\tau_1\} \cup \{\kappa_2 :\tau_2\} \cup \sigma &\preceq& \{\kappa_1 \sqcupplus \kappa_2 :\texttt{mer}(\tau_1,\tau_2)\} \cup \sigma \\
  \texttt{spt}(\tau,\slen{\kappa_1})=(\tau_1,\tau_2)&\Rightarrow&\{\kappa_1 \sqcupplus \kappa_2 :\tau\} \cup \sigma &\preceq& \{\kappa_1 :\tau_1\} \cup \{\kappa_2 :\tau_2\} \cup \sigma
    \end{array}
  \]
\end{center}
{\scriptsize
\[
\begin{array}{l}
\texttt{pmut}((c_1.i_1.i_2.c_2),n)=(c_1.i_2.i_1.c_2) \;\;\texttt{when}\;\slen{c_1}=n
\\
\texttt{mut}(\tnor{c},n)=\tnor{\texttt{pmut}(c,n)}
\qquad
\texttt{mut}(\tch{\overline{c}(m)},n)=\tch{\{\texttt{pmut}(c,n)|c\in\overline{c}(m)\}(m)}
\qquad
\texttt{mut}(\tau,n)=\tau\;\;[\texttt{owise}]
\\
\texttt{mer}(\tnor{c_1},\tnor{c_2})=\tnor{(c_1.c_2)}
\qquad
\texttt{mer}(\thad{\bigcirc},\thad{\bigcirc})=\thad{\bigcirc}
\qquad
\texttt{mer}(T\;\infty,T\;t)=T\;\infty
\\
\texttt{mer}(\tch{\overline{c_1}(m_1)},\tch{\overline{c_2}(m_2)})=\tch{(\overline{c_1}\times \overline{c_2})(m_1*m_2)}
\\
\texttt{spt}(\tnor{c_1.c_2},n)=(\tnor{c_1},\tnor{c_2}) \;\;\texttt{when}\;\slen{c_1}=n
\qquad
\texttt{spt}(\thad{t},n)=(\thad{t},\thad{t})
\\
\texttt{spt}(\tch{\{c_j.c|j\in [0,m)\wedge |c_j|=n\}(m)},n)=(\tch{\{c_j|j\in [0,m)\wedge |c_j|=n\}(m)},\tnor{c})
\end{array}
\]
}
\caption{Upgraded type environment partial order. We use set union ($\cup$) to describe the type environment concatenation with the empty set operation $\emptyset$. $i$ is a single bit either $0$ or $1$. The $.$ operation is a bitstring concatenation. $\times$ is the Cartesian product of two sets.
$T$ is either $\texttt{Nor}$, $\texttt{Had}$ or $\texttt{CH}$. }
  \label{fig:env-equiva}
}
\end{figure}

\begin{figure}
{\footnotesize
\begin{center}
 \[
  \begin{array}{r@{~}c@{~}l@{~}cl}
  &&\{\emptyset:q\} \cup \varphi &\equiv& \varphi\\
  q\equiv_{\slen{\kappa}}q' &\Rightarrow& \{\kappa:q\} \cup \varphi &\equiv& \{\kappa:q'\} \cup \varphi\\
  &&\{\kappa_1\sqcupplus s_1 \sqcupplus s_2 \sqcupplus \kappa_2 :q\} \cup \varphi &\equiv& \{\kappa_1\sqcupplus s_2 \sqcupplus s_1 \sqcupplus \kappa_2 : \texttt{mut}(q,\slen{\kappa_1})\} \cup \varphi\\
  &&\{\kappa_1 :q_1\} \cup \{\kappa_2 :q_2\} \cup \varphi &\equiv& \{\kappa_1 \sqcupplus \kappa_2 :\texttt{mer}(q_1,q_2)\} \cup \varphi \\
  \texttt{spt}(\tau,\slen{\kappa_1})=(\tau_1,\tau_2)&\Rightarrow&\{\kappa_1 \sqcupplus \kappa_2 :\tau\} \cup \sigma &\preceq& \{\kappa_1 :\tau_1\} \cup \{\kappa_2 :\tau_2\} \cup \sigma
    \end{array}
  \]
\end{center}
{\scriptsize
\[
\begin{array}{l}
\texttt{pmut}((c_1.d_1.d_2.c_2),n)=(c_1.d_2.d_1.c_2) \;\;\texttt{when}\;\slen{c_1}=n
\\
\texttt{mut}(\ket{c},n)=\ket{\texttt{pmut}(c,n)}
\qquad
\texttt{mut}(\sch{m}{z_j}{\ket{c_j}},n)=\sch{m}{z_j}{\ket{\texttt{pmut}(c_j,n)}}
\\
\texttt{mut}(\frac{1}{\sqrt{2^m}}(q_1 \bigotimes (\ket{0}+\alpha(r_n)\ket{1}) \bigotimes (\ket{0}+\alpha(r_{n+1})\ket{1}) \bigotimes q_2),n)
\\
\qquad\qquad
=\frac{1}{\sqrt{2^m}}(q_1 \bigotimes (\ket{0}+\alpha(r_{n+1})\ket{1}) \bigotimes (\ket{0}+\alpha(r_n)\ket{1}) \bigotimes q_2)
\quad\texttt{when\;}\slen{q_1}=n
\\
\texttt{mer}(\ket{c_1},\ket{c_2})=\ket{c_1}\ket{c_2}
\\
\texttt{mer}(\shad{2^n}{n\sminus 1}{\alpha(r_j)},\shad{2^m}{m\sminus 1}{\alpha(r_j)})=\shad{2^{n+m}}{n\splus m\sminus 2}{\alpha(r_j)}
\\
\texttt{mer}(\sch{n}{z_j}{\ket{c_j}},\schk{m}{z_k}{c_k})=\schi{n\cdot m}{z_j\cdot z_k}{c_j}\ket{c_k}
\\
\texttt{spt}(\ket{c_1.c_2},n)=(\ket{c_1},\ket{c_2}) \;\;\texttt{when}\;\slen{c_1}=n
\qquad
\texttt{spt}(q_1\bigotimes q_2,n)=(q_1,q_2) \;\;\texttt{when}\;\slen{q_1}=n
\\
\texttt{spt}(\sch{m}{z_j}{\ket{c_j}}\ket{c},n)=(\sch{m}{z_j}{\ket{c_j}},\ket{c})
\end{array}
\]
}
\caption{Upgraded quantum state equivalence. We use set union ($\cup$) to describe the state concatenation with the empty set operation $\emptyset$. $i$ is a single bit either $0$ or $1$. The $.$ operation is a bitstring concatenation. 
Term $\Msum^{n\cdot m}P$ is a summation formula that omits the indexing details.
Term $(\frac{1}{\sqrt{2^{n}}}\Motimes_{j=0}^{n\sminus 1}q_j) \bigotimes(\frac{1}{\sqrt{2^{m}}}\Motimes_{j=0}^{m\sminus 1}q_j)$ is equivalent to $\frac{1}{\sqrt{2^{n+m}}}\Motimes_{j=0}^{n\splus m \sminus 2}q_j$.
}
  \label{fig:state-equiva}
}
\end{figure}

\begin{figure}[t]
{
{\scriptsize
  \begin{mathpar}
    \inferrule[T-Mea]{ \Omega(y)=\qmode{j} \\ \Omega[x\mapsto \mmode],\sigma\uplus\{\kappa : \tch{\infty}\}\vdash_{\cmode} e \triangleright  \sigma'}{\Omega,\sigma\uplus \{y\srange{0}{j}\sqcupplus\kappa: \tau \} \vdash_{\cmode} \sexp{x}{\smea{y}}{e} \triangleright \sigma' }

    \inferrule[T-MeaN]{ \Omega(y)=\qmode{j} \\ \overline{c'}=\{c_2\;\bm{|}\;\tob{n}.c_2\in\overline{c}\wedge\slen{\tob{n}}=j\}
     \\\\ \Omega[x\mapsto \mmode],\sigma\uplus\{\kappa : \tch{\overline{c'}(\slen{\overline{c'}})}\}\vdash_{\cmode} e \triangleright  \sigma'}{\Omega,\sigma\uplus \{y\srange{0}{j}\sqcupplus\kappa: \tch{\overline{c}(m)}\} \vdash_{\cmode} \sexp{x}{\smea{y}}{e} \triangleright \sigma' }

    \inferrule[TA-CH]{\overline{c'}=\{(\llbracket \mu \rrbracket c_1).c_2\;\bm{|}\;c_1.c_2\in\overline{c}\wedge\slen{c_1}=\slen{\kappa}\}}
 {\Omega,\sigma\uplus\{\kappa\sqcupplus\kappa':\tch{\overline{c}(m)}\} \vdash_g \ssassign{\kappa}{}{\mu} \triangleright \sigma\uplus\{\kappa\sqcupplus\kappa':\tch{\overline{c'}(m)}\} }

\inferrule[T-IF]{ 
FV(b@x[j])=\kappa\sqcupplus x\srange{j}{j\,\splus\, 1}\\
FV(b@x[j])\cap FV(e) =\emptyset \\\\
\sigma(\kappa\sqcupplus x\srange{j}{j\,\splus\, 1}\sqcupplus \kappa_1)=\tch{\overline{c}(m)}
\\
     \Omega,\sigma \vdash_{\mmode} e \triangleright \{\kappa\sqcupplus x\srange{j}{j\,\splus\, 1}\sqcupplus \kappa_1:\tch{\overline{c'}(m)}\}} 
{\Omega,\sigma \vdash_g \sifq{b@x[j]}{e} \triangleright \{\kappa\sqcupplus x\srange{j}{j\,\splus\, 1}\sqcupplus \kappa_1:\tch{\overline{c''}(m)}\} }
  \end{mathpar}
}
{\footnotesize
\[
\begin{array}{l}
\overline{c''}=\{\tob{n}.1.c_2\;\bm{|}\;\tob{n}.d.c_1 \in \overline{c} \wedge \tob{n}.d.c_2 \in \overline{c'} \wedge b[\tob{n}/\kappa] \oplus d \wedge \slen{\tob{n}}=\slen{\kappa} \}
\\ \qquad\quad \cup \{\tob{n}.0.c_1\;\bm{|}\; \tob{n}.d.c_1 \in \overline{c} \wedge \neg (b[\tob{n}/\kappa] \oplus d) \wedge \slen{\tob{n}}=\slen{\kappa} \}
\end{array}
\]
}
}
  \caption{\qafny type system. $\llbracket \mu \rrbracket c$ is the \oqasm semantics of interpreting reversible expression $\mu$ in \Cref{fig:deno-sem}. Boolean expression $b$ can be $a_1=a_2$, $a_1 < a_2$ or \texttt{false}. $b[\tob{n}/\kappa]$ means that we treat $b$ as a \oqasm $\mu$ expression, replace qubits in array $\kappa$ with bits in bitstring basis $\tob{n}$, and evaluate it to a Boolean value.
$\sigma(\kappa)=\tau$ is an abbreviation of $\sigma(\kappa)=\{\kappa\mapsto \tau\}$. $FV(-)$ produces a locus by union all qubits appearing in $-$.}
  \label{fig:exp-sessiontypeb}
\end{figure}

\Cref{fig:env-equiva} and \Cref{fig:state-equiva} provide upgraded equivalence relations for type environments and quantum states.
The most significant improvement compared to the ones in \Cref{fig:env-equiv} and \Cref{fig:qafny-stateequiv} is the split of an $\tcht$ typed quantum state. Notice that splitting an $\tcht$ typed quantum state is equivalent to disentanglement, and we only provide one case for such splitting. The split function ($\texttt{spt}$) sees if every basis-ket in the $\tcht$ type has a common factor. If so, we can split the state by factoring in the common factor. 
For example, $\ket{00}+\ket{10}$ can be disentangled as $(\ket{0}+\ket{1})\otimes \ket{0}$, as the common factor $\ket{0}$ is factored out from the original formula $\ket{00}+\ket{10}$.

\Cref{fig:exp-sessiontypeb} provides some rules in the upgraded type system. Other than the functionality discussed in \Cref{appx:syntax-types}. The most significant difference is that we now need to track the basis-ket bitstrings appearing in $\tcht$, $\thadt$, and $\tcht$ typed values.
In rule \textsc{TA-CH}, the oracle $\mu$ is applied on $\kappa$ belonging to a locus $\kappa \sqcupplus \kappa'$. Correspondingly, the locus's type is $\tch{\overline{c}(m)}$, for each basis $c_1.c_2\in \overline{c}$, with $\slen{c_1}=\slen{\kappa}$, we apply $\mu$ on the $c_1$ and leave $c_2$ unchanged.
Here, we utilize the \oqasm semantics that describes transitions from a $\texttt{Nor}$-typed quantum value to another $\texttt{Nor}$ one, and we generalize it to apply the semantic function on every element in the \texttt{CH} type.
During the transition, the number of elements $m$ does not change.
Similarly, applying a partial measurement on range $y\srange{0}{j}$ of the locus $y\srange{0}{j}\sqcupplus\kappa$ in rule \textsc{T-MeaN}
can be viewed as an array filter, i.e., for an element, $c_1.c_2$ in set $\overline{c}$ of the type $\tch{\overline{c}(m)}$, with $\slen{c_1}=j$, we keep only the ones with $c_1=\tob{n}$ ($n$ is the measurement result) in the new set $\overline{c'}$ and recompute $\slen{\overline{c'}}$. In \qafny, the tracking procedure is to generate symbolic predicates that permit the production of the set $\overline{c'}(\slen{c'})$, not to produce such sets. If the predicates are not effectively trackable, we can always use $\infty$ to represent the set given in rule \textsc{T-Mea}.

In addition, the upgraded type system enforces equational properties of 
quantum qubit loci through a partial order relation over type environments, including subtyping, qubit position mutation, merge, and split quantum loci.
This property is enforced with the additional equivalence relations in \Cref{fig:env-equiva} and \Cref{fig:state-equiva}.
Based on the new \texttt{CH} type with the set $\overline{c_1}\times \overline{c_2}$,
the quantum conditional (\rulelab{T-If}) creates a new set based on $\overline{c_1}\times \overline{c_2}$, i.e., for each element $\tob{n}.d.c$ in the set, with $\slen{\tob{n}}=\slen{\kappa_1}$, we compute Boolean guard $b$ value by substituting qubit variables in $b$ with the bitstring basis $\tob{n}$. The result $b[\tob{n} / \kappa_1] \oplus d$ is true or not ($d$ represents the bit value for the qubit at $x\srange{j}{j\,\splus\, 1}$); if it is true, we replace the $c$ basis by an updated basis by applying the semantics of the conditional body if we can learn about its semantics; otherwise, we keep $c$ to be the same.
In short, the quantum conditional behavior can be understood as applying a conditional map function on an $m$ array of bitstrings, and we only apply the conditional body's effect on the second part of some bitstrings whose first part is checked to be true by computing the Boolean guard $b$.

\section{OQASM: An Assembly Language for Quantum Oracles}
\label{sec:vqir}

The \oqasm expression $\mu$ used in \Cref{fig:vqimp} places an additional wrapper on top of the \oqasm expression $\iota$ given in \Cref{fig:vqir}. Here, we first provide a step-by-step explanation of \oqasm; then show the additional wrapper in \Cref{sec:exp-mu}.

\oqasm is designed to express efficient quantum
oracles that can be easily tested and, if desired, proved
correct.
\oqasm operations leverage both the standard
computational basis and an alternative basis connected by the quantum
Fourier transform (QFT). 
\oqasm's type system tracks the bases of variables in
\oqasm programs, forbidding operations that would introduce
entanglement. \oqasm states are therefore efficiently
represented, so programs can be effectively tested and are simpler to
verify and analyze. In addition, \oqasm uses \emph{virtual qubits}
to support \emph{position shifting operations}, which support
arithmetic operations without introducing extra gates during
translation. All of these features are novel to quantum assembly
languages. 

This section presents \oqasm states and the language's syntax,
semantics, typing, and soundness results. As a running example, the QFT
adder~\cite{qft-adder} is shown in \Cref{fig:circuit-example}. The Coq
function \coqe{rz_adder} generates an \oqasm program that adds two
natural numbers \coqe{a} and \coqe{b}, each of length \coqe{n} qubits.

\begin{figure*}[t]
  \centering
  \begin{tabular}{c @{\qquad} c}

  \begin{minipage}[b]{.6\textwidth}
{\scriptsize
    \Qcircuit @C=0.5em @R=0.75em {
      \lstick{\ket{a_{n-1}}} & \qw & \ctrl{5} & \qw & \qw & \qw & \qw & \qw & \qw & \qw & \rstick{\ket{a_{n-1}}} \\
      \lstick{\ket{a_{n-2}}} & \qw & \qw & \ctrl{4} & \qw & \qw & \qw & \qw & \qw & \qw & \rstick{\ket{a_{n-2}}}\\
      \lstick{\vdots} & & & & & & & & & & \rstick{\vdots} \\
      \lstick{} & & & & & & & & & & \\
      \lstick{\ket{a_0}} & \qw & \qw & \qw & \qw & \qw & \qw & \ctrl{1} & \qw & \qw & \rstick{\ket{a_0}} \\
      \lstick{\ket{b_{n-1}}} & \multigate{5}{\texttt{QFT}} & \gate{\texttt{SR 0}} & \multigate{3}{\texttt{SR 1}} & \qw & \qw & \qw & \multigate{5}{\texttt{SR (n-1)}} & \multigate{5}{\texttt{QFT}^{-1}} & \qw & \rstick{\ket{a_{n-1} + b_{n-1}}} \\
      \lstick{} & & & & & \dots & & & & \\
      \lstick{\ket{b_{n-2}}} & \ghost{\texttt{QFT}} & \qw  &  \ghost{\texttt{SR 1}} & \qw & \qw & \qw & \ghost{\texttt{SR (n-1)}} & \ghost{\texttt{QFT}^{-1}} & \qw & \rstick{\ket{a_{n-2} + b_{n-2}}} \\
      \lstick{\vdots} & & & & & & & & & & \rstick{\vdots} \\
      \lstick{} & & & & & & & & & & \\
      \lstick{\ket{b_0}} & \ghost{\texttt{QFT}} & \qw & \qw & \qw & \qw & \qw & \ghost{\texttt{SR (n-1)}} & \ghost{\texttt{QFT}^{-1}}  & \qw & \rstick{\ket{a_0 + b_0}} 
      }
      }
  \subcaption{Quantum circuit}
  \end{minipage}
  \hfill\hfill
  \begin{minipage}[b]{.38\textwidth}
  \begin{coq}
  Fixpoint rz_adder' (a b:var) (n:nat) 
    := match n with 
       | 0 => ID (a,0)
       | S m => CU (a,m) (SR m b); 
                rz_adder' a b m
       end.
  Definition rz_adder (a b:var) (n:nat) 
    := Rev a ; Rev b ; $\texttt{QFT}$ b ;
       rz_adder' a b n;
       $\texttt{QFT}^{-1}$ b; Rev b ; Rev a.
  \end{coq}
  \subcaption{\oqasm metaprogram (in Coq)}
  \end{minipage}
  \end{tabular}
\caption{Example \oqasm program: QFT-based adder}
  \label{fig:circuit-example}
  \end{figure*}

\subsection{OQASM States} \label{sec:pqasm-states}

\begin{figure}[t]
  \small
  \[\hspace*{-0.5em}
\begin{array}{l>{$} p{1.2cm} <{$} c l}
      \text{Bit} & b & ::= & 0 \mid 1 \\
      \text{Natural number} & n & \in & \mathbb{N} \\
      \text{Real} & r & \in & \mathbb{R}\\
      \text{Phase} & \alpha(r) & ::= & e^{2\pi i r} \\
      \text{Basis} & \tau & ::= & \texttt{Nor} \mid \texttt{Phi}\;n \\
      \text{Unphased qubit} & \overline{q} & ::= & \ket{b} ~~\mid~~ \qket{r} \\
      \text{Qubit} & q & ::= &\alpha(r) \overline{q}\\
      \text{State (length $d$)} & \varphi & ::= & q_1 \otimes q_2 \otimes \cdots \otimes q_d
    \end{array}
  \]
  \caption{\oqasm state syntax}
  \label{fig:vqir-state}
\end{figure}

An \oqasm program state is represented according to the grammar in
\Cref{fig:vqir-state}. A state $\varphi$ of $d$ qubits is 
a length-$d$ tuple of qubit values $q$; the state models the tensor
product of those values. This means that the size of $\varphi$ is
$O(d)$ where $d$ is the number of qubits. A $d$-qubit state in a
language like \sqir is represented as a length $2^d$ vector of complex
numbers, $O(2^d)$ in the number of qubits. Our linear state
representation is possible because applying for any well-typed \oqasm
program on any well-formed \oqasm state never causes qubits to be
entangled.

A qubit value $q$ has one of two forms $\overline{q}$, scaled by a
global phase $\alpha(r)$. The two forms depend on the \emph{basis} $\tau$ that the qubit is in---it could be either \texttt{Nor} or \texttt{Phi}. A \texttt{Nor} qubit has form
$\ket{b}$ (where $b \in \{ 0, 1 \}$), which is a
computational basis value. 
A \texttt{Phi} qubit has the form $\qket{r} = \frac{1}{\sqrt{2}}(\ket{0}+\alpha(r)\ket{1})$, which is a value of the (A)QFT basis.
The number $n$ in \texttt{Phi}$\;n$ indicates the precision of the state $\varphi$.
As shown by~Beauregard \cite{qft-adder}, arithmetic on the computational basis can sometimes be more efficiently carried out on the QFT basis, which leads to the use of quantum operations (like QFT) when implementing circuits with classical input/output behavior.
 
\subsection{OQASM Syntax, Typing, and Semantics}\label{sec:oqasm-syn}

\begin{figure}[t]
\begin{minipage}[t]{0.5\textwidth}
{\small \centering

  $ \hspace*{-0.8em}
\begin{array}{llcl}
      \text{Position} & p & ::= & (x,n) \qquad   \text{Nat. Num}~n
                                  \qquad   \text{Variable}~x\\
      \text{Instruction} & \instr & ::= & \iskip{p} \mid \inot{p}
                                          \mid \irz[\lbrack -1 \rbrack]{n}{p} \mid \iseq{\instr}{\instr}\\
                & & \mid &  \isr[\lbrack -1 \rbrack]{n}{x} \mid \iqft[\lbrack -1 \rbrack]{n}{x} \mid \ictrl{p}{\instr}  \\
                      & & \mid & \ilshift{x} \mid \irshift{x} \mid \irev{x} 
    \end{array}
  $
}
  \caption{\oqasm syntax. For an operator \texttt{OP}, $\texttt{OP}^{\lbrack -1 \rbrack}$ indicates that the operator has a built-in inverse available.}
  \label{fig:vqir}
\end{minipage}
\hfill
\begin{minipage}[t]{0.38\textwidth}
{\scriptsize
\centering
\begin{tabular}{c@{$\quad=\quad$}c}
  \begin{minipage}{0.3\textwidth}

\Qcircuit @C=0.5em @R=0.5em {
    \lstick{} & \qw     & \multigate{4}{\texttt{SR m}} & \qw & \qw \\
    \lstick{} & \qw     & \ghost{\texttt{SR m}}           & \qw & \qw \\
    \lstick{} & \vdots & & \vdots & \\
    \lstick{} & & & & \\
    \lstick{} & \qw     & \ghost{\texttt{SR m}}           & \qw  & \qw
    }
  \end{minipage} & 
  \begin{minipage}{0.3\textwidth}
  \small
  \Qcircuit @C=0.5em @R=0.5em {
    \lstick{} & \qw     & \gate{\texttt{RZ (m+1)}} & \qw & \qw \\
    \lstick{} & \qw     & \gate{\texttt{RZ m}}          & \qw & \qw \\
    \lstick{} & & \vdots & & \\
    \lstick{} & & & & \\
    \lstick{} & & & & \\
    \lstick{} & \qw     & \gate{\texttt{RZ 1}}           & \qw  & \qw
    }
 
  \end{minipage} 
\end{tabular}
}
\caption{\texttt{SR} unfolds to a series of \texttt{RZ} instructions}
\label{fig:sr-meaning}
\end{minipage}
\end{figure}

\Cref{fig:vqir} presents \oqasm's syntax. An \oqasm program consists of
a sequence of instructions $\instr$. Each instruction applies an
operator to either a variable $x$, representing a group of qubits, or a \emph{position} $p$, which identifies a particular offset into a variable $x$. 

The instructions in the first row correspond to simple single-qubit
quantum gates---$\iskip{p}$, $\inot{p}$, and $\irz[\lbrack -1 \rbrack]{n}{p}$
 ---and instruction sequencing.
The instructions in the next row apply to whole variables: $\iqft{n}{x}$
applies the AQFT to variable $x$ with $n$-bit precision and
$\iqft[-1]{n}{x}$ applies its inverse.
If $n$ equals the size of $x$, then the AQFT operation is exact.
$\isr[\lbrack -1 \rbrack]{n}{x}$
applies a series of \texttt{RZ} gates (\Cref{fig:sr-meaning}). 
Operation $\ictrl{p}{\instr}$
applies instruction $\instr$ \emph{controlled} on qubit position
$p$. All of the operations in this row---\texttt{SR}, \texttt{QFT}, and \texttt{CU}---will be translated to multiple \sqir
gates. The function \coqe{rz_adder} in \Cref{fig:circuit-example}(b) uses
many of these instructions; e.g., it uses \texttt{QFT} and \texttt{QFT}$^{-1}$ and applies
\texttt{CU} to the $m$th position of variable \texttt{a} to control
instruction \texttt{SR m b}.

In the last row of \Cref{fig:vqir}, instructions $\ilshift{x}$,
$\irshift{x}$, and $\irev{x}$ are \emph{position shifting operations}.
Assuming that $x$ has $d$ qubits and $x_k$ represents the $k$-th qubit
state in $x$, $\texttt{Lshift}\;x$ changes the $k$-th qubit state to
$x_{(k + 1)\mmod d}$, $\texttt{Rshift}\;x$ changes it to
$x_{(k + d - 1)\mmod d}$, and \texttt{Rev} changes it to $x_{d-1-k}$. In
our implementation, shifting is \emph{virtual}, not physical. The \oqasm
translator maintains a logical map of variables/positions to concrete
qubits and ensures that shifting operations are no-ops, introducing no extra gates.

Other quantum operations could be added to \oqasm to
allow reasoning about a larger class of quantum programs while still
guaranteeing a lack of entanglement. 

\begin{figure}[t]
\begin{minipage}[b]{0.57\textwidth}
{\scriptsize
  \begin{mathpar}
    \inferrule[X]{\Omegaty(x)=\texttt{Nor} \\ n < \Omegasz(x)}{\Sigma;\Omega \vdash \inot{(x,n)}\triangleright \Omega}
  
    \inferrule[RZ]{\Omegaty(x)=\texttt{Nor} \\ n < \Omegasz(x)}{\Sigma;\Omega \vdash \irz{q}{(x,n)} \triangleright \Omega}

    \inferrule[SR]{\Omegaty(x)=\tphi{n} \\ m < n}{\Sigma;\Omega \vdash \texttt{SR}\;m\;x\triangleright \Omega}   

    \inferrule[QFT]{\Omegaty(x)=\texttt{Nor}\\n \le \Omegasz(x)}{\Sigma; \Omega \vdash \iqft{n}{x}\triangleright \Omega[x\mapsto \tphi{n}]}    
     
    \inferrule[RQFT]{\Omegaty(x)=\tphi{n}\\n \le \Omegasz(x)}{\Sigma; \Omega \vdash \iqft[-1]{n}{x}\triangleright \Omega[x\mapsto \texttt{Nor}]}             
    
    \inferrule[CU]{\Omegaty(x)=\texttt{Nor} \\ \texttt{fresh}~(x,n)~\instr \\\\ \Sigma; \Omega\vdash \instr\triangleright \Omega \\ \texttt{neutral}(\instr)}{\Sigma; \Omega \vdash \texttt{CU}\;(x,n)\;\instr \triangleright \Omega} 
     
    \inferrule[LSH]{\Omegaty(x)=\texttt{Nor}}{\Sigma; \Omega \vdash \texttt{Lshift}\;x\triangleright \Omega}

     \inferrule[SEQ]{\Sigma; \Omega\vdash \instr_1\triangleright \Omega' \\ \Sigma; \Omega'\vdash \instr_2\triangleright \Omega''}{\Sigma; \Omega \vdash \instr_1\;;\;\instr_2\triangleright \Omega''} 
    
  \end{mathpar}
}
  \caption{Select \oqasm typing rules}
  \label{fig:exp-well-typed}
\end{minipage}
\hfill
\hfill
\begin{minipage}[b]{0.4\textwidth}
{\footnotesize
\begin{center}\hspace*{-1em}
\begin{tikzpicture}[->,>=stealth',shorten >=1pt,auto,node distance=3.2cm,
                    semithick]
  \tikzstyle{every state}=[fill=black,draw=none,text=white]

  \node[state] (A)              {$\texttt{Nor}$};
  \node[state]         (C) [left of=A] {$\tphi{n}$};

  \path (A) edge [loop above]            node {$\Big\{\begin{array}{l}\texttt{ID},~\texttt{X},~\texttt{RZ}^{\lbrack -1 \rbrack},~\texttt{CU},\\
              \texttt{Rev},\texttt{Lshift},\texttt{Rshift}\end{array}\Big\}$} (A)
            edge   node [above] {\{$\texttt{QFT}\;n$\}} (C);
  \path (C) edge [loop above]            node {$\{\texttt{ID},~\texttt{SR}^{\lbrack -1 \rbrack}\}$} (C)
            edge  [bend right]             node {$\{\texttt{QFT}^{-1}\;n\}$} (A);
\end{tikzpicture}
\end{center}
}
\caption{Type rules' state machine}
\label{fig:state-machine}
\end{minipage}
\end{figure}

\myparagraph{Typing}
\label{sec:vqir-typing}

In \oqasm, typing is concerning a \emph{type environment}
$\Omega$ and a predefined \emph{size
  environment} $\Sigma$, which map \oqasm
variables to their basis and size (number of qubits), respectively.
The typing judgment is written $\Sigma; \Omega\vdash \instr \triangleright \Omega'$ which
states that $\instr$ is well-typed under $\Omega$ and $\Sigma$, and
transforms the variables' bases to be as in $\Omega'$ ($\Sigma$ is unchanged). 
$\Sigma$ is fixed because the number of qubits in execution is always fixed.
It is generated in the high-level language compiler, such as \sourcelang in \cite{oracleoopsla}.
The algorithm generates $\Sigma$ by taking an \sourcelang program and scanning through
all the variable initialization statements.
Select type rules are given in \Cref{fig:exp-well-typed}; 
the rules not shown (for \texttt{ID}, \texttt{Rshift}, \texttt{Rev}, \texttt{RZ}$^{-1}$, and \texttt{SR}$^{-1}$) are similar.

The type system enforces three invariants. First, it enforces that
instructions are well-formed, meaning that gates are applied to valid
qubit positions (the second premise in \rulelab{X}) and that any control qubit is distinct from the
target(s) (the \texttt{fresh} premise in
\rulelab{CU}).  This latter property enforces the quantum
\emph{no-cloning rule}.
For example, applying the \texttt{CU} in \code{rz\_adder'} (\Cref{fig:circuit-example}) is valid
because position \code{a,m} is distinct from variable \code{b}.

Second, the type system enforces that instructions leave affected
qubits on a proper basis (thereby avoiding entanglement). The
rules implement the state machine shown in
\Cref{fig:state-machine}. For example, $\texttt{QFT}\;n$ transforms a variable from \texttt{Nor} to
$\tphi{n}$ (rule \rulelab{QFT}), while $\texttt{QFT}^{-1}\;n$
transforms it from $\tphi{n}$ back to \texttt{Nor} (rule
\rulelab{RQFT}). Position shifting operations 
are disallowed on variables $x$ in
the \texttt{Phi} basis because the qubits that makeup $x$ are
internally related (see \Cref{appx:well-formed}) and cannot be rearranged. Indeed, applying a
\texttt{Lshift} and then a $\texttt{QFT}^{-1}$ on $x$ in \texttt{Phi}
would entangle $x$'s qubits.

Third, the type system enforces that the effect of position-shifting
operations can be statically tracked. The \texttt{neutral} condition of
\rulelab{CU} requires that any shifting within $\instr$ is restored by the time it
completes. 
For example, $\sseq{\ictrl{p}{(\ilshift{x})}}{\inot{(x,0)}}$ is not well-typed because knowing the final physical position of qubit $(x,0)$ would require statically knowing the value of $p$. 
On the other hand, the program $\sseq{\ictrl{c}{(\sseq{\ilshift{x}}{\sseq{\inot{(x,0)}}{\irshift{x}}})}}{\inot{(x,0)}}$ is well-typed 
because the effect of the \texttt{Lshift} is ``undone'' by an \texttt{Rshift} inside the body of the \texttt{CU}.

\myparagraph{Semantics}\label{sec:pqasm-dsem}

\begin{figure}[t]
{\scriptsize
\[
\begin{array}{lll}
\llbracket \iskip{p} \rrbracket\varphi &= \varphi\\[0.2em]

\llbracket \inot{(x, i)} \rrbracket\varphi &= \app{\uparrow\xsem(\downarrow\varphi(x,i))}{\varphi}{(x,i)}
& \texttt{where  }\xsem(\ket{0})=\ket{1} \qquad\, \xsem(\ket{1})=\ket{0}
\\[0.5em]

\llbracket \ictrl{(x,i)}{\instr} \rrbracket\varphi &=  \csem(\downarrow\varphi(x,i),\instr,\varphi)
&
\texttt{where  }
\csem({\ket{0}},{\instr},\varphi)=\varphi\quad\;\,
\csem({\ket{1}},{\instr},\varphi)=\llbracket \instr \rrbracket\varphi
\\[0.4em]

\llbracket \irz{m}{(x,i)} \rrbracket\varphi &= \app{\uparrow {\rsem}({m},\downarrow\varphi(x,i))}{\varphi}{(x,i)}
&\texttt{where  }{\rsem}(m,\ket{0})=\ket{0} \; \quad{\rsem}(m,\ket{1})=\alpha(\frac{1}{2^m})\ket{1}
\\[0.5em]

\llbracket \irz[-1]{m}{(x,i)} \rrbracket\varphi &= \app{\uparrow {\rrsem}({m},\downarrow\varphi(x,i))}{\varphi}{(x,i)}
 &\texttt{where  }{\rrsem}(m,\ket{0})=\ket{0}
\quad{\rrsem}(m,\ket{1})=\alpha(-\frac{1}{2^m})\ket{1}
\\[0.5em]

\llbracket \isr{m}{x} \rrbracket\varphi &
                                            \multicolumn{2}{l}{= \app{\uparrow \qket{r_i+\frac{1}{2^{m-i+1}}}}{\varphi}{\forall i \le m.\;(x,i)}
\qquad \texttt{when  }
\downarrow\varphi(x,i) = \qket{r_i}}\\[0.5em]

\llbracket \isr[-1]{m}{x} \rrbracket\varphi&\multicolumn{2}{l}{= \app{\uparrow \qket{r_i-\frac{1}{2^{m-i+1}}}}{\varphi}{\forall i \le m.\;(x,i)}
\qquad \texttt{when  }
\downarrow\varphi(x,i) = \qket{r_i}}\\[0.5em]

\llbracket \iqft{n}{x} \rrbracket\varphi &= \app{\uparrow\qsem(\Sigma(x),\downarrow\varphi(x),n)}{\varphi}{x}
& \texttt{where  }\qsem(i,\ket{y},n)=\bigotimes_{k=0}^{i-1}(\qket{\frac{y}{2^{n-k}}})
\\[0.5em]

\llbracket \iqft[-1]{n}{x} \rrbracket\varphi &=  \app{\uparrow\qsem^{-1}(\Sigma(x),\downarrow\varphi(x),n)}{\varphi}{x}
\\[0.5em]

\llbracket \ilshift{x} \rrbracket\varphi &= \app{{\psem}_{l}(\varphi(x))}{\varphi}{x}
&
\texttt{where  }{\psem}_{l}(q_0\otimes q_1\otimes \cdots \otimes q_{n-1})=q_{n-1}\otimes q_0\otimes q_1 \otimes \cdots
\\[0.5em]

\llbracket \irshift{x} \rrbracket\varphi &= \app{{\psem}_{r}(\varphi(x))}{\varphi}{x}
&
\texttt{where  }{\psem}_{r}(q_0\otimes q_1\otimes \cdots \otimes q_{n-1})=q_1\otimes \cdots \otimes q_{n-1} \otimes q_0
\\[0.5em]

\llbracket \irev{x} \rrbracket\varphi &= \app{{\psem}_{a}(\varphi(x))}{\varphi}{x}
&
\texttt{where  }{\psem}_{a}(q_0\otimes \cdots \otimes q_{n-1})=q_{n-1}\otimes \cdots \otimes q_0
\\[0.5em]

\llbracket \iota_1; \iota_2 \rrbracket\varphi &= \llbracket \iota_2 \rrbracket (\llbracket \iota_1 \rrbracket\varphi)
\end{array}
\]
}
{\scriptsize
$
\begin{array}{l}
\\[0.2em]
\downarrow \alpha(b)\overline{q}=\overline{q}
\qquad
\downarrow (q_1\otimes \cdots \otimes q_n) = \downarrow q_1\otimes \cdots \otimes \downarrow q_n
\\[0.2em]
\app{\uparrow \overline{q}}{\varphi}{(x,i)}=\app{\alpha(b)\overline{q}}{\varphi}{(x,i)}
\qquad \texttt{where  }\varphi(x,i)=\alpha(b)\overline{q_i}
\\[0.2em]
\app{\uparrow \alpha(b_1)\overline{q}}{\varphi}{(x,i)}=\app{\alpha(b_1+b_2)\overline{q}}{\varphi}{(x,i)}
\qquad \texttt{where  }\varphi(x,i)=\alpha(b_2)\overline{q_i}
\\[0.2em]
\app{q_x}{\varphi}{x}=\app{q_{(x,i)}}{\varphi}{\forall i < \Sigma(x).\;(x,i)}
\\[0.2em]
\app{\uparrow q_x}{\varphi}{x}=\app{\uparrow q_{(x,i)}}{\varphi}{\forall i < \Sigma(x).\;(x,i)}
\end{array}
$
}
\caption{\oqasm semantics}
  \label{fig:deno-sem}
\end{figure}

The semantics of an \oqasm program is a partial function
$\llbracket\rrbracket$ from
an instruction $\instr$ and input state $\varphi$ to an output state
$\varphi'$, written 
$\llbracket \instr \rrbracket\varphi=\varphi'$, shown in \Cref{fig:deno-sem}.

Recall that a state $\varphi$ is a tuple of $d$ qubit values,
modeling the tensor product $q_1\otimes \cdots \otimes q_d$. 
The rules implicitly map each variable $x$ to a
range of qubits in the state, e.g., 
$\varphi(x)$ corresponds to some sub-state $q_k\otimes \cdots \otimes q_{k+n-1}$
where $\Omegasz(x)=n$.
Many of the rules in \Cref{fig:deno-sem} update a \emph{portion} of a
state. $\app{q_{(x,i)}}{\varphi}{(x,i)}$ updates the $i$-th
qubit of variable $x$ to be the (single-qubit) state $q_{(x,i)}$, and
$\app{q_{x}}{\varphi}{x}$ to update variable $x$ according to
the qubit \emph{tuple} $q_x$.
$\app{\uparrow q_{(x,i)}}{\varphi}{(x,i)}$ and $\app{\uparrow q_{x}}{\varphi}{x}$ are similar, except that they also accumulate the previous global phase of $\varphi(x,i)$ (or $\varphi(x)$).
$\downarrow$ is to convert a qubit $\alpha(b)\overline{q}$ to an unphased qubit $\overline{q}$.

Function $\xsem$ updates the state of a single
qubit according to the rules for the standard quantum gate $X$.  
\texttt{cu} is a conditional operation
depending on the \texttt{Nor}-basis qubit $(x,i)$. 
\texttt{RZ} (or $\texttt{RZ}^{-1}$) is an z-axis phase rotation operation.
Since it applies to \texttt{Nor}-basis, it applies a global phase.
By \Cref{thm:sem-same}, when it is compiled to \sqir,
the global phase might be turned into a local one.
For example, to prepare the state $\sum_{j=0}^{2^n\sminus 1}(-i)^x\ket{x}$ \cite{ChildsNAND}, 
a series of Hadamard gates are applied, followed by several controlled-\texttt{RZ} gates on $x$,
where the controlled-\texttt{RZ} gates are definable by \oqasm.
\texttt{SR} (or
$\texttt{SR}^{-1}$) applies an $m+1$ series of \texttt{RZ} (or
$\texttt{RZ}^{-1}$) rotations where the $i$-th rotation
applies a phase of $\alpha({\frac{1}{2^{m-i+1}}})$
(or $\alpha({-\frac{1}{2^{m-i+1}}})$).
$\qsem$ applies an approximate quantum Fourier transform; $\ket{y}$ is an abbreviation of
$\ket{b_1}\otimes \cdots \otimes \ket{b_i}$ (assuming $\Omegasz(y)=i$) and $n$ is the degree of approximation.
If $n = i$, then the operation is the standard QFT\@.
Otherwise, each qubit in the state is mapped to $\qket{\frac{y}{2^{n-k}}}$, which is equal to $\frac{1}{\sqrt{2}}(\ket{0} + \alpha(\frac{y}{2^{n-k}})\ket{1})$ when $k < n$ and $\frac{1}{\sqrt{2}}(\ket{0} + \ket{1}) = \ket{+}$ when $n \leq k$ (since $\alpha(n) = 1$ for any natural number $n$).
$\qsem^{-1}$ is the inverse function of $\qsem$. 
Note that the input state to $\qsem^{-1}$ is guaranteed to have the form $\bigotimes_{k=0}^{i-1}(\qket{\frac{y}{2^{n-k}}})$ because it has type $\tphi{n}$.
$\psem_l$, $\psem_r$, and
$\psem_a$ are the semantics for \itext{Lshift}, 
\itext{Rshift}, and \itext{Rev}, respectively.

\subsection{OQASM Metatheory}\label{sec:metatheory}

\myparagraph{Soundness}
The following statement is proved: well-typed \oqasm programs are well-defined; i.e., the type system is sound concerning the semantics. 
Below is the well-formedness of an \oqasm state.

\begin{definition}[Well-formed \oqasm state]\label{appx:well-formed}\rm 
  A state $\varphi$ is \emph{well-formed}, written
  $\Sigma;\Omega \vdash \varphi$, iff:
\begin{itemize}
\item For every $x \in \Omega$ such that $\Omegaty(x) = \texttt{Nor}$,
  for every $k <\Omegasz(x)$, $\varphi(x,k)$ has the form
  $\alpha(r)\ket{b}$.

\item For every $x \in \Omega$ such that $\Omegaty(x) = \tphi{n}$ and $n \le \Omegasz(x)$,
  there exists a value $\upsilon$ such that for
  every $k < \Omegasz(x)$, $\varphi(x,k)$ has the form
  $\alpha(r)\qket{\frac{\upsilon}{ 2^{n- k}}}$.\footnote{Note that $\Phi(x) = \Phi(x + n)$, where the integer $n$ refers to phase $2 \pi n$; so multiple choices of $\upsilon$ are possible.}
\end{itemize}
\end{definition}

\noindent
Type soundness is stated as follows; the proof is by induction on $\instr$ and is mechanized in Coq.

\begin{theorem}\label{thm:type-sound-oqasm}\rm[\oqasm type soundness]
If $\Sigma; \Omega \vdash \instr \triangleright \Omega'$ and $\Sigma;\Omega \vdash \varphi$ then there exists $\varphi'$ such that $\llbracket \instr \rrbracket\varphi=\varphi'$ and $\Sigma;\Omega' \vdash \varphi'$.
\end{theorem}

\myparagraph{Algebra}
Mathematically, the set of well-formed $d$-qubit \oqasm states for a given $\Omega$ can be interpreted as a subset $\hsp{S}^d$ of a $2^d$-dimensional Hilbert space $\hsp{H}^d$ \footnote{A \emph{Hilbert space} is a vector space with an inner product that is complete with respect to the norm defined by the inner product. $\hsp{S}^d$ is a sub\emph{set}, not a sub\emph{space} of $\hsp{H}^d$ because $\hsp{S}^d$ is not closed under addition: Adding two well-formed states can produce a state that is not well-formed.}. The semantics function $\llbracket \rrbracket$ can be interpreted as a $2^d \times 2^d$ unitary matrix, as is standard when representing the semantics of programs without measurement~\cite{PQPC}.
Because \oqasm's semantics can be viewed as a unitary matrix, correctness properties extend by linearity from $\hsp{S}^d$ to $\hsp{H}^d$---an oracle that performs addition for classical \texttt{Nor} inputs will also perform addition over a superposition of \texttt{Nor} inputs. The following statement is proved: $\hsp{S}^d$ is closed under well-typed \oqasm programs.

Given a qubit size map $\Sigma$ and type environment $\Omega$, the set of \oqasm programs that are well-typed concerning $\Sigma$ and $\Omega$ (i.e., $\Sigma;\Omega \vdash \instr \triangleright \Omega'$) form an algebraic structure $(\{\instr\},\Sigma, \Omega,\hsp{S}^d)$, where $\{\instr\}$ defines the set of valid program syntax, such that there exists $\Omega'$, $\Sigma;\Omega \vdash \instr \triangleright \Omega'$ for all $\instr$ in $\{\instr\}$; $\hsp{S}^d$ is the set of $d$-qubit states on which programs $\instr\in \{\instr\}$ are run, and are well-formed ($\Sigma;\Omega \vdash \varphi$) according to \Cref{appx:well-formed}.
From the \oqasm semantics and the type soundness theorem, for all $\instr \in \{\instr\}$ and $\varphi \in \hsp{S}^d$, such that $\Sigma;\Omega \vdash \instr \triangleright \Omega'$ and $\Sigma;\Omega \vdash \varphi$, and $\llbracket \instr \rrbracket\varphi=\varphi'$, $\Sigma;\Omega' \vdash \varphi'$, and $\varphi' \in \hsp{S}^d$. Thus, $(\{\instr\},\Sigma, \Omega,\hsp{S}^d)$, where $\{\instr\}$ defines a groupoid.

The groupoid can be certainly extended to another algebraic structure $(\{\instr'\},\Sigma,\hsp{H}^d)$, where $\hsp{H}^d$ is a general $2^d$ dimensional Hilbert space $\hsp{H}^d$ and $\{\instr'\}$ is a universal set of quantum gate operations.
Clearly, the following is true: $\hsp{S}^d \subseteq \hsp{H}^d$ and $\{\instr\} \subseteq \{\instr'\}$, because sets $\hsp{H}^d$ and $\{\instr'\}$ can be acquired by removing the well-formed ($\Sigma;\Omega \vdash \varphi$) and well-typed ($\Sigma;\Omega \vdash \instr \triangleright \Omega'$) definitions for $\hsp{S}^d$ and $\{\instr\}$, respectively.
$(\{\instr'\},\Sigma,\hsp{H}^d)$ is a groupoid because every \oqasm operation is valid in a traditional quantum language like \sqir. The following two theorems are to connect \oqasm operations with operations in the general Hilbert space: 

 \begin{theorem}\label{thm:subgroupoid}\rm
   $(\{\instr\},\Sigma, \Omega,\hsp{S}^d) \subseteq (\{\instr\},\Sigma,\hsp{H}^d)$ is a subgroupoid.
 \end{theorem}

\begin{theorem}\label{thm:sem-same}\rm
Let $\ket{y}$ be an abbreviation of $\bigotimes_{m=0}^{d-1} \alpha(r_m) \ket{b_m}$ for $b_m \in \{0,1\}$.
If for every $i\in [0,2^d)$, $\llbracket \instr \rrbracket\ket{y_i}=\ket{y'_i}$, then $\llbracket \instr \rrbracket (\sum_{i=0}^{2^d-1} \ket{y_i})=\sum_{i=0}^{2^d-1} \ket{y'_i}$.
\end{theorem}

The following theorems are proved as corollaries of the compilation correctness theorem from \oqasm to \sqir (\cite{oracleoopsla}). 
\Cref{thm:subgroupoid} suggests that the space $\hsp{S}^d$ is closed under the application of any well-typed \oqasm operation.
\Cref{thm:sem-same} says that \oqasm oracles can be safely applied to superpositions over classical states.\footnote{Note that a superposition over classical states can describe \emph{any} quantum state, including entangled states.}

\begin{figure}[t]
  {\scriptsize
    \begin{mathpar}
      \inferrule[ ]{}{\inot{(x,n)}\xrightarrow{\text{inv}} \inot{(x,n)}}
    
      \inferrule[  ]{}{\texttt{SR}\;m\;x\xrightarrow{\text{inv}} \texttt{SR}^{-1}\;m\;x}
  
      \inferrule[ ]{}{\iqft{n}{x} \xrightarrow{\text{inv}}  \iqft[-1]{n}{x}}   
  
      \inferrule[ ]{}{\texttt{Lshift}\;x\xrightarrow{\text{inv}} \texttt{Rshift}\;x} 
       
      \inferrule[ ]{\instr \xrightarrow{\text{inv}} \instr'}{\texttt{CU}\;(x,n)\;\instr \xrightarrow{\text{inv}} \texttt{CU}\;(x,n)\;\instr'} 
  
      \inferrule[ ]{\instr_1 \xrightarrow{\text{inv}} \instr'_1 \\ \instr_2 \xrightarrow{\text{inv}} \instr'_2}{\instr_1\;;\;\instr_2\xrightarrow{\text{inv}} \instr'_2\;;\;\instr'_1} 
      
    \end{mathpar}
  }
  \caption{Select \oqasm inversion rules}
  \label{fig:exp-reversed-fun}
\end{figure}

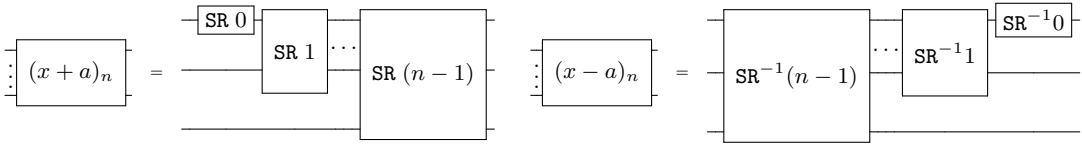
\begin{figure}[t]
{\tiny
\hspace*{-2em}
\centering
\begin{tabular}{c@{$\quad=\quad$}c@{\qquad}c@{$\quad=\quad$}c}
  \begin{minipage}{0.25\textwidth}
  \footnotesize
  \Qcircuit @C=0.25em @R=0.35em {
    & \qw & \multigate{3}{(x+a)_n} & \qw \\
    & \vdots & & \\
    & & & \\
    & \qw & \ghost{(x+a)_n} & \qw \\
    }
  \end{minipage}
&
\begin{minipage}{.45\textwidth}
\footnotesize
  \Qcircuit @C=0.35em @R=0.55em {
     & \qw & \gate{\texttt{SR}\;0} & \multigate{3}{\texttt{SR}\;1} & \qw & \qw & \qw & \multigate{5}{\texttt{SR}\;(n-1)} & \qw  \\
      & & & & & \dots & & &  \\
      & \qw & \qw  &  \ghost{\texttt{SR}\; 1} & \qw & \qw & \qw & \ghost{\texttt{SR}\;(n-1)} & \qw \\
      & & & & & & & &  \\
     & & & & & & & &  \\
   & \qw & \qw & \qw & \qw & \qw & \qw & \ghost{\texttt{SR}\;(n-1)}  & \qw 
    }
\end{minipage}
&  
\begin{minipage}{0.25\textwidth}
  \footnotesize
  \Qcircuit @C=0.25em @R=0.35em {
    & \qw & \multigate{3}{(x-a)_n} & \qw \\
    & \vdots & & \\
    & & & \\
    & \qw & \ghost{(x+a)_n} & \qw \\
    }
  \end{minipage}
&
\begin{minipage}{.45\textwidth}
\footnotesize
  \Qcircuit @C=0.35em @R=0.55em {
    & \qw & \multigate{5}{\texttt{SR}^{-1} (n-1)} & \qw & \qw & \qw & \multigate{3}{\texttt{SR}^{-1} 1} & \gate{\texttt{SR}^{-1} 0} & \qw \\
    &     &                                  &     & \dots &   &                              &                      &   \\
    & \qw & \ghost{\texttt{SR}^{-1} (n-1)}        & \qw & \qw   & \qw & \ghost{\texttt{SR}^{-1} 1} & \qw & \qw  \\
      & & & & & & & &  \\
     & & & & & & & &  \\
    & \qw & \ghost{\texttt{SR}^{-1} (n-1)} & \qw & \qw & \qw & \qw & \qw & \qw 
    }
\end{minipage}
\end{tabular}
}
\caption{Addition/subtraction circuits are inverses}
\label{fig:circuit-add-sub}
\end{figure}

\oqasm programs are easily invertible, as shown by the rules in \Cref{fig:exp-reversed-fun}.
This inversion operation is useful for constructing quantum oracles; for example, the core logic in the QFT-based subtraction circuit is just the inverse of the core logic in the addition circuit (\Cref{fig:exp-reversed-fun}).
This allows us to reuse the proof of addition in the proof of subtraction.
The inversion function satisfies the following properties:

 \begin{theorem}\label{thm:reversibility}\rm[Type reversibility]
    For any well-typed program $\instr$, such that $\Sigma; \Omega \vdash \instr \triangleright \Omega'$, its inverse $\instr'$, where $\instr \xrightarrow{\text{inv}} \instr'$, is also well-typed and $\Sigma;\Omega' \vdash \instr' \triangleright \Omega$. Moreover, $\llbracket \instr ; \instr' \rrbracket \varphi=\varphi$.
 \end{theorem}

\subsection{Explaining the Oracle Expression $\mu$ in Qafny }\label{sec:exp-mu}

The \oqasm expression $\mu$ used in \Cref{fig:vqimp} places an additional wrapper on top of the \oqasm expression $\iota$ given in \Cref{fig:vqir}. The wrapper is given as:

{
\begin{center}
$
\mu \triangleq \lambda \overline{p} .\, \iota
$
\end{center}
}

Here $\overline{p}$ is a list of positions, and $\iota$ is an \oqasm instruction in \Cref{fig:vqir}. Here, we require all qubit positions mentioned in $\iota$ to be in $\overline{p}$. Ranges and loci are directly related to positions. Given a range $x\srange{n}{m}$, it is equal to a list of positions: $(x,n),(x,n\splus 1),...,(x,m\sminus 1)$, while a locus is a list of disjoint ranges and it can be understood as concatenations of different ranges in the locus. For simplicity, we can understand the above $\mu$ definition as $\lambda \kappa .\, \iota$ by rearranging the positions as a locus. 
In the predicate $FV(\Omega,\mu)=\kappa$ in rule \rulelab{TA-CH} in \Cref{fig:exp-sessiontype},
function $FV(\Omega,\mu)$ returns $\kappa$, which means that the $\mu$ lambda expression is given as $\lambda \kappa .\, \iota$.
Given a $\mu$ expression as $\lambda \kappa .\, \iota$, 
to evaluate an oracle expression $\ssassign{\kappa'}{}{\mu}$, we actually do the following:

{
\begin{center}
$
\ssassign{\kappa'}{}{\lambda \kappa . \, \iota} \longrightarrow \denote{\iota[\kappa'/\kappa]}
$
\end{center}
}

Here, we replace the positions $\kappa$ in $\iota$ with qubits mentioned in locus $\kappa'$ and then evaluate the expression $\iota[\kappa'/\kappa]$. In the paper, we also utilize some oracle expressions from OQIMP \cite{oracleoopsla}, the arithmetic operation language based on \oqasm. The expression $ \qass{y\srange{0}{n}\;}{\;y\srange{0}{n}\,\splus\, 1}$ is actually a syntactic sugar for the expression: 
$\qass{y\srange{0}{n}\;}{\;(\lambda u\srange{0}{n} . \,\qass{u\;}{\;u\, \splus\, 1})}$. Here, the expression $\qass{u\;}{\;u\, \splus\, 1}$ is an OQIMP expression, meaning we apply the function $\lambda \,x. \,x\splus 1$ to a qubit array $u$, where $u$ is a $n$-length qubit array.
In addition, the expression $\qass{y\srange{0}{n}\;}{a^{2^j}\cdot y\srange{0}{n}\;\mmod\; N}$ is a syntactic sugar for the expression: $\qass{y\srange{0}{n}\;}{\;(\lambda u\srange{0}{n} . \, \qass{u\;}{\;a^{2^j}\cdot u\;\mmod\; N})}$.

Sometimes, in OQIMP, we will need additional ancillary qubits to guarantee that an arithmetic operation is reversible, a requirement in quantum oracle computation. For example,
if we want to express a modulo expression in OQIMP, we need to write $\lambda u\srange{0}{n}\sqcupplus w\srange{0}{n}.\,\qass{u\;}{\;(u \;\mmod\; m,w)}$, meaning that $u$ and $w$ are $n$-length qubit array. we compute the modulo operation $u \;\mmod\; m$ and put the result in $u$, while $w$ is an additional ancillary qubits required for the computation.
Then, the oracle operation for the modulo function in \qafny is written as:

{
\begin{center}
$
\qass{x\srange{0}{n}\sqcupplus y\srange{0}{n}\;}{\;(\lambda u\srange{0}{n}\sqcupplus w\srange{0}{n}.\,\qass{u\;}{\;(u \;\mmod\; m,w)})}
$
\end{center}
}

In \qafny, the expression computes $x \;\mmod\; m$ and then assigns the value to the range $x\srange{0}{n}$, while $y$ is an array of additional ancillary qubits required for the computation.
In the Shor's example in \Cref{appx:shors}, the oracle application ($\qass{y\srange{0}{n}\;}{\;a^{2^j}\cdot y\srange{0}{n}\;\mmod\; N}$) requires hidden ancillary qubits, but builtin reversed circuits will clean up those qubits in the modulo multiplication itself,
i.e., before the modulo multiplication function ends, all ancillary qubits are reversed to their initial values $\ket{0}$.
All oracle examples in the paper do not have ancillary qubits, or they only have hidden ones that will be cleaned up before the oracle execution ends. Thus, \qafny treats them as nonexistence and leaves them for \oqasm to handle.

 \end{document}